\documentclass[11pt,a4paper]{article}
\pdfoutput=1
\usepackage{amssymb,amsmath,amsfonts, mathtools, mathrsfs}
\usepackage[utf8]{inputenc} 
\usepackage[dvipsnames]{xcolor}

\newif\ifnatbibsort\natbibsorttrue

\relax

\ifnatbibsort\RequirePackage[numbers,sort&compress]{natbib}\else\RequirePackage[numbers,compress]{natbib}\fi
\RequirePackage[colorlinks=true
,urlcolor=blue
,anchorcolor=blue
,citecolor=blue
,filecolor=blue
,linkcolor=blue
,menucolor=blue
,pagecolor=blue
,linktocpage=true
,pdfproducer=medialab
,pdfa=true
]{hyperref}

\usepackage{graphicx}
\usepackage{caption}
\usepackage{tensor}
\usepackage{subcaption}
\usepackage{enumerate}
\usepackage{dsfont}
\usepackage{verbatim}
\usepackage{amsfonts,euscript,amssymb,stmaryrd,braket}
\usepackage{tikz}
\usetikzlibrary{arrows,decorations.markings,patterns}
\usepackage{slashed}

\def\clock{{\count0=\time
		\divide\count0 60
		\ifnum\count0<10 0\fi\the\count0
		\multiply\count0 -60 \advance\count0 \time
		:\ifnum\count0<10 0\fi \the\count0
}}
\newcommand{\timestamp}{{\small\vbox{\hbox{\tt\jobname.tex}
			\hbox{\the\day/\the\month/\the\year, \clock}}}}

\newcommand{\bea}{\begin{eqnarray}}
\newcommand{\eea}{\end{eqnarray}}

\newcommand{\be}{\begin{equation}}
\newcommand{\ee}{\end{equation}}

\makeatletter
\let\old@startsection=\@startsection
\let\oldl@section=\l@section
\renewcommand{\@startsection}[6]{\old@startsection{#1}{#2}{#3}{#4}{#5}{#6\mathversion{bold}}}
\renewcommand{\l@section}[2]{\oldl@section{\mathversion{bold}#1}{#2}}
\makeatother

\numberwithin{equation}{section}

\usepackage{color}

\begin{document}
	\renewcommand{\thefootnote}{\arabic{footnote}}

	\overfullrule=0pt
	\parskip=2pt
	\parindent=12pt
	\headheight=0in \headsep=0in \topmargin=0in \oddsidemargin=0in

	\vspace{ -3cm} \thispagestyle{empty} \vspace{-1cm}
	\begin{flushright} 
		\footnotesize
		\textcolor{red}{\phantom{print-report}}
	\end{flushright}

\begin{center}
	\vspace{1.2cm}
	{\Large\bf \mathversion{bold}
	On shape dependence of holographic entanglement entropy}
	\\
	\vspace{.2cm}
	\noindent
	{\Large\bf \mathversion{bold}
	in AdS$_4$/CFT$_3$ with Lifshitz scaling and hyperscaling violation}

	\vspace{0.8cm} {
		Giacomo Cavini$^{\,a,}$\footnote[1]{cavinig90@gmail.com},
		Domenico Seminara$^{\,a,}$\footnote[2]{seminara@fi.infn.it},
		Jacopo Sisti$^{\,b,}$\footnote[3]{jsisti@sissa.it}
		and Erik Tonni$^{\,b,}$\footnote[4]{erik.tonni@sissa.it}
	}
	\vskip  0.7cm
	
	\small
	{\em
		$^{a}\,$Dipartimento di Fisica, Universit\'a di Firenze and INFN Sezione di Firenze, Via G. Sansone 1, 50019 Sesto Fiorentino,
		Italy
		\vskip 0.05cm
		$^{b}\,$SISSA and INFN Sezione di Trieste, via Bonomea 265, 34136, Trieste, Italy 
	}
	\normalsize
	
\end{center}

\vspace{0.3cm}
\begin{abstract} 
We study the divergent terms and the finite term in the expansion of the holographic entanglement entropy as the ultraviolet cutoff vanishes
for smooth spatial regions having arbitrary shape,
when the gravitational background is a four dimensional asymptotically Lifshitz spacetime with hyperscaling violation,
in a certain range of the hyperscaling parameter.
Both static and time dependent backgrounds are considered. 
For the coefficients of the divergent terms and for the finite term,
analytic expressions valid for any smooth entangling curve are obtained.
The analytic results for the finite terms are checked 
through a numerical analysis focussed on disks and ellipses.
\end{abstract}

\newpage

\tableofcontents

\newpage
\section{Introduction}
\label{intro}

Understanding entanglement in quantum systems is a challenge that has attracted a lot of research
in quantum gravity, condensed matter theory and quantum information during the last decade 
(see e.g. the reviews \cite{Amico:2007ag,Eisert:2008ur,specialissue,Solodukhin:2011gn,Rangamani:2016dms}). 
Furthermore, recently some experimental groups have conducted pioneering experiments 
to capture some features of quantum entanglement \cite{Islam1, Kaufman794, Lukin256}.

The entanglement entropy describes the bipartite entanglement of pure states.
Considering a quantum system whose Hilbert space is bipartite, 
i.e. $\mathcal{H} = \mathcal{H}_A \otimes \mathcal{H}_B$,
and denoting by $\rho$ the state of the whole system, one first defines the reduced density matrix
$\rho_A \equiv \textrm{Tr}_{\mathcal{H}_B} \rho$ on $\mathcal{H}_A $
by tracing out the degrees of freedom corresponding to $\mathcal{H}_B$.
The entanglement entropy is the Von Neumann entropy of $\rho_A$,
namely $S_A \equiv - \textrm{Tr}_{\mathcal{H}_A}(\rho_A \log \rho_A)$ 
\cite{Bombelli:1986rw, Srednicki:1993im, Callan:1994py,Holzhey:1994we,Vidal:2002rm,Calabrese:2004eu}.
Similarly, we can introduce $S_B \equiv - \textrm{Tr}_{\mathcal{H}_B}(\rho_B \log \rho_B)$
for the reduced density matrix $\rho_B \equiv \textrm{Tr}_{\mathcal{H}_A} \rho$ on $\mathcal{H}_B$.
When $\rho= | \Psi\rangle \langle \Psi|$ is a pure state, $S_A = S_B$.
The entanglement entropy satisfies highly non trivial inequalities (e.g. the strong subadditivity conditions). 
In this manuscript we only consider bipartitions of the Hilbert space associated to spatial bipartitions
$A \cup B$ of a constant time slice of the spacetime.

In quantum field theories, a positive and infinitesimal ultraviolet (UV) cutoff is introduced to regularise the divergences
of the model at small distances. 
The entanglement entropy is power like divergent as the UV cutoff vanishes and 
the leading divergence of its series expansion usually scales like the area of the boundary of $A$ (area law of the entanglement entropy).
Nonetheless, in some interesting quantum systems like conformal field theories in one spatial dimension
and $d$ dimensional systems with a Fermi surface, 
a logarithmic violation of this area law occurs \cite{Wolf:2006zzb,Gioev:2006zz}.
Furthermore, many condensed matter systems exhibit a critical behaviour with anisotropic scaling characterised by the Lifshitz exponent $\zeta$ \cite{PhysRevLett.35.1678,PhysRevB.23.4615,PhysRevB.69.224415,PhysRevB.69.224416,Ardonne:2003wa} and hyperscaling violation \cite{Fisher:1986zz}.

In this manuscript we are interested to explore some aspects of the entanglement entropy in quantum gravity models
in the presence of Lifshitz scaling and hyperscaling violation. 
The most developed approach to quantum gravity is based on the  AdS/CFT correspondence, 
where a string theory defined in a $(d+1)$ dimensional asymptotically Anti de Sitter (AdS$_{d+1}$)
spacetime is related through a complicated duality to a $d$ dimensional Conformal Field Theory (CFT$_{d}$) on the boundary 
of the gravitational asymptotically AdS spacetime \cite{Maldacena:1997re,Witten:1998qj,Gubser:1998bc,Aharony:1999ti}. 
This duality is formulated in general dimensions and each dimensionality has peculiar features. 
In this manuscript we consider the case of AdS$_4$/CFT$_3$.
We mainly employ Poincar\'e coordinates to describe the gravitational spacetimes:
denoting by $z$ the holographic coordinate,  the boundary of the gravitational spacetime 
is identified by $z=0$ and the points in the bulk have $z>0$.
According to the holographic dictionary, the gravitational dual of the UV
cutoff of the CFT is an infinitesimal cutoff $\varepsilon$ in the holographic direction, 
namely $z\geqslant \varepsilon>0$.
Within the AdS/CFT correspondence,
gravitational backgrounds capturing the anisotropic Lifshitz scaling and the hyperscaling violation
have been introduced in \cite{Kachru:2008yh,Balasubramanian:2008dm, Taylor:2008tg}
and in \cite{Charmousis:2010zz,Gouteraux:2011ce,Huijse:2011ef,Ogawa:2011bz,Dong:2012se} respectively.


\begin{figure}[t!]
	\vspace{-.2cm}
	\hspace{-1.1cm}
	\begin{center}
		\includegraphics[width=1\textwidth]{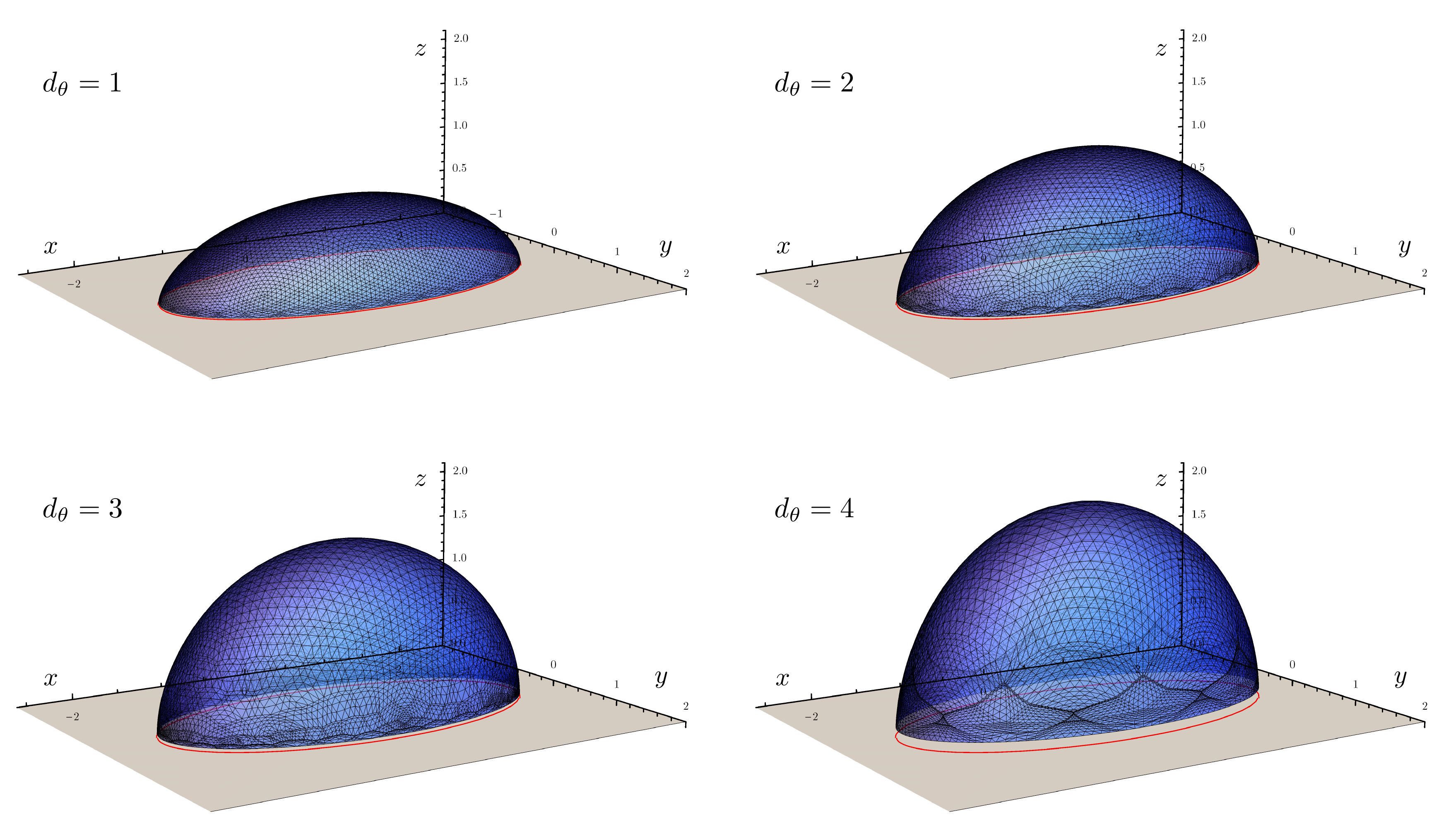}
	\end{center}
	\vspace{-.0cm}
	\caption{\small
		Minimal area surface obtained with Surface Evolver whose area provides the holographic entanglement entropy 
		of an ellipse $A$ delimited by the red curve. 
		The minimal surface is embedded in a constant time slice of the 
		four dimensional hyperscaling violating Lifshitz spacetime (\ref{hyperscaling4}),
		whose metric depends on the hyperscaling parameter $d_\theta$.
	}
	\label{fig:intro3Dplot}
\end{figure}

A fundamental result in the ongoing construction of the holographic dictionary 
is the Ryu-Takayanagi formula, that provides the gravitational prescription
to compute the leading order (in the large $N$ expansion) of the entanglement entropy
of a spatial region $A$ in the dual CFT \cite{Ryu:2006bv,Ryu:2006ef}. 
Given a spatial bipartition $A \cup B$ of a constant time slice of the static spacetime where the CFT is defined, 
the holographic entanglement entropy is
\be
\label{RTformula}
S_A = \frac{\mathcal{A}[ \hat{\gamma}_{A,\varepsilon}]}{4 G_{\textrm{\tiny N}}}
\ee
where $\mathcal{A}[ \hat{\gamma}_{A,\varepsilon}]$ is the area 
of the codimension two hypersurface $ \hat{\gamma}_{A,\varepsilon}$ 
obtained by restricting to $ z\geqslant \varepsilon$
the minimal area hypersurface $\hat{\gamma}_{A}$ on a constant time slice
anchored to $\partial A$ (often called entangling hypersurface).
The covariant generalisation of (\ref{RTformula}) 
has been introduced by Hubeny, Rangamani and Takayanagi \cite{Hubeny:2007xt}
and it requires to extremise the area of the codimension two hypersurfaces 
$\gamma_A$ constrained only by the condition $\partial \gamma_A = \partial A$.
%
These prescriptions for the holographic entanglement entropy satisfy 
the strong subadditivity property \cite{Headrick:2007km,Wall:2012uf}.
The covariant formula allows to study the temporal evolution of holographic entanglement entropy 
in time dependent gravitational backgrounds, like the ones describing the formation of black holes.
For instance, the Vaidya metrics provide simple models for the black hole formation
where the holographic entanglement entropy has been largely studied
\cite{AbajoArrastia:2010yt,Balasubramanian:2011ur,Balasubramanian:2011at,Allais:2011ys,Callan:2012ip,Liu:2013iza,Liu:2013qca,Hubeny:2013hz}.

The holographic entanglement entropy formula (\ref{RTformula}) satisfies interesting properties
that have been deeply explored during the last decade (see e.g. \cite{Hayden:2011ag, Hubeny:2013gta, Hubeny:2013gba, Headrick:2014cta})
in order to identify some constraints for the CFTs having a holographic dual description. 
For instance, whenever $A$ is made by two or more disjoint regions,
a characteristic feature of the holographic entanglement entropy is the occurrence of 
transitions between different types of surfaces providing the extremal area configuration
\cite{Hubeny:2007re,Tonni:2010pv,Headrick:2010zt}.
These transitions occur in the regime of classical gravity and they are smoothed out by quantum corrections 
\cite{Faulkner:2013ana}.
Indeed, they have not been observed e.g. for the entanglement entropy of disjoint intervals
in some CFT$_2$ models with central charge bigger than 1.
\cite{Calabrese:2009ez,Calabrese:2010he,Coser:2013qda,DeNobili:2015dla,Coser:2015dvp}.

A riformulation of the holographic entanglement entropy formula (\ref{RTformula}) 
has been recently proposed through particular flows \cite{Freedman:2016zud}
and exploring the various features of the holographic entanglement entropy
through this approach is very insightful \cite{Headrick:2017ucz,Cui:2018dyq}.

The quantitative analysis of the dependence of the holographic entanglement entropy  (\ref{RTformula})
on the shape of the region $A$ is an important task that is also very difficult 
whenever the shape of $A$ does not display particular symmetries 
\cite{Graham:1999pm, Solodukhin:2008dh,Hubeny:2012ry,Astaneh:2014uba}.
For this reason, spheres or infinite strips are usually considered because 
the symmetry of these regions allows to obtain analytic results or to make the numerical analysis easier.
Analytic results for domains with generic smooth shapes
have been found for the divergent terms in the expansion of the holographic entanglement entropy
as the UV cutoff vanishes. The divergent terms depend only on the part of the 
minimal hypersurface $\hat{\gamma}_{A}$ close to the conformal boundary.

In AdS$_4$/CFT$_3$, analytic results for generic smooth shapes have been obtained also for the finite term, 
which depends on the entire minimal surface $\hat{\gamma}_{A}$. 
These results are based on the Willmore functional in AdS$_4$ \cite{Babich,Mazzeo}  
and on a more general functional in asymptotically AdS$_4$ spacetimes  \cite{Fonda:2015nma}.
The shape dependence of the holographic entanglement entropy in AdS$_4$/CFT$_3$
has been studied also numerically in \cite{Fonda:2014cca,Fonda:2015nma} 
by employing the software {\it Surface Evolver}, developed by Ken Brakke \cite{brakke,brakke2}.

When the dual CFT has a physical boundary and proper boundary conditions are imposed,
we have a Boundary Conformal Field Theory (BCFT) \cite{Cardy:1986gw, Cardy:1989ir, Cardy:2004hm}
and a holographic duality (AdS/BCFT correspondence) for these models has been studied in  
\cite{Takayanagi:2011zk,Fujita:2011fp,Nozaki:2012qd}.
In AdS$_4$/BCFT$_3$, both analytic and numerical results have been obtained for 
the holographic entanglement entropy of regions with generic shape 
\cite{Seminara:2017hhh,Seminara:2018pmr}
by extending the above mentioned methods developed for AdS$_4$/CFT$_3$.


Gravitational backgrounds depending on the Lifshitz scaling and on the hyperscaling violation exponents have been largely explored
\cite{Goldstein:2009cv,Gubser:2009qt,Iizuka:2011hg,Narayan:2012hk,Ammon:2012je,Bhattacharya:2012zu,Alishahiha:2012qu,Bueno:2012sd,Gath:2012pg,Gouteraux:2012yr,Christensen:2013rfa,Christensen:2013lma,Hartong:2016nyx}.
The holographic entanglement entropy has been also studied, 
both in static backgrounds \cite{Ogawa:2011bz,Dong:2012se, Tonni:2010pv, Shaghoulian:2011aa, Alishahiha:2015goa} 
and in Vaidya spacetimes \cite{Keranen:2011xs,Liu:2013iza,Liu:2013qca,Alishahiha:2014cwa,Fonda:2014ula}.
We remark that spherical regions and infinite strips are the only smooth regions considered in these studies.


In this manuscript we explore the shape dependence of 
the holographic entanglement entropy in  four dimensional gravitational backgrounds 
having a non trivial Lifshitz scaling (characterised by the parameter $\zeta$)
and a hyperscaling violation exponent $\theta$
(we find it more convenient to employ the parameter $d_{\theta}\equiv d-1-\theta$).

Our analysis is restricted to $d=3$ and holds for smooth entangling curves $\partial A$,
which can be also made by disjoint components.
We consider $1\leqslant  d_\theta \leqslant 5$ for the sake of simplicity,
although the method can be adapted to higher values of $d_\theta$.
We study both the divergent terms and the finite
term in the expansion of the holographic entanglement entropy as $\varepsilon \to 0$.
Both analytic results and numerical data will be presented. 
For instance,  in Fig.\;\ref{fig:intro3Dplot} we show the minimal area surface obtained with Surface Evolver
whose area provides the holographic entanglement entropy of an elliptic region through (\ref{RTformula}), 
in the case where the gravitational background is a constant time slice of the 
four dimensional hyperscaling violating Lifshitz spacetime (\ref{hyperscaling4}),
whose geometry is characterised only by the hyperscaling parameter $d_\theta$.


The manuscript is organised as follows. 
The main results about the finite term in the expansion of the holographic entanglement entropy as $\varepsilon \to 0$
for a generic static gravitational background are presented in Section\;\ref{sec HEE},
where also some important special cases like the 
four dimensional hyperscaling violating Lifshitz spacetime (hvLif$_4$) defined in (\ref{hyperscaling4})
and the asymptotically hvLif$_4$ black hole are explicitly discussed. 
In Section\;\ref{sec: integral along entangling curve} we show that  the finite term in the expression for the area of a minimal
submanifold  
anchored 
on the boundary  reduces to an integral over 
their intersection when the bulk geometry possesses a conformal Killing vector generating dilatations.
In Section\;\ref{sec:time-dependent} we study the finite term of the holographic entanglement entropy for
time dependent backgrounds having $1< d_\theta < 3$.
In Section\;\ref{sec:particular regions} we discuss explicitly the infinite strip,
the disk and the ellipse. 
Some conclusions are drawn in Section\;\ref{sec:conclusions}.
In Appendices\;\ref{app:NEC}, \ref{appxUVbehav}, \ref{appxderivaionfa2}, \ref{app:integral entangling curve}, \ref{Apptime} 
and \ref{appendix_hard_wall_solution}
we provide the technical details underlying the results presented in the main text.


 \section{Holographic entanglement entropy in asymptotically hvLif$_4$ backgrounds}
\label{sec HEE}
In this manuscript  we consider four dimensional gravitational backgrounds $ \mathcal{M}_{4}$
that depend on the hyperscaling violation exponent $\theta$ 
and on the Lifshitz scaling exponent $\zeta\geqslant 1$. 
In Poincar\'e coordinates where $z>0$ denotes the holographic coordinate, 
these backgrounds have a boundary at $z=0$ and their asymptotic behaviour
as $z \to 0^+$ is given by the following metric, that defines the 
four dimensional hyperscaling violating Lifshitz spacetimes (hvLif$_4$) \cite{Gouteraux:2011ce,Huijse:2011ef,Dong:2012se}
\begin{equation}
\label{hyperscaling4}
ds^{2}=\dfrac{R^{d_\theta}_{\text{\tiny AdS}}}{z^{d_{\theta}}} 
\left(-\frac{z^{-2(\zeta-1)}}{R_{\text{\tiny AdS}}^{-2(\zeta-1)}}\,dt^{2} +dz^{2}+d\boldsymbol{x}^2 \right)
\end{equation}
where $d\boldsymbol{x}^2 \equiv dx^{2}+dy^{2}$ and $ d_{\theta}\equiv 2-\theta $.
The length scale  $R_{\text{\tiny AdS}}$ is the  analog of the AdS radius.
The spacetime (\ref{hyperscaling4}) is a solution of the equations of motion coming from a gravitational action 
containing gauge fields and a dilaton field \cite{Charmousis:2010zz}. 
When $d_\theta=2$ and $\zeta=1$, the background (\ref{hyperscaling4}) becomes AdS$_4$ in Poincar\'e coordinates.
In this manuscript we set $R_{\text{\tiny AdS}}$ to one for simplicity, 
although it plays a crucial role in the dimensional analysis.

In order to deal only with  geometries admitting physically sensible dual field theories,  
the allowed values of the parameters in \eqref{hyperscaling4} must satisfy some constraints on the putative energy momentum tensor 
computed via Einstein equations\footnote{In general $\Lambda= -d(d-1)/(2R^2_{\textrm{\tiny AdS}})$ in $d+1$ dimensional spacetimes. 
Here $d=3$; hence $\Lambda=-3/R^2_{\textrm{\tiny AdS}}$.}
$G_{MN}-\Lambda g_{MN}=T_{MN}$.
In particular the Null Energy Condition (NEC)\footnote{The NEC is insensible to the cosmological constant; 
indeed for a null vector $G_{MN}V^M V^N=T_{MN}V^M V^N$.} is required, 
namely $T_{MN}V^M V^N\geqslant 0$  for any (future directed) null vector $V^M$.  
The NEC  translates into the following constraints for $d_\theta$ and  $\zeta$ \cite{Dong:2012se}
\be
\label{NEC}
\left\{\begin{array}{l}
(d_{\theta}+\zeta)(\zeta-1)\geqslant 0 \\
\rule{0pt}{.5cm}
d_{\theta}(d_{\theta}+2\zeta-4)\geqslant 0\,. \\
\end{array} \right.
\ee
We refer to Appendix\,\ref{app:NEC} for a detailed discussion of the NEC and its consequences.

In this section we focus on static backgrounds;  
hence we can restrict our attention to the three dimensional Euclidean section 
$\mathcal{M}_{3}$  obtained by taking a constant time slice  of the asymptotically hvLif$ _{4} $ bulk manifold $ \mathcal{M}_{4}$.  
This submanifold is naturally endowed with a metric $ g_{\mu\nu} $ such that
\begin{equation}
\label{statichyperscaling}
\left. ds^{2}\right|_{t=\mathrm{const}} \equiv \,
g_{\mu\nu} \, dx^\mu dx^\nu 
\;\;\xrightarrow{z\,\to\, 0}\;\;
\dfrac{1}{z^{d_{\theta}}} \big(dz^{2}+dx^{2}+dy^{2} \big)\,.
\end{equation}

Given a two dimensional spatial region $A$ in a constant time slice of the CFT$_3$ at $z=0$, 
its holographic entanglement entropy is given by (\ref{RTformula}).
Thus, first we must consider the class of two dimensional surfaces $\gamma_A$ 
embedded in $\mathcal{M}_{3} $ whose boundary curve belongs to the plane $ z=0 $ 
and coincides with the entangling curve, i.e. $ \partial\gamma_A = \partial A$. 
Then, among these surfaces, we have to find the one having minimal area, that provides 
the holographic entanglement entropy according to the formula (\ref{RTformula}).
We will denoted by $\hat\gamma_A$ the extremal surfaces of the area functional,
without introducing a particular notation for the global minimum.

Considering the unit vector $ n^{\mu} $ normal to $\gamma_A$, 
the induced metric $h_{\mu\nu}$ on $\gamma_A $ and the extrinsic curvature $K_{\mu\nu}$ 
are given in terms of $n_\mu$ respectively by 
\begin{equation}
\label{fundamentalform}
h_{\mu\nu}=g_{\mu\nu}-n_{\mu}n_{\nu} 
\;\;\qquad \;\;
K_{\mu\nu}=\tensor{h}{_\mu^\alpha}\tensor{h}{_\nu^\beta}\,\nabla_{\alpha}n_{\beta}
\end{equation}
being $ \nabla_{\alpha} $ the torsionless covariant derivative compatible with $ g_{\mu\nu}$.

In our analysis, we find  convenient to introduce an auxiliary conformally equivalent three dimensional space
$\widetilde{\mathcal{M}}_{3}$ given by $\mathcal{M}_3$ with the same boundary at $z=0$, but equipped 
with the metric $ \tilde{g}_{\mu\nu}$, 
which is asymptotically flat as $z\to 0$ and Weyl related to $g_{\mu\nu}$, i.e.
\begin{equation}
\label{varphi def}
g_{\mu\nu}=e^{2\varphi}\,\tilde{g}_{\mu\nu}
\end{equation}
where $ \varphi $ is a function of the coordinates. 
The surface $ \gamma_A $ can be also viewed as a submanifold of $ \widetilde{\mathcal{M}}_{3} $. 
Denoting by $ \tilde{n}_{\mu} $ the unit normal vector to $ \gamma_A$ embedded in  
$\widetilde{\mathcal{M}}_{3}$, it is straightforward to find that $ n_{\mu}=e^{\varphi}\tilde{n}_{\mu}$. 
The first and second fundamental form $ \tilde{h}_{\mu\nu} $ and $ \widetilde{K}_{\mu\nu} $ \
of $ \gamma_A\subset\widetilde{\mathcal{M}}_{3} $ can be written in terms of the same quantities for
$ \gamma_A\subset \mathcal{M}_{3} $ (defined in (\ref{fundamentalform})) as follows
\begin{equation}
\label{weylexcurvature}
h_{\mu\nu}=e^{2\varphi}\,\tilde{h}_{\mu\nu} 
\;\;\qquad \;\;
K_{\mu\nu}=e^{\varphi}\bigl(\widetilde{K}_{\mu\nu}+\tilde{h}_{\mu\nu}\tilde{n}^{\lambda}\partial_{\lambda}\varphi\bigr)\,.
\end{equation}
The two induced area elements $ d\mathcal{A}=\sqrt{h}\,d\Sigma $ (of $ \gamma_A\subset \mathcal{M}_{3}$) and $d\tilde{\mathcal{A}}=\sqrt{\tilde{h}}\, d\Sigma $ (of $ \gamma_A\subset \widetilde{\mathcal{M}}_{3} $),
where $ d\Sigma$ is  a shorthand notation for $d\sigma_{1}d\sigma_{2}$ with $ \sigma_{i} $ some local coordinates on $ \gamma_A $,
are related as $ d\mathcal{A}=e^{2\varphi}d\tilde{\mathcal{A}}$.

Since  $ \gamma_A\subset \mathcal{M}_{3}$ extends up to the boundary plane at $ z=0,$  its  area  functional
\begin{equation}
\label{areafunctional}
\mathcal{A}[\gamma_A]\,
=
\int_{\gamma_A} \!\! \sqrt{h}\, d\Sigma
\end{equation} 
diverges when $ d_{\theta}\geqslant 1 $ 
because of the behaviour \eqref{statichyperscaling} 
near the conformal boundary.
%
%
The holographic entanglement entropy (\ref{RTformula}) is proportional to the area of the 
global minimum among the local extrema $\hat \gamma_A$ of  \eqref{areafunctional} anchored to the entangling curve $\partial A$.
These surfaces are obtained by solving the condition  of vanishing mean curvature
\begin{equation}
\label{extremalcondit}
\textrm{Tr} K=0
\end{equation}
with the Dirichlet boundary condition  $\partial\gamma_A=\partial A$. 
In terms of the second fundamental form defined by the embedding in $\widetilde{\mathcal{M}}_3$, 
the extremal area condition \eqref{extremalcondit} reads
\be
\label{traceminimalcondition}
\textrm{Tr} \widetilde{K}
=
-\,2 \,\tilde n^\lambda \partial_\lambda \varphi
\qquad \Longleftrightarrow \qquad
\textrm{Tr}\widetilde{K}
=
d_{\theta}\,\dfrac{\tilde{n}^{z}}{z}
\ee
where in the last step we choose $e^{2 \varphi } = 1/z^{d_\theta}$, as suggested by the asymptotic form (\ref{statichyperscaling}).
%


\subsection{Divergent terms}
\label{sec:UVstructure}

In our analysis we consider only smooth entangling curves $\partial \gamma_A$.
Furthermore, we restrict to two dimensional surfaces $\gamma_A$ that
intersect  orthogonally the spatial boundary at $z=0$ of $\mathcal{M}_3$;
and the extremal surfaces $\hat{\gamma}_A$ anchored to smooth entangling curves 
enjoy this property. 
In the following we discuss  the divergent contributions 
in the expansion of the holographic entanglement entropy (\ref{RTformula}) as $\varepsilon \to 0$.


Since $\gamma_A$ reaches the boundary and $d_\theta \geqslant 1$,
its area is divergent; hence we have to introduce a UV cutoff plane at $z=\varepsilon$
and evaluate the functional \eqref{areafunctional} on the part of $\gamma_A$ above the cutoff plane,
i.e. on $\gamma_{A,\varepsilon} \equiv \gamma_A \cap \{z \geqslant \varepsilon\}$.
The series expansion of $\mathcal{A}[\gamma_{A,\varepsilon}]$ as $\varepsilon \to 0$
contains divergent terms, a finite term and vanishing terms as $\varepsilon \to 0$.
By exploiting the techniques discussed in \cite{Babich, Graham:1999pm,Mazzeo} 
in  Appendix\;\ref{appxUVbehav} we study the surface $ \gamma_{A,\varepsilon}$,
singling out the structure of the divergences in the expansion of $\mathcal{A}[\gamma_{A,\varepsilon}]$ as $\varepsilon\to 0$.
In the following we report only the results of this analysis.
Let us stress that some of these results hold also for surfaces $\gamma_A$ that are not minimal.

The leading divergence of $\mathcal{A}[\gamma_{A,\varepsilon}]$ as $\varepsilon\to 0$ is given by
\begin{equation}
\label{divercont1}
\mathcal{A}[ \gamma_{A,\varepsilon}]=\dfrac{P_{A}}{(d_{\theta}-1)\,\varepsilon^{d_{\theta}-1}}
+\dots 
\;\; \qquad \;\;
d_\theta \ne 1
\end{equation}
where $P_A$ is  the perimeter of the entangling curve $\partial A$, as pointed out in \cite{Dong:2012se, Ogawa:2011bz, Huijse:2011ef}.
This leading divergence provides the area law of the holographic entanglement entropy 
for the asymptotically hvLif$_4$ backgrounds. 
When $d_{\theta} =1 $,  the leading divergence is logarithmic 
\begin{equation}
\label{leadinglog}
\mathcal{A}[ \gamma_{A,\varepsilon}]=P_{A}\log (P_A / \varepsilon)+O(1)
\;\; \qquad \;\;
d_\theta=1\,.
\end{equation}
The apparent dimensional mismatch between the two sides of \eqref{leadinglog} is due to our choice to set $R_{\textrm{\tiny AdS}}=1$.
The subleading terms in these expansions depend on the value of $d_\theta$
and we find it worth considering the ranges given by $2n+1<d_\theta<2n+3$, 
being $n \geqslant 0$ a positive integer. 
When $1<d_\theta<3$, after the leading divergence (\ref{divercont1}), a finite term occurs
\begin{equation}
\label{leadingdiver}
\mathcal{A}[\gamma_{A,\varepsilon}]
=
\frac{P_{A}}{(d_{\theta}-1)\,\varepsilon^{d_{\theta}-1}}- \mathcal{F}_{A}+O(\varepsilon) 
\;\;\qquad \;\;
1<d_\theta<3\,.
\end{equation}

At this point, let us restrict our analysis to extremal surfaces $\hat{\gamma}_A$.
When $\gamma_A = \hat\gamma_A$ is the minimal surface, 
in (\ref{leadingdiver}) we adopt the notation $\mathcal{F}_{A}=F_A$ for the finite term
(see Section\;\ref{fine_term}).

When $ d_{\theta}=3 $, the subleading term diverges logarithmically \cite{Dong:2012se, Ogawa:2011bz, Huijse:2011ef}.
In particular, for a generic smooth entangling curve we find
\begin{equation}
\label{2divlog}
\mathcal{A}[\hat \gamma_{A,\varepsilon}]
= 
\frac{P_{A}}{2\varepsilon^{2}}+\frac{\log\varepsilon}{8}\int_{\partial A} \!\! k^{2}(s)\,ds +O(1) 
\;\;\qquad \;\;
d_{\theta}=3
\end{equation}
where $k(s)$ is the geodesic curvature of $\partial \hat\gamma_A$ and $s$ parameterises the entangling curve. 
When $A$ is a disk of radius $R$, the geodesic curvature $k(s)=1/R$ is constant,
and the coefficient of the logarithmic divergence for this region has been considered also in \cite{Fonda:2014ula}.

In the range  $ 3<d_{\theta}<5 $, the subleading divergence is power like;
hence the finite term $\mathscr{F}_A$ is not changed by a global rescaling of the UV cutoff.  
The expansion of the area of $\hat\gamma_{A,\varepsilon}$ reads
\begin{equation}
\label{2div}
\mathcal{A}[\hat\gamma_{A,\varepsilon}]
=
\dfrac{P_{A}}{(d_{\theta}-1)\varepsilon^{d_{\theta}-1}}+\dfrac{C_A}{\varepsilon^{d_{\theta}-3}} - \mathscr{F}_A+ O(\varepsilon)  
\;\;\qquad \;\;
 3<d_{\theta}<5
\end{equation}
where the coefficient $ C_A $ is given by
	\begin{equation}
\label{coeff_second _div}
C_A
=-\,
\frac{(d_{\theta}-2)}{2(d_{\theta}-3)(d_{\theta}-1)^{2}}\int_{\partial A}\!\!\! k^{2}(s)\,ds\,.
\end{equation}

For $d_\theta=5$, a finite term in the expansion as $\varepsilon \to 0$ is not well defined because 
a logarithmic divergence occurs. In particular, we obtain
\begin{equation}
\label{log_div_5}
\mathcal{A}[\hat\gamma_{A,\varepsilon}]
=
\frac{P_A}{4 \varepsilon^4}
-\frac{3}{64 \varepsilon^2}
\int_{\partial A}\!\! k(s)^2\, ds
+\frac{\log \varepsilon}{2048}\, \int_{\partial A} \! \Big( 9 \,k(s)^4-16 \,k'(s)^2\Big) ds  +O(1)\,.
\end{equation}
The pattern outlined above seems to repeat also for higher values of $d_\theta$:
when $ d_{\theta}=2n+1$ is an odd integer with $n\geqslant 0$,
one finds power like divergences $O(1/\varepsilon^{2n- 2k})$ 
with integer $k \in [0,n-1]$
and a logarithmic divergence.
Instead, in the range $2n+1<d_\theta<2n+3$ only power like divergencies 
$O(1/\varepsilon^{d_\theta-1- 2k})$ with integer  $k \in [0,n]$ occur.

In Appendix\,\ref{appxUVbehav} we provide the derivations of the results reported above 
and we also discuss their extensions to the class of surfaces
that intersect orthogonally the boundary plane at $z=0$,
which includes the extremal surfaces.

 \subsection{Finite term}
 \label{fine_term}

 In this subsection we investigate the  finite term  in \eqref{leadingdiver}
 for surfaces $\gamma_A$ that can be also non extremal
 and in \eqref{2div} only for  $\hat{\gamma}_A$.
 The main result of this manuscript is their expression
 as (finite) geometrical  functionals over the two dimensional surface $\gamma_A$ 
 (or  $\hat{\gamma}_A$ for $\mathscr{F}_A$) 
 viewed as a submanifold of $\widetilde{\mathcal{M}}_3$.
The procedure to obtain the finite terms extends the one developed 
in \cite{Babich, Mazzeo} for AdS$_4$ and in \cite{Fonda:2015nma} for asymptotically AdS$_4$ spacetimes. 
Since the specific details of this analysis depend on the type of divergences occurring in the expansion of the area functional as $\varepsilon \to 0$, we will treat the regimes $ 1<d_{\theta}<3 $ and  $ 3<d_{\theta}<5 $ separately.
In the following we report only the main results, collecting
all the technical details of their derivation in Appendix\,\ref{appxderivaionfa2}.

 When $1<d_\theta<3$,  the only divergence in the  expansion of area functional $\mathcal{A}[\gamma_{A,\varepsilon}]$
 is the area law term \eqref{divercont1};
 hence our goal is to write an expression for the finite term $\mathcal{F}_{A}$ in (\ref{leadingdiver}).
In Appendix\;\ref{appendix_1-3} we adapt  the analysis performed in  \cite{Fonda:2015nma} to this case, finding
\bea
 \label{finitegeneralsurface}
 \mathcal{F}_{A}
 &=&
 \frac{2}{d_{\theta}(d_{\theta}\!-\!1)}
  \int_{\gamma_A}
 \hspace{-.2cm}
  e^{2\phi}
  \Bigl(2\tilde{h}^{\mu\nu}\partial_{\nu}\phi \,\partial_{\mu}\varphi\!-\!
  \dfrac{d_{\theta}(d_{\theta}\!-\!1)}{2}\,e^{2(\varphi-\phi)}\!
  +\!\widetilde{\nabla}^{2}\varphi\!-\!\tilde{n}^{\mu}\tilde{n}^{\nu}\widetilde{\nabla}_{\mu}\widetilde{\nabla}_{\nu}\varphi\!+\!(\tilde{n}^{\lambda} \partial_{\lambda} \varphi)^{2} \Bigr)d \tilde{\mathcal{A}}
  \nonumber
  \\
  \rule{0pt}{.7cm}
  & &
 + \,  \frac{1}{2\,d_{\theta}(d_{\theta}\!-\!1)}
 \left[\,\int_{\gamma_A }e^{2\phi}\big(\textrm{Tr} \widetilde{K}\big)^{2}d\tilde{\mathcal{A}}
\,+\!
 \int_{\gamma_A} e^{2\phi} \big(\textrm{Tr} K\big)^{2}d\mathcal{A}
 \,\right]
\eea
where $\varphi$ is the same conformal factor defined in \eqref{varphi def}, while $\phi$ is chosen so that
$e^{-2\phi} g_{\mu\nu}$ is asymptotically AdS$_4$.
 In our explicit calculations we have employed the simplest choice for 
 $\varphi$ and $\phi$, namely 
 $\varphi = -\tfrac{d_\theta}{2} \log z$ and $\phi = \tfrac{2-d_\theta}{2} \log z$.

 In the special case of $d_\theta=2$,  the field $\phi$ can be chosen to vanish (see \eqref{alpha})
 and this leads us to recover the result obtained in \cite{Fonda:2015nma} as a special case of our analysis.

When the functional (\ref{finitegeneralsurface}) is evaluated on an extremal surfaces $\hat\gamma_A$,
the forms \eqref{extremalcondit} and \eqref{traceminimalcondition} of the extremality condition
imply respectively  that the last term in (\ref{finitegeneralsurface}) does not occur
and that the term containing $(\tilde{n}^{\lambda} \partial_{\lambda} \varphi)^{2}$ 
can be written in terms of $(\textrm{Tr} \widetilde{K})^{2}$.
Finally we can write
 \bea
 \label{finite term}
 F_{A}
&=&
 \frac{2}{d_{\theta}(d_{\theta} -1)}
  \int_{\hat\gamma_A}
\!\!
  e^{2\phi}
 \biggl(
 2\,\tilde{h}^{\mu\nu}\partial_{\nu}\phi \,\partial_{\mu}\varphi
 +\widetilde{\nabla}^{2}\varphi
 -\tilde{n}^{\mu}\tilde{n}^{\nu}\, \widetilde{\nabla}_{\mu}\widetilde{\nabla}_{\nu}\varphi
 \\
 & & \hspace{5cm}
 -\,\dfrac{d_{\theta}(d_{\theta} -1)}{2}\;e^{2(\varphi-\phi)}
 + \frac{1}{2}(\textrm{Tr} \widetilde{K})^{2}
 \biggr)\, d \tilde{\mathcal{A}}\,.
 \nonumber
\eea

 The regime  $3<d_\theta<5$ is more challenging because the expansion of the area functional
 $\mathcal{A}[\hat\gamma_{A,\varepsilon}]$ as $\varepsilon \to 0$
 contains two power like divergent terms (see \eqref{2div}).
 Let us remind that the structure of this expansion is dictated by the geometry of  the entangling curve 
 only for extremal surfaces 
 (in this case the coefficient of the subleading divergent term is \eqref{coeff_second _div}).
For non extremal surfaces the structure of the divergent terms does not depend only on the geometry of the entangling curve,
but also on the  surface (see e.g.\,\eqref{A_exp_ortog}).

In Appendix\;\ref{app_deriv_finte_3-5} we find that the finite term in (\ref{2div}) for minimal surfaces reads
\be
 \label{finiteminsurf2}
\mathscr{F}_A=F_{A}
+\dfrac{2}{d_{\theta}^{3}(d_{\theta}-3)(d_{\theta}-1)} 
\int_{\hat\gamma_{A}} \!\! e^{2\psi}\Bigl((\textrm{Tr} \widetilde{K})^{2} f
-\tilde{h}^{\mu\nu}\partial_{\nu}\varphi \,\partial_{\mu}(\textrm{Tr} \widetilde{K})^2\Bigr)d\tilde{\mathcal{A}}
\ee
being
\be
 \label{f}
 f= 
 \tilde{n}^{\mu}\tilde{n}^{\nu} \, \widetilde{\nabla}_{\mu}\widetilde{\nabla}_{\nu}\varphi
 -\widetilde{\nabla}^{2}\varphi-2(\tilde{n}^{\lambda} \partial_{\lambda} \varphi)^{2}
 -2\tilde{h}^{\mu\nu}\partial_{\mu}\psi \, \partial_{\nu}\varphi
\ee
where $F_A$ is defined in \eqref{finite term}. 
In \eqref{finiteminsurf2} we have introduced a third conformal factor
$e^{2\psi}$ that scales as $z^{4-d_\theta}$ when we approach the boundary  at $z=0$. 
The scaling of  $e^{2\psi}$  with $z$ (for small $z$) is fixed by requiring that  the boundary terms  in  \eqref{pinodaniele}  match the divergence of order $1/\varepsilon^{d_\theta-3}$ appearing in \eqref{2div} (see  \eqref{boundaryterm1} and
\eqref{beta} for details). 

\subsection{HvLif$_{4}$}
\label{sec vacuum}

The simplest gravitation geometry to consider is hvLif$_{4}$, whose metric reads
\begin{equation}
\label{hyperscaling4bis}
ds^{2}=\dfrac{1}{z^{d_{\theta}}} \left(-\,z^{-2(\zeta-1)}dt^{2} +dz^{2}+dx^{2}+dy^{2}\right)
\end{equation}
namely (\ref{hyperscaling4}) with the length scale $R_{\text{\tiny AdS}}$ set to one.
In this background $\widetilde{g}_{\mu\nu} =\delta_{\mu\nu}$; hence
the general formulae \eqref{finite term} and  \eqref{finiteminsurf2} 
take a compact and elegant  form.
In Appendix\;\ref{hvLif} some details about these simplifications are provided.



When $1<d_\theta<3$, 
the expression \eqref{finite term} reduces to 
\begin{equation}
\label{Fa1}
F_{A}=\dfrac{1}{d_{\theta}-1}\int_{\hat\gamma_{A}}
\! \! \frac{(\tilde{n}^{z})^{2}}{z^{d_{\theta}}}\;d\tilde{\mathcal{A}}
\end{equation}
where we remind that $\tilde n^z$ is the $z$-component of the normal vector to $\hat\gamma_A$ in $\widetilde{\mathcal{M}}_3$. 
By employing the extremality condition (\ref{traceminimalcondition}), one can write $F_{A}$   in terms 
of the second fundamental form in $\widetilde{\mathcal{M}}_3$ as follows
\begin{equation}
\label{Fa2}
F_{A}=
\frac{1}{d_{\theta}^{2}(d_{\theta}-1)}\int_{\hat\gamma_{A}}\!\!
\frac{(\textrm{Tr} \widetilde{K})^{2}}{z^{d_{\theta}-2}}\;d\tilde{\mathcal{A}}\,.
\end{equation}
This functional is a deformation of the Willmore functional parameterised by $1<d_\theta<3$.
In the special case of $d_\theta=2$ the functional (\ref{Fa2}) becomes the well known Willmore functional, as expected from the
analysis of $F_A$ in AdS$_4$ performed in \cite{Babich, Mazzeo}.

As a consistency check, we can show that in the limit $d_\theta \rightarrow 3$ the functional \eqref{Fa1} reproduces the logarithmic divergence \eqref{2divlog}.
This can be done by first plugging \eqref{normal} and \eqref{inducedmetric} in \eqref{Fa1}, then expanding about $z=0$ 
and finally using \eqref{coefficientsu}. 
We find
\bea
\label{log_wilm1}
F_{A}&=& 
\frac{1}{d_\theta-1}\int_\varepsilon^{z_\text{\tiny max}} \!\! dz 
\int_{\partial \hat \gamma_{A,\varepsilon}} \!\! ds  \left[\frac{ k^2(s)}{(d_\theta-1)^2\, z^{d_\theta-2}}
+ \mathcal{O}\big(z^{d_\theta-3} \big) \right]  
\\
\rule{0pt}{.7cm}
& \rightarrow& \, -\,\frac{\log \varepsilon }{8} \int_{\partial A} \!\!  k^2(s)\, ds + \mathcal{O}(1)
\hspace{2.5cm}
d_\theta \rightarrow 3
\eea
which is  the logarithmic contribution occurring in \eqref{2divlog}.

In the regime $3<d_\theta<5$, 
the expression for $\mathscr{F}_A$ in (\ref{finiteminsurf2}) specified for (\ref{hyperscaling4bis}) on a constant time slice
becomes (see Appendix\;\ref{hvLif} for details)
\begin{equation}
\label{Fa2vacuum}
\mathscr{F}_A
=
-\,\frac{1}{(d_{\theta}-1)(d_{\theta}-3)}\int_{\hat\gamma_A}
\left[\,
\frac{3(\tilde{n}^{z})^{4}}{z^{d_{\theta}}}
-\frac{2\,\tilde{n}^{z}}{z^{d_{\theta}-2}} \,
\tilde{h}^{z\mu}\,\partial_{\mu}\bigg(\dfrac{\tilde{n}^{z}}{z}\bigg)
\right]d\tilde{\mathcal{A}}
\end{equation}
where both the integrals are convergent;
indeed, the former integrand scales as $z^{4-d_\theta}$, while the latter one as $z^{6-d_\theta}$.
Following the same steps that lead to \eqref{log_wilm1}, we find that the expansion near to the boundary of \eqref{Fa2vacuum} gives
\begin{equation}
\mathscr{F}_A  
=
-\int_\varepsilon^{z_\text{\tiny max}} \!\! dz 
\int_{\partial\hat\gamma_{A,\varepsilon}} \!\!ds 
\left\{\frac{ \left[(9d_\theta-2 d^2_\theta -13) k(s)^4-2 (d_\theta -1)^2 k(s)
	k''(s)\right]}{(d_\theta -3)^2 (d_\theta -1)^5\, z^{d_\theta-4}} 
+\mathcal{O}(z^{6-d_\theta}) \right\}\,.
\end{equation}

Taking the limit $d_\theta \rightarrow 5$, we find the logarithmic divergent term
\begin{equation}
\label{log_wilm2}
\mathscr{F}_A \,\to\,  -\frac{\log \varepsilon }{2048}\int_{\partial A}\, \!\! \Big[16 \,k(s) \,k''(s)+9\, k(s)^4 \Big] ds + \mathcal{O}(1)
\;\; \qquad \;\; 
d_\theta \rightarrow 5
\end{equation}
which becomes the logarithmic divergent term occurring in \eqref{log_div_5}, after a partial integration.

	\subsection{Asymptotically hvLif$_4$ black hole}
\label{sec static BH}

Another static background of physical interest is the asymptotically hvLif$_4$ black hole,
whose metric reads \cite{Dong:2012se,Alishahiha:2012qu,Bueno:2012sd}
\begin{equation}
\label{BH_metric}
ds^2= \frac{1}{z^{d_\theta}} \left( \!-\,z^{-2(\zeta-1)} f(z)dt^2+\frac{dz^2}{f(z)}+dx^2+dy^2 \right) 
\;\;\qquad\;\;
f(z)\equiv1-(z/z_h)^{d_\theta+\zeta}
\end{equation} 
where the parameter $z_h$ corresponds to the horizon, which determines the black hole temperature \cite{Dong:2012se}
	\be
	\label{temp_BH}
	T=\frac{|d_\theta+\zeta|}{4\pi z_h^\zeta}\,.
	\ee
Unlike hvLif$_4$, where the Lifshitz exponent $\zeta$ occurs only in the $g_{tt}$ component of the metric, 
in (\ref{BH_metric}) it enters also in $f(z)$; 
hence the minimal surface $\hat{\gamma}_A$ depends on $\zeta$.

For $1<d_\theta<3$,
specialising the general formula \eqref{finite term} to the black hole metric \eqref{BH_metric}, 
for the finite term of the holographic entanglement entropy we find
\begin{equation}
\label{Fa_BH}
F_A=\frac{1}{(d_\theta-1)}
\int_{\hat \gamma_A} \frac{1}{z^{d_\theta}} \left[ (d_\theta-1)(f(z)-1)-\frac{z f'(z)}{2}+(\tilde n^z)^2 \left( 1+\frac{z f'(z)}{2 f(z)} \right) \right] d\tilde{\mathcal{A}}\,.
\end{equation} 
This functional reduces to \eqref{Fa1} when $f(z) = 1$ identically, as expected. 
For simplicity, here we do not consider the case $3<d_\theta<5$, but the corresponding computation 
to obtain $\mathscr{F}_A$ is very similar to the one leading to (\ref{Fa_BH}).

In the regime where the size of the domain $A$ is very large with respect to the black hole horizon scale $z_h$,
the extremal surface can be approximated by a cylinder $\hat \gamma_A^{\text{\tiny cyl}}$ 
with horizontal cross section $\partial A$ and the second base  located at $z=z_* \sim z_h$. 
Within  this approximation, the functional \eqref{Fa_BH} simplifies to
\bea
\label{large_BH1}
F_A^{\text{\tiny cyl}} 
& =&
\frac{d_\theta  [ f(z_*)-1] + 1}{(d_\theta-1)\,z_*^{d_\theta}}\;\text{Area}(A)
+\frac{P_A}{d_\theta-1} \, \int_0^{z_*} \!  \left[ f(z)-\frac{z f'(z)}{2} -1 \right] \frac{dz}{z^{d_\theta}}
\nonumber
\\
& =&
\frac{1-(z_*/z_h)^{d_\theta+\zeta}\,d_\theta}{z_*^{d_\theta}(d_\theta-1)}\;\text{Area}(A)
+ \frac{(d_\theta +\zeta -2)\, z_*^{1-d_\theta}}{2 (\zeta +1)(d_\theta-1)} 
\left(\frac{z_*}{z_h}\right)^{d_\theta +\zeta } \! P_A
\eea
where we used that $\tilde n^z=\sqrt{f(z_*)}$ on the base and $\tilde n^z =0$ on the vertical part of $\hat \gamma_A^{\text{\tiny cyl}}$.
In the special case of $d_\theta =2$, the expression (\ref{large_BH1}) 
reduces to the corresponding result of \cite{Fonda:2015nma}.
Taking the limit $z_* \rightarrow z_h$ of (\ref{large_BH1}), we find 
\begin{equation}
\label{large_BH2}
F_A^{\text{\tiny cyl}} = -\,\frac{\text{Area}(A)}{z_h^{d_\theta}}+\dots\;.
\end{equation}	
By using \eqref{temp_BH}, this relation can be written as
$F_A^{\text{\tiny cyl}} \simeq -\,T^{d_\theta/\zeta} \text{Area}(A)$ (up to a numerical coefficient), 
which tells us that $-F_A^{\text{\tiny cyl}}$ approaches the thermal entropy in this limit.

 \section{ Finite term as an integral along the entangling curve}
 \label{sec: integral along entangling curve}

This section is devoted to show that the finite term in the expansion of the entanglement entropy for the case hvLif$_{d+1}$
can be written as an integral over the entangling $(d-2)$ dimensional hypersurface. 
This analysis  extends the result obtained in  \cite{Mazzeo} for AdS$_4$.
In Appendix\;\ref{app:integral entangling curve}  we show that the same result can be obtained  through a procedure on the area functional
that is similar to the one leading to the Noether theorem.

The geometry of this spacetime is given by (\ref{hyperscaling4}) 
with $d\boldsymbol{x}^2 = \sum_{i=1}^{d-1} dx_i^2$, $R_{\text{\tiny AdS}} =1$ and $d_\theta = d-1-\theta$.
This spacetime possesses a conformal Killing vector  generating the following transformation
 \be
 \label{deformeddilatations}
 t\mapsto \lambda^{1-\zeta} t\qquad  z\mapsto \lambda  z\qquad  \qquad  \boldsymbol{x} \mapsto \lambda  \boldsymbol{x}\qquad 
  \ee 
  under which the metric changes as $ds^2\mapsto \lambda^{2-d_\theta} ds^2$, being $d_\theta \geqslant 1$.
  
  An amusing consequence of the existence of this conformal Killing vector is the possibility to write the finite term 
  (whenever a logarithmic divergence does not occur)
  as an integral over the entangling hypersurface 
  independently of the number of divergent terms appearing  in the expansion of the area 
  and of the spacetime dimensionality.  
This can be shown by considering the variation of the induced area element  for an infinitesimal transformation generated by the infinitesimal  parameter $\lambda=1+\epsilon+\cdots$. From the scaling law of the metric, we find
  \be
  \label{scaling}
  \delta_\epsilon \big(\sqrt{h}\,\big)
  = 
  \epsilon\; \frac{(2-d_\theta) m}{2}\; \sqrt{h}
  \ee
 where $m$ is the dimension of the minimal hypersurface. Namely, if we perform the transformations
 \eqref{deformeddilatations} the volume of the hypersurface scales  as $\mathcal{V}\rightarrow \lambda^{\frac{m(2-d_{\theta})}{2}}\mathcal{V}.$
 
 Since the transformation (\ref{deformeddilatations}) can be also viewed as an infinitesimal diffeomorphism generated by a  
 conformal Killing vector field $V_{\mu}$  
 acting on the bulk, its action on the induced metric can be cast into the following form
 \be
 \delta h_{ab}
 =
\big(\nabla_\mu V_\nu+\nabla_\nu V_\mu \big)
\frac{\partial x^\mu}{\partial \sigma^a}\frac{\partial x^\nu}{\partial \sigma^b}
=
D_a V_b+D_b V_a +K^{(i)}_{ab} (n_{(i)}\cdot V)
 \ee
  where $\sigma^a$ are the coordinates on the minimal surface,
  $D_a$ is the induced covariant derivative on $\gamma_A$,
  the vector field $V_a= V_\mu \partial_a   x^{\mu}$ 
   is the pullback of $V_\mu$ on $\gamma_A$,
   $n_{(i)}$ are the normal vectors  to the minimal surface 
   and $K^{(i)}_{ab}$ the associated extrinsic curvature
   (the dot corresponds to the scalar product given by the bulk metric). 
   Then, the variation of the volume form can be written as
 \be
 \label{quasidiffeo}
  \delta_\epsilon \big(\sqrt{h}\, \big)
  =
  \frac{1}{2}\sqrt{h} \, h^{ab}\delta_\epsilon h_{ab}
  =
  \frac{\epsilon}{2} \sqrt{h} \, \Big(2 \,D_a V^a+ K^{(i)} (n_{(i)}\cdot V)\Big)
  =
\epsilon\, \sqrt{h} \,(D_a V^a)
 \ee
 where in the last step the extremality condition has been employed.  
 If we compare \eqref{scaling} and \eqref{quasidiffeo}, we find 
 \be
 \label{puddu}
 \sqrt{h}=\frac{2}{(2-d_\theta)m} \sqrt{h} (D_a V^a)
 \ee
which can be integrated over $\hat \gamma_{A,\varepsilon}$, finding 
 \be
 \label{Areaasboundaryintegral}
\mathcal{A}[ \hat{\gamma}_{A,\varepsilon}]
 =
 \frac{2}{(2-d_\theta)m}\int_{\hat \gamma_{A,\varepsilon}} \!\!\!
 \sqrt{h} (D_a V^a)\,d^m
 \sigma
 =  
 \frac{2}{(2-d_\theta)m}\int_{\partial \hat \gamma_{A,\varepsilon}} \!\!\!
 \sqrt{h}  (b_a V^a)\,d^{m-1}
 \xi
\ee
where $ b^{a} $ is the unit vector normal  to $ \partial\hat \gamma_{A,\varepsilon}$ along the surface $\hat \gamma_{A,\varepsilon}$,
and  $\xi^j$ denote the coordinates on the boundary of the minimal hypersurface. 
Actually, identities similar to \eqref{puddu} and \eqref{Areaasboundaryintegral} hold 
if the manifold admits a vector of constant divergence. The conformal Killing vector generating dilatations is just an example of this type.
The above analysis is valid in any dimension and for generic codimension of the minimal submanifold. To complete our analysis we need
to know the behavior of the vector $b_a$ close to the boundary.   In the present paper, we have performed this analysis  only for the case of interest,
i.e. $d=3$ and $m=2$ (see Appendix\;\ref{appxUVbehav}), but it can be extended to more general situations by means of the same  techniques.


For $d=3$ and $m=2$, by plugging the expansion  \eqref{svil} into  \eqref{Areaasboundaryintegral}, 
for the finite term  we find
	\begin{equation}
\label{Fa=udtheta}
F_{A}=
-\dfrac{d_{\theta}+1}{d_{\theta}-2}
\int_{\partial A} \!\!\!
\big( \boldsymbol{x}_A\cdot\widetilde{N} \big)
\,\mathcal{U}_{d_\theta+1}\,ds
\;\;\qquad\;\;
d_\theta\ne 2
\end{equation}
where $\mathcal{U}_{d_\theta+1}$ is the first non analytic term encountered in the expansion \eqref{svil}, 
$\boldsymbol{x}_A$ is  a shorthand notation for the parametric representation $\boldsymbol{x}_A\equiv (x(s), y(s))$ of the entangling curve and the vector  $\widetilde N$ is  the unit normal to  this curve in the plane $z=0$ in $\widetilde{\cal M}_3$ (see also Appendix\;\ref{appxUVbehav}).

Further remarks about \eqref{Fa=udtheta} are in order. 
The representation  \eqref{Fa=udtheta} for the finite term holds  for any $d_\theta \ne 2$ and there is no restriction on the range of $d_\theta$.  Even though  the expression \eqref{Fa=udtheta} may suggest that $F_A$ is completely characterized 
by the local behaviour of the extremal surface near the boundary,
it turns out that the coefficient $\mathcal{U}_{d_\theta+1}$ cannot be determined only by solving perturbatively \eqref{extremalcondit} about $z=0$
(see Appendix\;\ref{appxUVbehav}); hence it depends on the whole minimal surface $\hat \gamma_A$.


\section{ Time dependent backgrounds for $ 1<d_{\theta}<3$}
\label{sec:time-dependent}

When the gravitational background is time dependent, 
the covariant prescription for the holographic entanglement entropy introduced in \cite{Hubeny:2007xt}
must be employed.
The class of surfaces $\gamma_A$ to consider is defined only by the constraint $\partial \gamma_A = \partial A$;
hence $\gamma_A$ is not restricted to lay on a slice of constant time, as in the static gravitational spacetimes. 

In this section we study the finite term in the expansion of the holographic entanglement entropy 
in time dependent asymptotically hvLif$_{4}$ backgrounds.
A crucial difference with respect to the case of static backgrounds
is that surfaces in four dimensional spacetimes  have two normal directions identified by the unit normal vectors $n^{(i)}_N$
(with $i=1,2$, whose squared norm $\epsilon_i=g^{MN} n^{(i)}_M n^{(i)}_N$ is either $+1$ or $-1$) 
and therefore two extrinsic curvatures $K_{MN}^{(i)}$.
In this analysis we need to extend the result obtained in \cite{Fonda:2015nma} 
by including the Lifshitz scaling and the hyperscaling violation. 
The technical details of this computation are discussed in Appendix\;\ref{Apptime} and in the following we report only the final results.

In the range $1< d_\theta<3$,
for surfaces $\gamma_A$ that intersect orthogonally the boundary,
the expansion (\ref{leadingdiver}) holds with the finite term given by 
\bea
\label{Fa time-dep}
\mathcal{F}_A
&= &
c_1\int_{\gamma_A } \!\! e^{2\phi}
\Bigg[ \,
2\,\tilde{h}^{MN}\partial_M\varphi \, \partial_N\phi
-\sum_{i=1}^{2}\epsilon_i \,\tilde{n}^{(i)M} \tilde{n}^{(i)N}\Bigl(\widetilde{D}_M \widetilde{D}_N\varphi-\widetilde{D}_M\varphi \widetilde{D}_N\varphi\Bigr)+\widetilde{D}^2\varphi
\hspace{1cm}
\\ 
& &
\hspace{2.2cm}
 +\,\frac{1}{4}\sum_{i=1}^2\epsilon_i\bigl(\textrm{Tr} \widetilde{K}^{(i)}\bigr)^{2}\, \Bigg] d \tilde{\mathcal{A}}\,
-\int_{\gamma_A}  \!\!\!  e^{2\varphi}\,d\mathcal{\widetilde{A}}\,
-\frac{c_1}{4}\sum_{i=1}^2\epsilon_i
\int_{\gamma_A }  \!\!\! e^{2\phi}\bigl(\textrm{Tr} K^{(i)}\bigr)^{2}d\mathcal{A}\,.
\nonumber
\eea

Specialising this expression to extremal surfaces $\hat{\gamma}_A$, that satisfy $\textrm{Tr} K^{(i)} =0$
and for which $c_1$ is given in (\ref{alpha}), we find 
\bea
\label{Fa time-dep extremal surf}
F_A
&= &
\int_{\hat \gamma_A } \dfrac{2\,e^{2\phi}}{d_{\theta}(d_{\theta}-1)}
\Bigg[\,
2\,\tilde{h}^{MN}\partial_M\varphi \,\partial_N\phi
-\sum_{i=1}^{2}\epsilon_i \, \tilde{n}^{(i)M} \tilde{n}^{(i)N}\widetilde{D}_M \widetilde{D}_N\varphi
\\
& &\hspace{3cm}
+\,\widetilde{D}^2\varphi-\dfrac{d_{\theta}(d_{\theta}\!-\!1)}{2}\,e^{2(\varphi-\phi)} 
+\frac{1}{2}\sum_{i=1}^2\epsilon_i\bigl(\textrm{Tr} \widetilde{K}^{(i)}\bigr)^{2}\Bigg]
d \tilde{\mathcal{A}}\,.
\nonumber
\eea

In the special case of $d_\theta=2$, the expressions (\ref{Fa time-dep}) and (\ref{Fa time-dep extremal surf}) simplify to the ones
obtained in  \cite{Fonda:2015nma} for time dependent asymptotically AdS$_4$ backgrounds.
In the final part of Appendix\;\ref{Apptime} we show that (\ref{Fa time-dep extremal surf}) becomes (\ref{finite term}) for static backgrounds.  

The temporal evolution of the holographic entanglement entropy in the presence of Lifshitz scaling and hyperscaling violation exponents
has been studied in \cite{Keranen:2011xs, Liu:2013iza, Liu:2013qca, Alishahiha:2014cwa, Fonda:2014ula}
by considering infinite strips and disks. 
It would be interesting to extend this numerical analysis to non spherical finite domains, 
also to check the analytic expression (\ref{Fa time-dep extremal surf}).
 

 \section{Some particular regions}
 \label{sec:particular regions}

In the previous sections we discussed expressions for the finite term in the expansion of the 
holographic entanglement entropy that hold for any smooth region $A$, independently of its shape.
In this section we test these expressions by considering 
infinite strips (Section\;\ref{stripcase}), disks (Section\;\ref{diskcase}) and ellipses (Section\;\ref{ellipsecase}).


\subsection{Strip}
\label{stripcase}

The spatial region $A=\left\{(x,y): \lvert x\rvert\leqslant \ell/2,\lvert y\rvert\leqslant  L/2\right\}$ in the limit of $\ell \ll L$
can be seen as an infinite strip that is invariant under translations along the $y$-direction.
The occurrence of this symmetry leads to a drastic simplification 
because the search of the minimal area surface $\hat{\gamma}_A$ can be restricted to the class of surfaces $\gamma_A$
invariant under translations along the $y$-direction, which are fully characterised by the profile $z=z(x)$
of a section at $y= \textrm{const}$.

 \subsubsection{HvLif$_4$}
 \label{sec_strip_hvLif}

Considering the hvLif$_4$ gravitational background given by (\ref{hyperscaling4bis}),
in the regime  $\ell \ll L$ the area functional evaluated on the surfaces $\gamma_A$ characterised by the profile $z=z(x)$
simplifies to 
\begin{equation}
\label{area_func_strip}
\mathcal{A}[\gamma_A]= L \int_{-\ell/2}^{\ell/2} 
 \frac{\sqrt{1+(z')^2}}{z^{d_\theta}}\; dx\,.
\end{equation}
Since the coordinate $x$ is cyclic, its conjugate momentum is conserved, namely
\begin{equation}
\label{conservedmomentum}
\dfrac{d}{dx}\bigg(
\dfrac{1}{z^{d_{\theta}}}\dfrac{1}{\sqrt{1+(z')^2}}
\bigg)
=0
\;\;\qquad \Longrightarrow \qquad \;\;
\frac{1}{z^{d_{\theta}} \sqrt{1+(z')^2}}=\dfrac{1}{z_{\ast}^{d_{\theta}}}
\end{equation}
where in the integration
we have denoted by $ z_{*}\equiv z(0) $ the value of the function $z(x)$ corresponding to the tip of the surface, where $z'(0)=0$.
The parameter $z_*$ can be also expressed in terms of the width of the strip $\ell$ as follows
\begin{equation}
\label{stripwidth}
\dfrac{\ell}{2}
=\int_{0}^{z_{\ast}} \dfrac{dz}{z'}
=\int_{0}^{z_{\ast}}\!\!
\dfrac{dz}{\sqrt{\bigl(z_{\ast}/z\bigr)^{2d_{\theta}}-1}}
\,=\,
\frac{\sqrt{\pi}\;\Gamma\big((1+1/d_\theta)/2\big)}{
	\Gamma\big(1/(2d_{\theta})\big)}
	\;z_{\ast}\,.
\end{equation} 
By integrating the conservation law \eqref{conservedmomentum}, for the profile $x(z)$ one finds
\begin{equation}
\label{stripsolution}
x(z)=\dfrac{\ell}{2}-\dfrac{z_{*}}{d_{\theta}+1}\bigg(\dfrac{z}{z_{*}}\bigg)^{d_{\theta}+1}\!\!
\,_2F_{1}\bigg(
\dfrac{1}{2}\,,\dfrac{1}{2}+\dfrac{1}{2d_{\theta}} \,;\dfrac{3}{2}+\dfrac{1}{2d_{\theta}}\,;(z/z_{*})^{2d_{\theta}} \!\bigg)\,.
\end{equation}

The most direct approach to obtain $\mathcal{A}[\hat{\gamma}_{A,\varepsilon}]$ 
consists in evaluating (\ref{area_func_strip}) on the profile (\ref{stripsolution}).
This calculation has been done in \cite{Dong:2012se} and the corresponding expansion as $\varepsilon \to 0$
has been obtained. 
In the following we reproduce the finite term of this expansion 
by specialising the expressions \eqref{Fa1} and \eqref{Fa2vacuum} to the strip
(for the latter formula, the computation is reported in Appendix \ref{app strip 3-5}).

Let us first consider
the tangent and normal vectors to the surfaces anchored to the boundary of the infinite strip
that are characterised by the profile $z=z(x)$. They read
\begin{equation}
\tilde{t}_{1}^{\mu}=\left(\dfrac{z'}{\sqrt{1+(z')^2}}\,,\dfrac{1}{\sqrt{1+(z')^2}}\,,0\right) 
\;\;\;\; 
\tilde{t}^{\mu}_{2}=\big(0,0,1\big)
\;\;\;\; 
\tilde{n}^{\mu}=\left(\dfrac{-1}{\sqrt{1+(z')^2}}\,,\dfrac{z'}{\sqrt{1+(z')^2}}\,,0\right).
\end{equation}

For $1< d_\theta <3$, we can plug the component $\tilde{n}^{z}$ into \eqref{Fa1}, 
that holds for the minimal surface $\hat{\gamma}_A$,
finding that the finite term of the holographic entanglement entropy becomes
\begin{equation}
\label{Fastrip}
F_{A}
=
\dfrac{1}{d_{\theta}-1}\int_{\hat \gamma_{A}}\dfrac{dx\,dy}{z^{d_{\theta}}\sqrt{1+(z')^2}} 
\,=\,
\dfrac{4}{(d_{\theta}-1)\, z_{\ast}^{d_{\theta}}}
\int^{L/2}_{0} \!\! \int_{0}^{\ell /2}\!\!dxdy
\,=\,
\dfrac{L\, \ell}{(d_{\theta}-1)\, z_{\ast}^{d_{\theta}}}
\end{equation}
where \eqref{conservedmomentum} has been used in the last step. 
By employing  (\ref{stripwidth}), the expression (\ref{Fastrip})
can be written as \cite{Dong:2012se}
\begin{equation}
\label{finitestripterm}
F_{A}
=
\dfrac{L\,\ell^{1-d_\theta}}{d_{\theta}-1}\,
\Bigg(
\frac{2\,\sqrt{\pi}\;\Gamma\big((1+1/d_\theta)/2\big)}{
	\Gamma\big(1/(2d_{\theta})\big)}
	\Bigg)^{d_{\theta}}.
\end{equation}

We have obtained this result for $1<d_{\theta}<3$, but it turns out to be valid for any $d_\theta>1$
(in Appendix \ref{app strip 3-5} we have checked that (\ref{finitestripterm}) is recovered also by specialising to the strip the general formula (\ref{Fa2vacuum}) that holds for $3<d_{\theta}<5$).
In fact all the subleading divergences can be expressed recursively in terms of the geodesic curvature of $\partial A$ and its derivatives  (see Appendix\;\ref{appxUVbehav}); and this quantity trivially vanishes for the straight line.

We find it instructive  to specialise the method discussed in Section\;\ref{sec: integral along entangling curve} to the infinite strip. 
The analytic profile (\ref{stripsolution}) allows us to determine the scalar function $u(z,s)$ used in Appendix\;\ref{appxUVbehav} 
to describe the minimal surface:  $u(z,s) = \ell/2 - x(z)$.
	By expanding this result in powers of $z$ and by comparing the expansion with \eqref{svil}, one finds the following coefficient 
	\begin{equation}
	\label{u global coef}
	\mathcal{U}_{d_{\theta}+1}=\dfrac{1}{(d_{\theta}+1)\, z_\ast^{d_{\theta}}}\,.
	\end{equation}

	The expression \eqref{Fa=udtheta} must be slightly modified for the infinite strip because in this case we evaluate the finite ratio 
	$\mathcal{A}/L$ and the scaling in the direction along which the strip is infinitely long is not considered.
	Thus, the ratio $\mathcal{A}/L$ scales like $\mathcal{A}/L \rightarrow \lambda^{1-d_{\theta}}\mathcal{A}/L$ under (\ref{deformeddilatations}).
	As a consequence, for the infinite strip \eqref{Fa=udtheta} has to be replaced with
\begin{equation}
	\label{Fa=udtheta_strip}
	F_{A}=-\dfrac{d_{\theta}+1}{d_{\theta}-1}
	\int_{\partial A}
	\! \big( \boldsymbol{x}_A\cdot\widetilde{N}\big)\, \mathcal{U}_{d_{\theta}+1} \,ds\,.
\end{equation}
	Plugging \eqref{u global coef} into \eqref{Fa=udtheta_strip} and using that $ \boldsymbol{x}_A\cdot\tilde{N}=-\ell/2$,  
	we recover \eqref{Fastrip}, and therefore also (\ref{finitestripterm}),
	which is the result found in \cite{Dong:2012se} for the infinite strip in a generic number of spacetime dimensions.

\subsubsection{Asymptotically hvLif$_4$ black hole}

We find it worth considering also the finite term of the holographic entanglement entropy of an infinite strip $A$ 
when the gravitational background is given by the asymptotically hvLif$_4$ black hole (\ref{BH_metric}).
This can be done by adapting the procedure described in Section\;\ref{sec_strip_hvLif} for hvLif$_4$.

The area functional restricted to the class of surfaces $\gamma_A$ that are 
invariant under translations along the $y$-direction
(which are fully determined by the profile $z=z(x)$ of any section at $y=\textrm{const}$) reads
\begin{equation}
\label{area_func_strip_BH}
\mathcal{A}[\gamma_A]
=
 L \int_{-\ell/2}^{\ell/2}  \frac{1}{z^{d_\theta}} \,\sqrt{1+\frac{(z')^2}{f(z)}}\; dx
\end{equation}
that simplifies to (\ref{area_func_strip}) when $f(z) =1$ identically, as expected. 
Since $x$ is a cyclic coordinate in (\ref{area_func_strip_BH}), 
one obtains the following conservation law
\begin{equation}
\label{conser_law_BH}
z^{d_\theta} \,\sqrt{1+\frac{(z')^2}{f(z)}}=z_*^{d_\theta}
\end{equation} 
being  $(z,x)=(z_*,0)$ the coordinates of the tip of the profile of the minimal surface $\hat \gamma_A$,
where $z'(0)=0$ holds. 
We also need the unit vector $\tilde n^\mu$ normal to the surface, whose components read
\be
\label{normal_strip_BH}
\tilde{n}^{\mu}
=\big(\tilde n^z,\tilde n^x,\tilde n^y\big)
=
\Bigg(
\dfrac{f(z)}{\sqrt{f(z)+(z')^{2}}} \, ,-\dfrac{z'}{\sqrt{f(z)+(z')^{2}}} \, ,0
\Bigg)\,.
\ee

Now we can specialise \eqref{Fa_BH}, which holds for minimal surfaces, to the strip by employing (\ref{normal_strip_BH}),
finding that
\be
\label{Fa_BH_strip}
F_A 
=
\frac{2L}{z_*^{d_\theta}(d_\theta-1)}\int_{0}^{\ell/2}  
\Bigg[ 
\left((d_\theta-1)(f(z)-1)-\frac{z f'(z)}{2}\right)\frac{z_*^{ 2 d_\theta}}{z^{2d_\theta}}+  f(z)+\frac{z f'(z)}{2}  
\Bigg] dx
\ee
where the emblacking factor $f(z)$ is given in (\ref{BH_metric}).
By employing the conservation law \eqref{conser_law_BH}, it is straightforward to write
 (\ref{Fa_BH_strip}) as an integral in $z$ between $0$ and $z_\ast$.
 Notice that, by setting $\zeta=1$ and $d_\theta=2$ in (\ref{Fa_BH_strip}), we recover the result obtained in \cite{Fonda:2015nma}.


 \subsection{Disk}
 \label{diskcase}

 In this subsection we study the holographic entanglement entropy of a disk $A$ with radius $R$
 when the gravitational background is hvLif$_4$ (Section\;\ref{sec disk hvLif}) or 
 the asymptotically hvLif$_4$ black hole (Section\;\ref{sec disk bh}).
 Fixing the origin of the Cartesian coordinates $(x,y,z>0)$ in the center of $A$, 
 the rotational symmetry of $A$ about the $z$-axis implies that 
 $\hat{\gamma}_A$ belongs to the subset of surfaces $\gamma_A$ displaying this rotational symmetry;
hence it is more convenient to adopt cylindrical coordinates $(z,\rho,\phi)$, 
 where $(\rho,\phi)$ are polar coordinates in the plane at $z=0$.
In these coordinates the entangling curve is given by $(\rho=R\,,\phi)$ in the plane at $z=0$.

 \subsubsection{HvLif$_4$}
 \label{sec disk hvLif}

When the gravitational background is hvLif$_4$
(now it is convenient to express the metric (\ref{hyperscaling4bis}) in cylindrical coordinates), 
the area functional for the surfaces 
invariant under rotations about the $z$-axis that 
are defined by their radial profile $z=z(\rho)$
and that are anchored to the entangling curve $(\rho,\phi)=(R,\phi)$ (i.e. such that $z(R)=0$)
reads 
\be
\label{area_func_disk}
\mathcal{A}[\gamma_A]\,=\, 2\pi\! \int_{0}^R  \frac{\sqrt{1+ (z')^2} }{z^{d_\theta}} \,\rho \,d\rho
\ee
where $z' = \partial_\rho z(\rho)$.
Imposing the vanishing of the first variation of the functional (\ref{area_func_disk})
leads to the following second order ordinary differential equation 
 \begin{equation}
 \label{eqmotosphere}
 \dfrac{z''}{1+(z')^2}+\dfrac{z'}{\rho}+\dfrac{d_{\theta}}{z}=0
 \end{equation}
 where  the boundary conditions $ z(R)=0 $ and $ z'(0)=0 $ hold. 
  It is well known that, in the special case of $d_{\theta}=2$, the hemisphere $z(\rho)=\sqrt{R^{2}-\rho^{2}}$
  is a solution of (\ref{eqmotosphere}) \cite{Ryu:2006bv,Ryu:2006ef}.
  For $d_{\theta}\neq 2$, the solution of (\ref{eqmotosphere}) has been studied numerically in \cite{Fonda:2014ula}.

 %
%

In the following we provide the finite term in the expansion of the holographic entanglement entropy for disks
by specialising \eqref{Fa1} and \eqref{Fa2vacuum} to these domains. 
In terms of the cylindrical coordinates, the unit tangent and normal vectors to $\hat{\gamma}_A$ read
\be
 \label{disk_vectors}
 \tilde{t}^{\mu}_{\rho}
 =\bigg( \dfrac{z'}{\sqrt{1+(z')^2}} \, , \dfrac{1}{\sqrt{1+(z')^2}} \, ,0 \bigg) 
 \qquad 
 \tilde{t}^{\mu}_{\phi}=\big( 0,0,1 \big)
 \qquad
 \tilde{n}^{\mu}=\bigg( \dfrac{1}{\sqrt{1+(z')^2}} \,, \dfrac{-\,z'}{\sqrt{1+(z')^2}} \, ,0\bigg)
\ee
 where $z=z(\rho)$ satisfies (\ref{eqmotosphere}).
 We remark that only the component $\tilde{n}^z$ occurs in \eqref{Fa1} and \eqref{Fa2vacuum}.
 Thus, from (\ref{disk_vectors}), we easily find that for $1<d_\theta<3 $ the expression \eqref{Fa1} becomes 
\be
 \label{fasfera1}
F_{A} = \dfrac{2\pi}{d_{\theta}-1}\int_{0}^{R}\! \dfrac{\rho\, d\rho}{z^{d_\theta}\sqrt{1+(z')^2}}\,. 
\ee
In the regime $3<d_\theta<5$, we have that \eqref{Fa2vacuum} gives
\be
 \label{fasfera2}
\mathscr{F}_A  
=  
\frac{2\pi}{(d_\theta-1)(d_\theta-3)} \int_0^R 
 \frac{2 \big[ (d_\theta-1)+ z \,z'/\rho \big] (z')^2  -3}{z^{d_\theta} \big[1+(z')^2 \big]^{3/2}}
\, \rho \,d\rho
\ee
 where \eqref{eqmotosphere} has been used to rewrite $z''$.

Even though  \eqref{eqmotosphere} is invariant under the scale transformation $(z,\rho)\rightarrow \lambda (z,\rho)$, 
 the expressions in \eqref{fasfera1} and \eqref{fasfera2} do not enjoy this invariance. 
 However, since the metric scales as $ds^2\mapsto \lambda^{2-d_\theta} ds^2$, 
 it is straightforward to observe that 
 \begin{equation}
 \label{FascalingR}
 F_{A}(R)= R^{2-d_{\theta}} \,F_{A} \big|_{R=1} 
 \;\;\qquad \;\;
  \mathscr{F}_{A}(R)= R^{2-d_{\theta}}\,\mathscr{F}_{A} \big|_{R=1}\,. 
 \end{equation}
Thus, the finite term in the holographic entanglement entropy  
decreases with the radius for $d_{\theta}>2$, while it increases for $d_{\theta}<2$. 
The case $d_{\theta}=2$ corresponds to AdS$_4$, which is scale invariant, 
and $F_A = 2\pi$ for a disk, independently of the radius $R$, as expected.


\begin{figure}[t!]
	\vspace{-.2cm}
	\hspace{-.5cm}
		\includegraphics[width=1\textwidth]{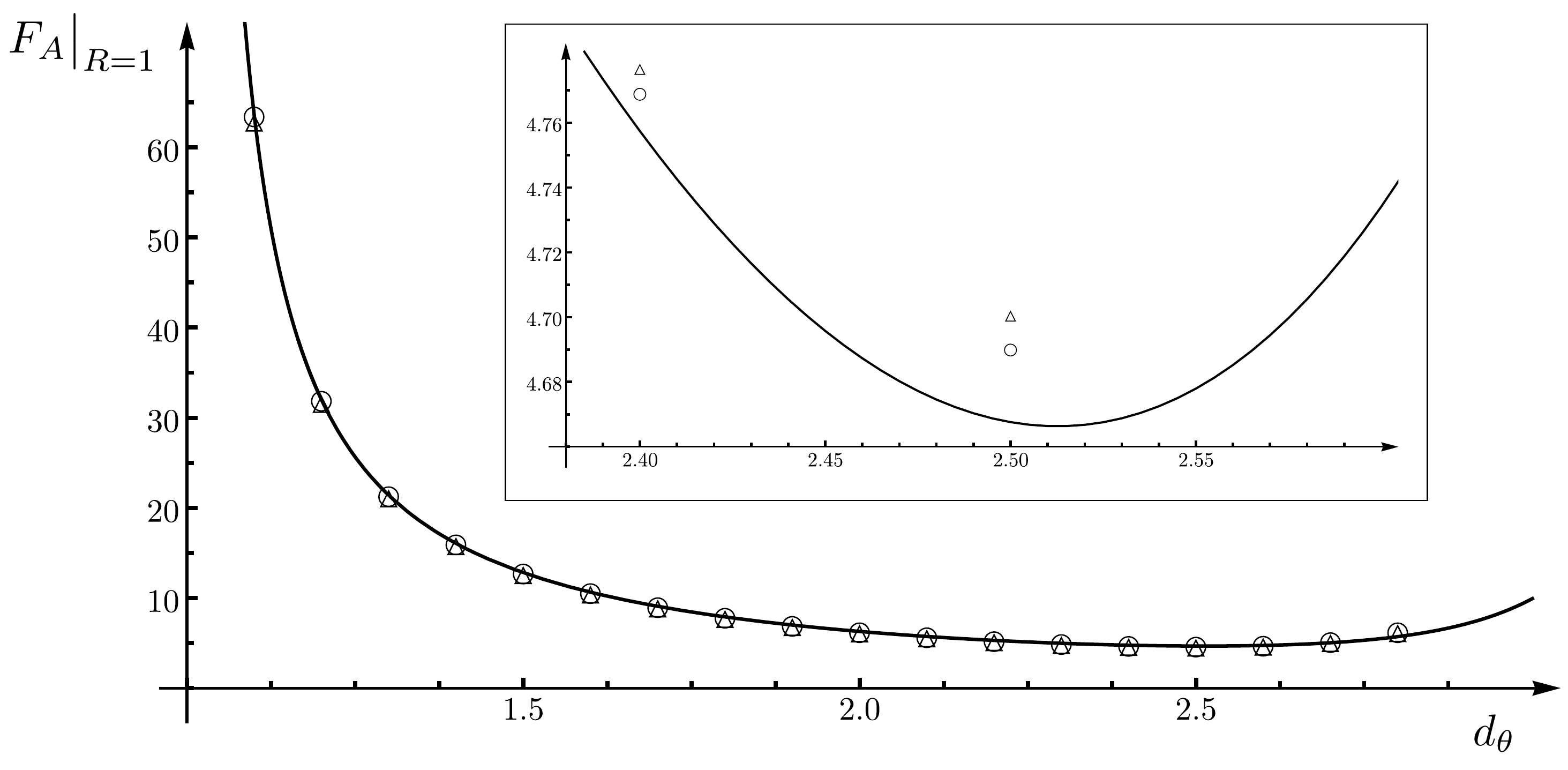}
	\vspace{-.2cm}
	\caption{\small
		Finite term $F_{A}$  in terms of $1< d_\theta < 3$ for minimal surfaces anchored to
		a disk of radius $R=1$ in the hvLif$_{4}$ geometry (\ref{hyperscaling4bis})
		at $t =\textrm{const}$.
		The solid line is found by first solving numerically (with Wolfram Mathematica) the differential equation \eqref{eqmotosphere}
		and then plugging the resulting radial profile into \eqref{fasfera1}.
		The data points labelled by the empty circles and the empty triangles  
		have been obtained with Surface Evolver through the two formulas in \eqref{FaSurfaceEvolver} respectively.
		The inset contains a zoom close to the minimum of the curve, that corresponds to 
		$(d_\theta, F_A) \simeq (2.52\,, 4.67)$.
	}
	\label{F_A_plot_sphere}
\end{figure}

  \begin{figure}[t!]
	\vspace{-.2cm}
	\hspace{-.5cm}
		\includegraphics[width=1\textwidth]{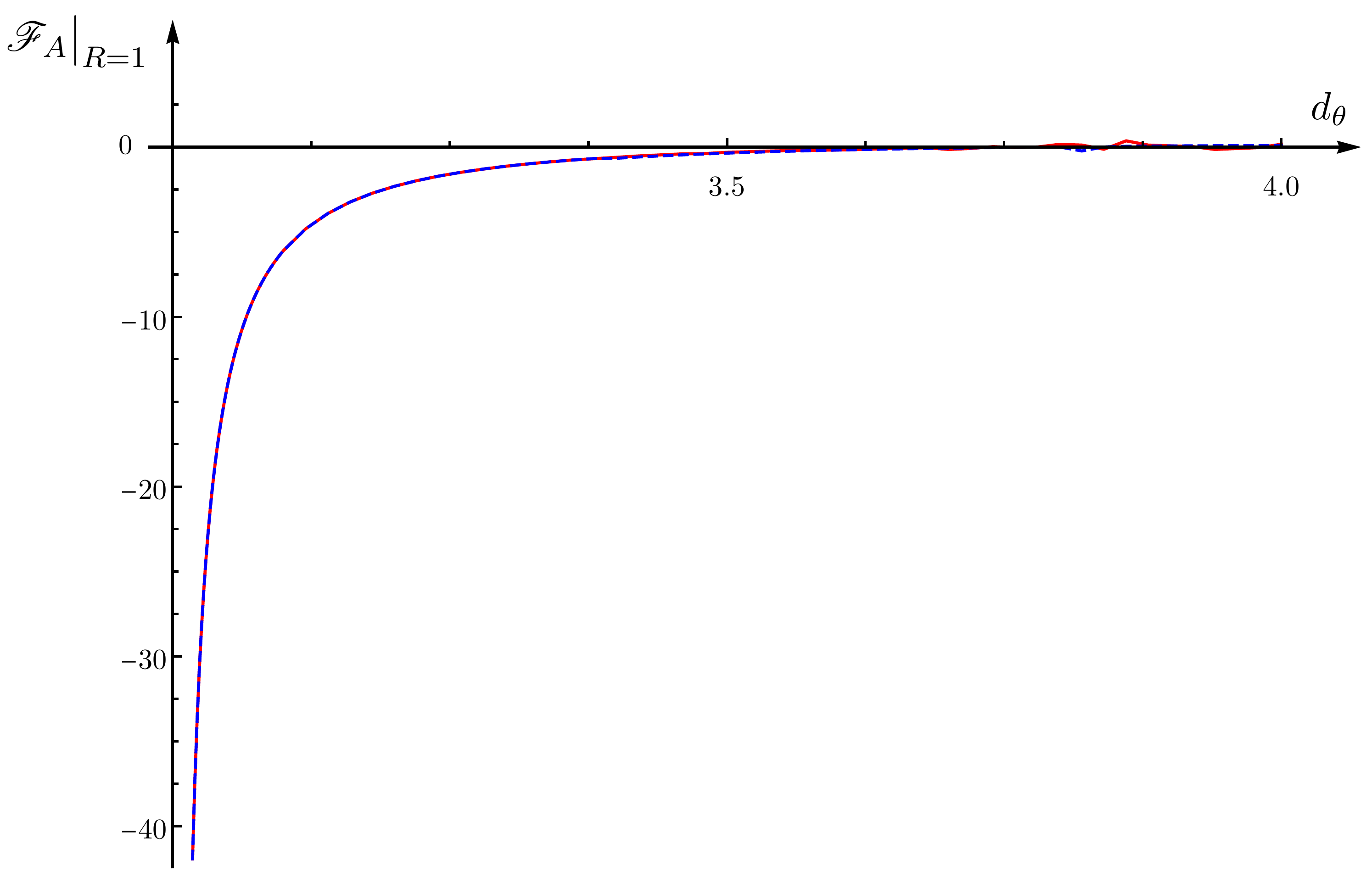}
	\vspace{-.0cm}
	\caption{\small
		Finite term $\mathscr{F}_A$  in terms of $3< d_\theta < 5$ for minimal surfaces anchored to
		a disk of radius $R=1$ in the hvLif$_{4}$ geometry (\ref{hyperscaling4bis})
		at $t =\textrm{const}$.
		The two curves have been obtained by first solving numerically (with Wolfram Mathematica) the differential equation \eqref{eqmotosphere}
		and then plugging the resulting profile either in \eqref{fasfera2} (solid red line) 
		or into \eqref{area_func_disk} (dashed blue line), once the area law term has been subtracted. 
	}
	\label{F_A_plot_sphere_2}
\end{figure}

In our numerical analysis we have employed  {\it Wolfram Mathematica} and {\it Surface Evolver} \cite{brakke,brakke2}.
Wolfram Mathematica has been used to solve numerically ordinary differential equations, 
which can be written whenever the symmetry of $A$ and of the gravitational background 
allows to parameterise $\gamma_A$ only in terms of a function of a single variable. 
In this manuscript, this is the case for the disk. 
Instead, Surface Evolver is more versatile in our three dimensional gravitational backgrounds 
(on a constant time slice) because it provides an approximation of the minimal surface $\hat{\gamma}_A$ 
through triangulated surfaces without implementing any particular parameterisation of the surface. 
In particular, once the three dimensional gravitational background has been introduced, 
given the UV cutoff $\varepsilon$ and the entangling curve $\partial A$,
only the trial surface (a rough triangulation that fixes the topology of the expected minimal surface)
has to be specified as initial data for the evolution. 
This makes Surface Evolver suitable to study the holographic entanglement entropy in AdS$_4$/CFT$_3$
for  entangling curve of generic shape, as already done in \cite{Fonda:2014cca,Fonda:2015nma,Seminara:2017hhh,Seminara:2018pmr}
(we refer the interested reader to these manuscripts for technical details about the application of Surface Evolver in this context).
We remark that, besides the position of the vertices of the triangulated surface, 
Surface Evolver can provide also the unit vectors normal to the triangles composing the triangulated surface. 
This information can be used to evaluate numerically the expressions discussed in Section\;\ref{fine_term}.

  Let us denote by $\hat{\gamma}_{A, \textrm{\tiny SE}}$ the best approximation of the minimal surface 
  obtained with Surface Evolver and by $\mathcal{A}_{\textrm{\tiny SE}}$ its area, 
  which depends on the value of $\varepsilon$ adopted in the numerical analysis.  
  These data allow to compute the finite term in the expansion of the holographic entanglement entropy in two ways:
  by subtracting the area law term from $\mathcal{A}_{\textrm{\tiny SE}}$
  or by plugging   the numerical data provided by Surface Evolver
  into the general formulas discussed in Section\;\ref{fine_term}.
  For $1<d_\theta < 3$, these two ways to find the finite term are given by
\be
 \label{FaSurfaceEvolver}
 F_{A,\textrm{\tiny SE}}
 \equiv
 -\Big(\mathcal{A}_{\textrm{\tiny SE}} -P_{A}/\varepsilon^{d_{\theta}-1}\Big) 
 \;\;\;\; \qquad \;\;\;\;
 \widetilde{F}_{A,\textrm{\tiny SE}} 
 \equiv 
 F_{A}\big|_{\hat{\gamma}_{A, \textrm{\tiny SE}}}
\ee
where $F_A$ is the expression in (\ref{finite term}).
In the range $3<d_\theta < 5$ we can write expressions similar to the ones in (\ref{FaSurfaceEvolver}) 
starting from (\ref{2div}) and (\ref{finiteminsurf2}).

 %

In Fig.\;\ref{F_A_plot_sphere} we show the finite term $F_{A}$ for a disk of radius $R=1$ 
as a function of the effective dimensionality $d_\theta$, in the range $1<d_{\theta}<3$,
when the gravitational background is hvLif$_4$.
The solid black curve has been found with Mathematica,
by solving numerically \eqref{eqmotosphere} first and then plugging the resulting radial profile for the minimal surface into \eqref{fasfera1}.
The data points have been found with Surface Evolver by using $F_{A,\textrm{\tiny SE}}$ (empty circles) 
and $\widetilde{F}_{A,\textrm{\tiny SE}}$ (empty triangles), introduced in \eqref{FaSurfaceEvolver}.
The very good agreement between the data points and the continuous curve provides 
a non trivial check both of the analytic formula \eqref{Fa1} 
and of the procedure implemented in Surface Evolver, that is sensible to the value of $d_\theta$.
For $d\simeq 3$ our numerical analysis fails; hence in Fig.\;\ref{F_A_plot_sphere}
we have reported only the reliable results.

An interesting feature that can be observed in Fig.\;\ref{F_A_plot_sphere} is the occurrence of a minimum for $F_{A}$ 
corresponding to $(d_\theta, F_A) \simeq (2.52\,, 4.67)$.
When the gravitational background is AdS$_4$, the bound $F_{A} \geqslant 2\pi$ holds for any entangling curve 
and the inequality is saturated for the disks \cite{Fonda:2015nma}.
From Fig.\;\ref{F_A_plot_sphere} we notice that, for hyperscaling violating theories,
$F_A$ assumes also values lower than $2\pi$ for certain $d_\theta$.

In Fig.\;\ref{F_A_plot_sphere_2} the finite term $\mathscr{F}_A$ for a disk of radius $R=1$ is shown in terms of $d_\theta$, in the range $3<d_\theta<5$,
when the gravitational background is hvLif$_4$.
The radial profile $z(\rho)$ for the minimal surface has been obtained by solving numerically  the equation of motion \eqref{eqmotosphere}.
Then, the finite term has been obtained  by plugging this result 
either into the area functional regularised by subtracting the divergent terms (solid red line)
or into the analytic expression \eqref{fasfera2} (dashed blue line).
In the figure we have reported only the reliable numerical data.


\subsubsection{Asymptotically hvLif$_4$ black hole}
 \label{sec disk bh}

 It is worth studying the holographic entanglement entropy of a disk of radius $R$ 
 when the gravitational background is the black hole (\ref{BH_metric}).
 By adopting the cylindrical coordinates, we can find the minimal surface among 
 the surfaces $\gamma_A$ invariant under rotations about the $z$-axis,
 characterised by their radial profile $z(\rho)$ such that $z(R)=0$, 
 as in Section\;\ref{sec disk hvLif}.
 The area functional for this class of surfaces reads
\begin{equation}
\label{area_func_BH_disk}
\mathcal{A}[\gamma_A]\,=\,
 2\pi\!
  \int_{0}^R  \frac{1}{z^{d_\theta}} \,\sqrt{1+\frac{(z')^2}{f(z)}} \,\rho \,d\rho\,.
\end{equation}

Under the rescaling $(z,\rho) \rightarrow \lambda (z,\rho)$, 
we have that $z_h \rightarrow \lambda z_h$, $R \rightarrow \lambda R$
and $\mathcal{A}[\gamma_A] \rightarrow \lambda^{2-d_\theta} \mathcal{A}[\gamma_A]$
for (\ref{area_func_BH_disk}).
This  rescaling leaves invariant both the equation of motion and the shape of the extremal surface $\hat \gamma_A$.

The unit vector normal to $\hat{\gamma}_A$ reads
\begin{equation}
\label{n_vector disk BH}
	\tilde n^\mu
	= \big(\tilde n^z,\tilde n^\rho, \tilde n^\phi \big)
	=\left( \,\dfrac{f(z)}{\sqrt{f(z)+(z')^{2}}} \, ,-\dfrac{z'}{\sqrt{f(z)+(z')^{2}}} \,, 0\right)
\end{equation}
where $z(\rho)$ satisfies the equation of motion coming from (\ref{area_func_BH_disk}).
By employing the component $\tilde n^z$ in (\ref{n_vector disk BH}),
we can specialise \eqref{Fa_BH} to this case, finding that for $1<d_\theta < 3$
the finite term of the holographic entanglement entropy of a disk in the black hole geometry (\ref{BH_metric})
is proportional to  
\begin{equation}
\label{Fa_BH_disk}
F_A=
\frac{2\pi}{d_\theta-1}\int_{0}^R \!
\bigg[ (d_\theta-1)(f(z)-1)-\frac{z f'(z)}{2}+\dfrac{f^2(z)}{f(z)+(z')^{2}} \left( \!1+\frac{z f'(z)}{2 f(z)} \right) \!\bigg]
\,\frac{\sqrt{1+(z')^2/f(z)}}{z^{d_\theta}}
\, \rho \,d\rho\,.
\end{equation} 
This expression scales like $F_A \rightarrow \lambda^{2- d_\theta} F_A$
under the rescaling introduced above. 

The radial profile characterising the minimal area surface $\hat{\gamma}_A$ 
can be found by solving the second order ordinary differential equation obtained 
by extremising the area functional (\ref{area_func_BH_disk}).
This can be done numerically for any $d_\theta$ (e.g. with Wolfram Mathematica).
Then, the finite term $F_A$ for $1< d_\theta <3$ 
can be found by plugging the resulting profile into the integral \eqref{area_func_BH_disk} properly regularised 
and subtracting the leading divergence (\ref{divercont1}),

 \begin{figure}[t!]
 	\vspace{-.2cm}
 	\hspace{-1.1cm}
 	\begin{center}
 		\includegraphics[width=1\textwidth]{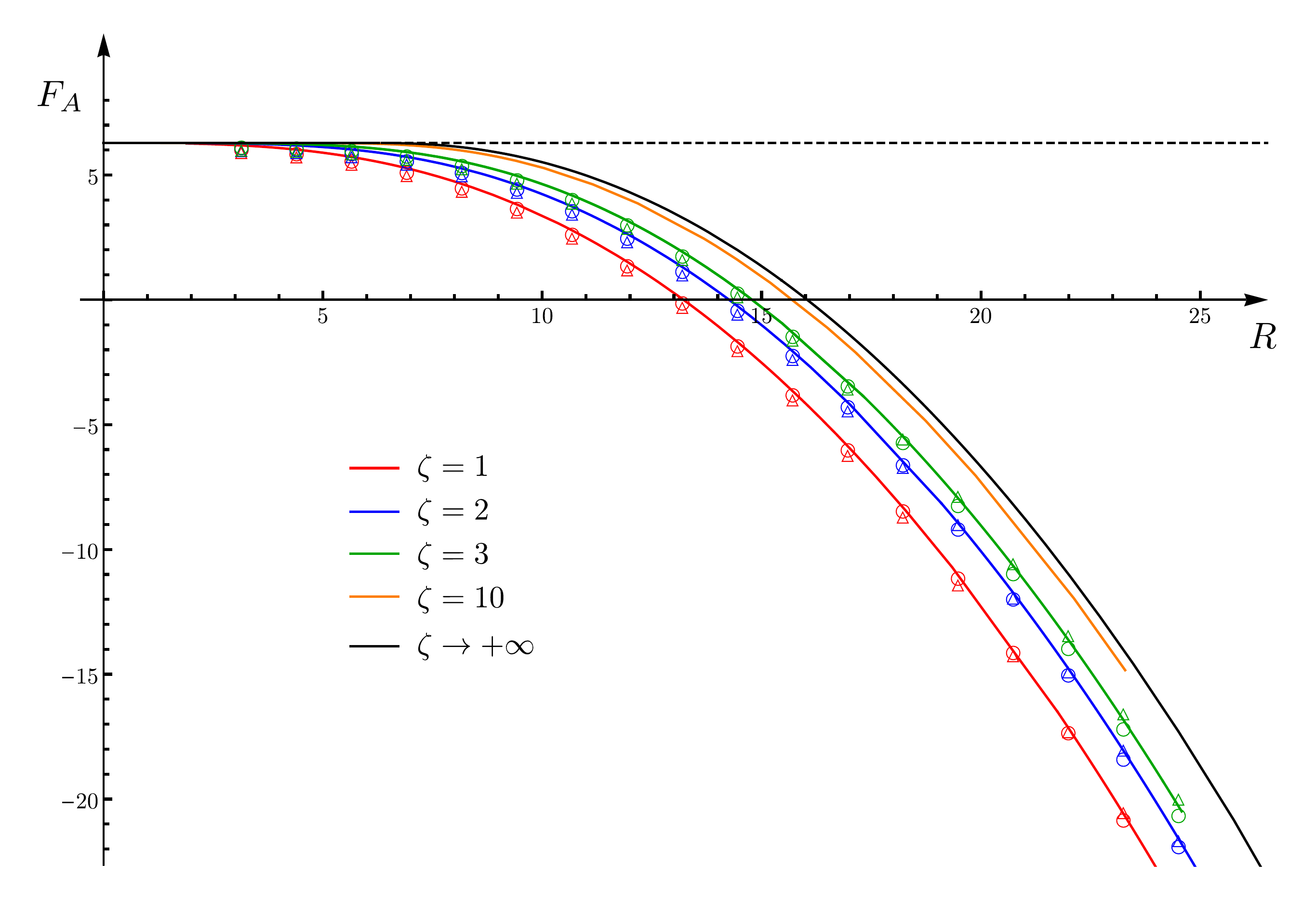}
 	\end{center}
 	\vspace{-.5cm}
 	\caption{\small
 		Finite term $F_{A}$ for minimal surfaces anchored to
		a disk of radius $R$ when the bulk metric is the black hole \eqref{BH_metric}, 
		with $d_\theta=2$, different values of $\zeta$ and the horizon set to $z_h=1$.
		The solid black curve corresponds to the analytic solution (\ref{hardwall_FA}) described in Section\;\ref{sec hard wall},
 		while the remaining coloured solid lines have been obtained by evaluating \eqref{Fa_BH_disk} on the minimal surface 
		whose radial profile has been found
		by solving numerically the equation of motion of (\ref{area_func_BH_disk}).
		The data points labelled by the empty circles and the empty triangles  
		have been obtained with Surface Evolver through the two formulas in \eqref{FaSurfaceEvolver} respectively.
		The horizontal black dashed line corresponds to $F_A =2\pi$, that gives the finite term of the holographic entanglement entropy
		of disks when the gravitational background is  AdS$_4$.
 	}
 	\label{F_A_plot_BH_varyingzeta}
 \end{figure}

In order to check our results, we have studied  
the finite term $F_A$  as a function of the radius $R$ for different values of $\zeta$,
where the gravitational background given by the black hole \eqref{BH_metric} with 
fixed $d_\theta=2$ and the black hole horizon set to $z_h=1$.
The results are shown in Fig.\,\ref{F_A_plot_BH_varyingzeta}, 
where the same quantity has been computed by employing 
analytic expressions and numerical methods based either on Mathematica or on Surface Evolver,
finding a remarkable agreement. 
For very small regions, $F_A$ tends to $2\pi$ as in the AdS$_4$ and, in particular, it is independent on $\zeta$. For very large regions we expect to obtain the behaviour  \eqref{large_BH2}, indepedent of $\zeta$, 
while for intermediate sizes $F_A$ depends on $\zeta$ in a non trivial way.

Let us remark that, in Fig.\,\ref{F_A_plot_BH_varyingzeta}, the curves having $d_\theta =2$ and different $\zeta$ 
tend to accumulate toward a limiting curve as $\zeta$ increases.
In Section\;\ref{sec hard wall} we provide the analytic expression of this limiting curve.

\begin{figure}[t!]
	\vspace{-.2cm}
	\hspace{-1.1cm}
	\begin{center}
		\includegraphics[width=1\textwidth]{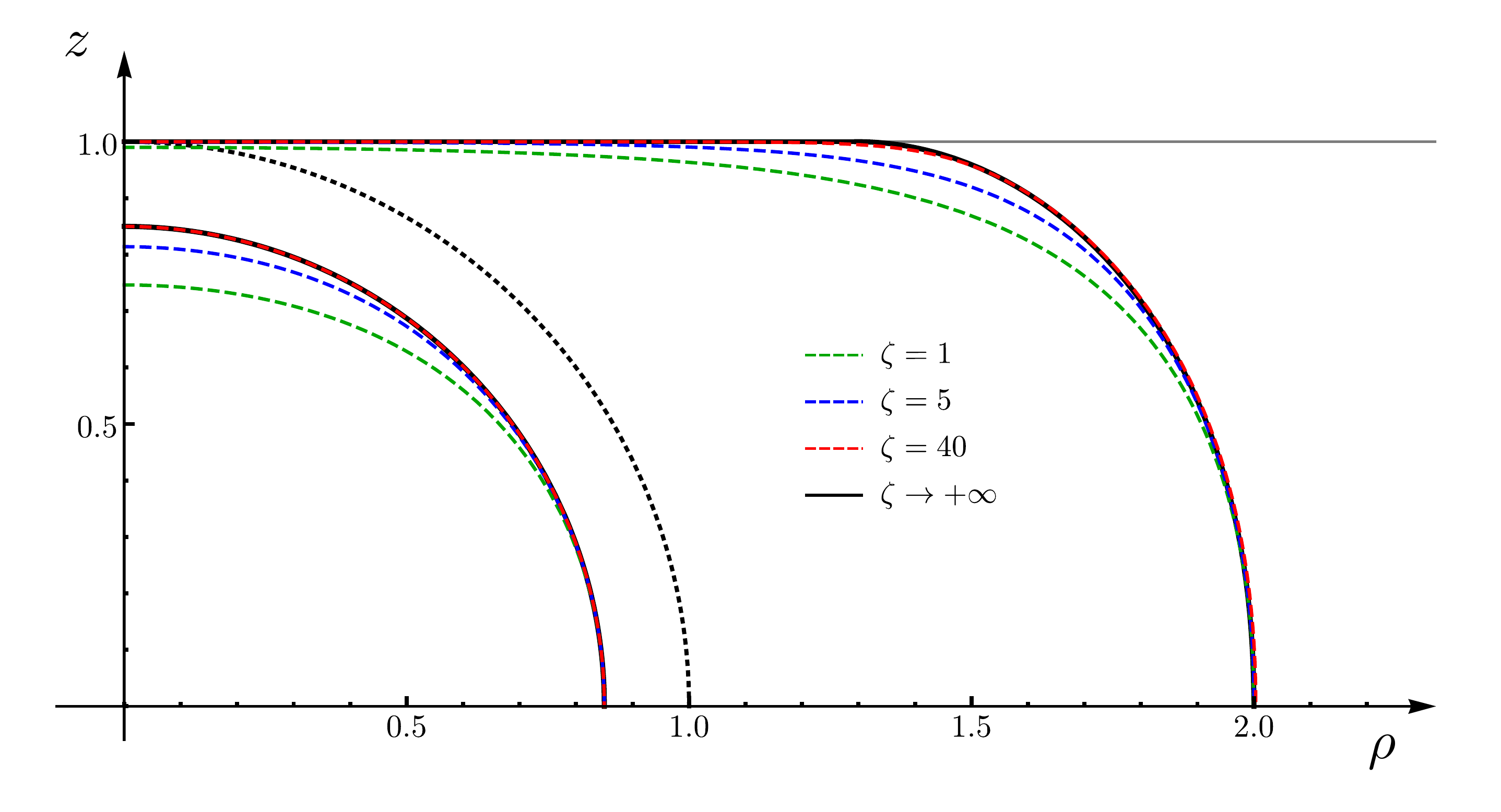}
	\end{center}
	\vspace{-.5cm}
	\caption{\small
		Radial profiles of minimal surfaces anchored to disks with $R=0.85$ and $R=2$ 
		in the black hole background \eqref{BH_metric} for $d_\theta=2$ and different values of $\zeta$. 
		The grey horizontal line is the black hole horizon at $z_h=1$.
		The solid black lines correspond to the asymptotic regime $\zeta \rightarrow +\infty$: 
		when $R \leqslant z_h$ they are hemispheres $z(\rho)=\sqrt{R^{2}-\rho^{2}}$,
		otherwise they are given by \eqref{profile_har_wall}.
		The coloured dashed lines, that correspond to some finite values of $\zeta$,
		are radial profiles obtained numerically with Mathematica. 
	}
	\label{fig_hard_wall}
\end{figure}

\subsubsection{Analytic solution for $d_\theta =2$ and $\zeta \to \infty$}	
\label{sec hard wall}

	Analytic solutions for the minimal  surfaces anchored to the disk with radius $R$ can be found  for  the black hole background 
	\eqref{BH_metric} in the asymptotic regime given by $d_\theta=2$ and large $\zeta$. 
	In this limit the original black hole geometry collapses to AdS$_4$ for $z \leqslant z_h$, with an event horizon located at $z=z_h$. 
	The  horizon prevents  the minimal surface from entering  the region $z>z_h$.
	
	When $R/z_h\leqslant1$,  the minimal surface  is provided by the usual hermisphere, that in cylindrical coordinates reads $z(\rho)=\sqrt{R^{2}-\rho^{2}}$. When $R/z_h>1$, the extremal surface consists of two branches: a non trivial profile connecting the conformal boundary to the horizon and a flat disk that lies on the horizon.  The detailed procedure to construct analytically this minimal surface is given in  Appendix\;\ref{appendix_hard_wall_solution}
	and below we summarize the main results.

	In cylindrical coordinates, the profile of the minimal surface  for $R/z_h>1$ is parametrically defined by 
	\begin{equation}
	\label{profile_har_wall}
	(z,\rho)=
	\Bigg\{
	\begin{array}{ll}
	R\, e^{q_{+,k}(\hat z)} (\hat z ,1) \qquad  \;& 0<\hat z<k^{1/4} \\
	\rule{0pt}{.5cm}
	(z_h,\rho) \qquad \;  & 0<\rho< z_h/k^{1/4} 
	\end{array}
	\end{equation} 
	where $\hat z= z/\rho$ and $k$ is an integration constant whose value as function of $R /z_h$ is  determined by 
	the following condition
	\begin{equation}
	\label{Rvsk}
	\frac{R}{z_h} = \frac{e^{q_{+,k}(k^{1/4})}}{k^{1/4}}\,. 
	\end{equation}
	The function $q_{+,k}(\hat z)$ is one of the two functions emerging from the integration of the differential equation for the extremal
	surface (see Appendix\;\ref{appendix_hard_wall_solution}). They  both  can be written in terms of elliptic integrals of different kinds:
	\begin{equation}
	\label{qpm}
	q_{\pm,k} (\hat z) =
	\frac{1}{2} \log (1+\hat z^2)
	\pm
	\kappa \, 
	\sqrt{\frac{1-2\kappa^2}{\kappa^2-1}}\,
	\Big[\,
	\Pi \big(1-\kappa^2, \Omega(\hat z) | \kappa^2 \big)
	-
	\mathbb{F}\big(\Omega(\hat z) | \kappa^2\big)
	\Big]
	\end{equation}
	with
	\begin{equation}
	\Omega(\hat z)
	\,\equiv\, 
	\arcsin \bigg(
	\frac{\hat z/\hat z_m}{\sqrt{1+\kappa^2({\hat z^2/\hat z^2_m-1)}}}
	\bigg)
	\qquad 
	\kappa \equiv
	\sqrt{\frac{1+\hat z_m^2}{2+\hat z_m^2}}
	\end{equation}
	where  $\hat z_m^2= (k+\sqrt{k(k+4)})/2$.
	
	In Fig.\,\ref{fig_hard_wall}, we have plotted the profile
	of the minimal surfaces in the limit $\zeta \rightarrow +\infty$ for two different radii $R=0.85$ and $R=2$ (continuous black lines). 
	In the former case the solution is the hemisphere, while in the latter one it is given by the profile \eqref{profile_har_wall}. 
	As a consistency check, we have obtained numerically (with Mathematica) the radial profiles for finite values of $\zeta$ (coloured dashed lines),
	finding that they approach the analytical solution as $\zeta$ increases.

	We can now compute the  finite term $F_A$  for this family of surfaces and the result reads
	\begin{equation}
	\label{hardwall_FA}
	F_A = 
	\left\{
	\begin{array}{ll}
	2\pi \hspace{4cm} & \mbox{when} \qquad R\leqslant z_h \\
	\rule{0pt}{.5cm}
	\displaystyle
	2\pi \left( \mathcal{F}_k(k^{1/4})-\frac{1}{2\sqrt{k}}  \right) & \mbox{when} \qquad R > z_h 
	\end{array}
	\right.
	\end{equation}
	with
	\begin{equation}
	\label{F_A int_evaluated}
	\mathcal{F}_k(\hat z)
	\,\equiv \,
	\frac{\sqrt{k(1+\hat z^2) -\hat z^4}}{\sqrt{k}\,\hat z}
	\,-\,
	\frac{\mathbb{F} ( \arcsin ( \hat z/\hat z_m )\,|-\hat z_m^2-1 )
		-
		\mathbb{E} ( \arcsin ( \hat z/\hat z_m )\,|-\hat z_m^2-1 )
	}{
		\hat z_m}
	\end{equation}
	where $\mathbb{F}$ and $\mathbb{E}$ are the first and second elliptic integral respectively.
	The curve (\ref{hardwall_FA}) is a continuous function of $R$.

	The solid black curve in Fig.\,\ref{F_A_plot_BH_varyingzeta} 
	has been obtained by a parametric plot employing \eqref{Rvsk} and \eqref{hardwall_FA} (with $z_h=1$)
	for $R > 1$, while $F_A=2\pi$ for $R < 1$.

\subsection{Ellipses}
\label{ellipsecase}

 \begin{figure}[t!]
	\vspace{-.2cm}
	\hspace{-1.1cm}
	\begin{center}
		\includegraphics[width=1\textwidth]{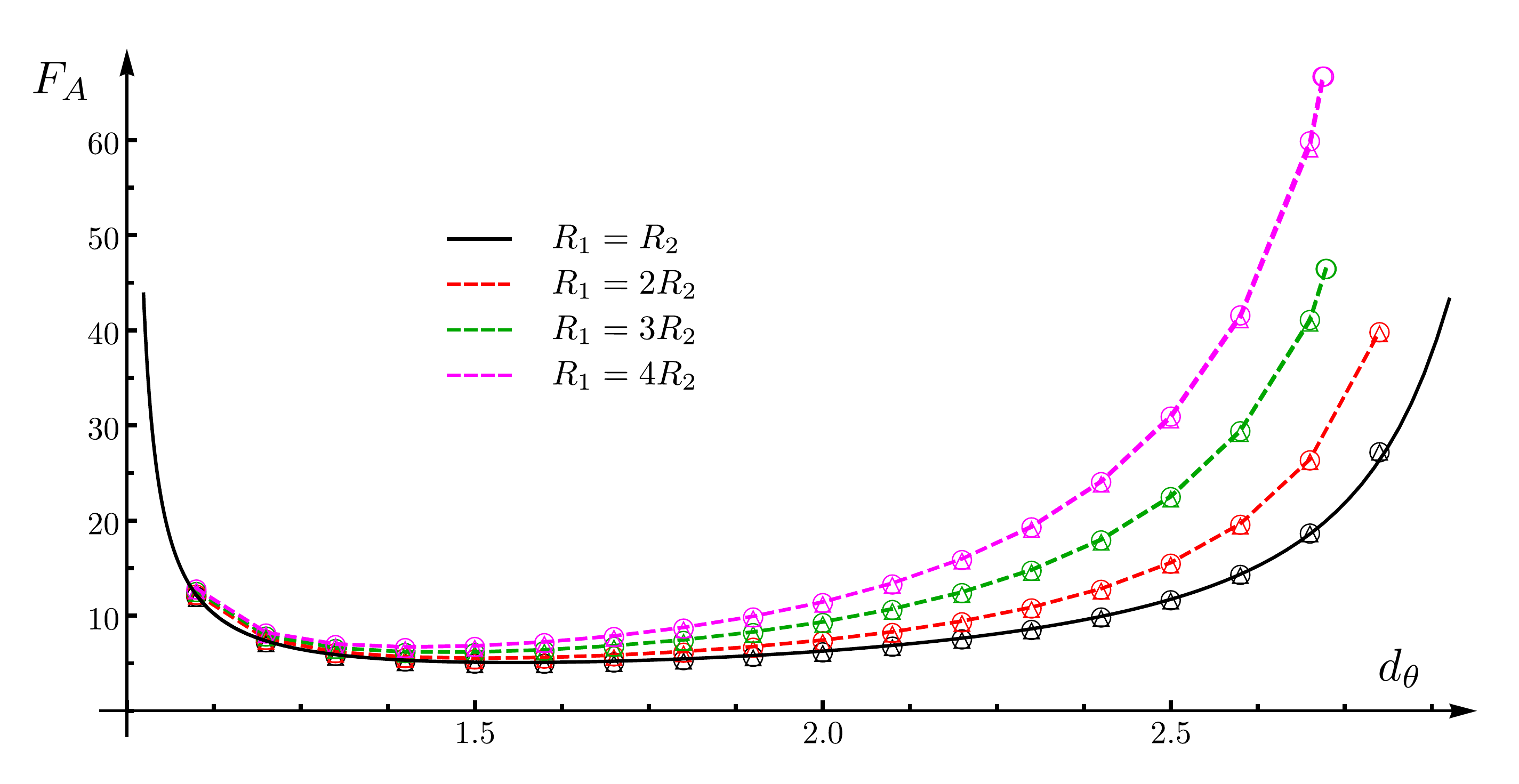}
	\end{center}
	\vspace{-.5cm}
	\caption{\small
		Finite term $F_A$  in terms of $d_\theta$ in the range $1<d_\theta<3$ 
		for minimal surfaces in hvLif$_4$ anchored to ellipses $A$ having fixed perimeter $P_A =1$.
		Different colours correspond to ellipses with different eccentricity.
		The data points have been obtained with Surface Evolver
		in the two ways described in (\ref{FaSurfaceEvolver}) (the markers have been assigned as in the previous figures).
		The solid black curve, that corresponds to the disk, is the curve reported in Fig.\;\ref{F_A_plot_sphere}
		multiplied by $(P_A/(2\pi R))^{2-d_\theta}$. 
	}
	\label{ellipses}
\end{figure}

The main feature of the analytic expressions obtained in Section\,\ref{sec HEE} and Section\,\ref{sec:time-dependent} 
for the finite term of the holographic entanglement entropy
is that they hold for any smooth shape of the entangling curve. 
In order to evaluate these formulas for explicit domains, 
one needs to know the entire minimal surface $\hat{\gamma}_A$ and this task is usually very difficult, 
in particular when the entangling curve does not display some useful symmetry. 
Surface Evolver can be employed to study numerically $\hat{\gamma}_A$
for a generic smooth entangling curve $\partial A$,
as already done in some asymptotically AdS$_4$ backgrounds
\cite{Fonda:2014cca,Fonda:2015nma,Seminara:2017hhh,Seminara:2018pmr}.


In this subsection we consider the finite term of the holographic entanglement entropy of ellipses
when the gravitational spacetime is hvLif$_4$ in \eqref{finiteminsurf2} or
the asymptotically hvLif$_4$ black hole \eqref{BH_metric}.

In Fig.\,\ref{ellipses}, we show the finite term $F_A$ of elliptic regions having the same perimeter $P_A=1$
as a function of the effective dimension $1<d_\theta<3$, when the bulk is hvLif$_4$.
Ellipses with different eccentricity $e$ have been considered
(we recall that $e= \sqrt{1-(R_1/R_2)^2} \in [0,1)$, being $R_1 \leqslant R_2$ the semi-axis of the ellipse).
The numerical data have been obtained with Surface Evolver 
and $F_A$ has been found through the two different methods described in \eqref{FaSurfaceEvolver}.
In particular, the empty circles and the empty triangles correspond respectively to $F_{A,\textrm{\tiny SE}}$ and $\widetilde{F}_{A,\textrm{\tiny SE}}$
(the coloured dashed lines just join the data points).
The solid black line gives the finite term for disks and it has been obtained by using Mathematica
(it is the same curve shown in Fig.\,\ref{F_A_plot_sphere},multiplied by the factor $(P_A/(2\pi R))^{2-d_\theta}$).

\begin{figure}[t!]
	\vspace{-.2cm}
	\hspace{-.8cm}
	\includegraphics[width=1.08\textwidth]{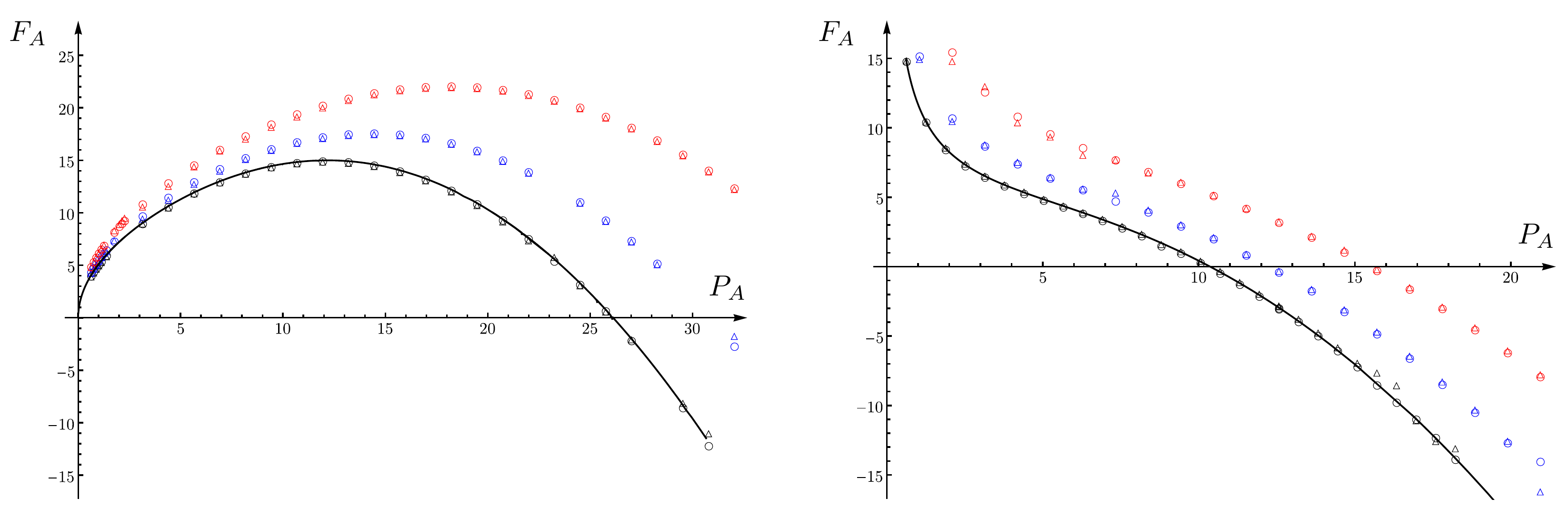}
	\vspace{-.4cm}
	\caption{\small
		Finite term $F_A$ in terms of the perimeter $P_A$ for minimal surfaces in the 
		asymptotically hvLif$_4$ black hole (\ref{BH_metric})
		anchored to ellipses $A$. The Lifshitz exponent is fixed to $\zeta=1.5$,
		while $d_\theta=1.5$ in the left panel and $d_\theta=2.5$ in the right panel.
		Different colours correspond to ellipses with different eccentricity: disk (black), $R_2=2 R_1$ (blue) and $R_2=3 R_1$ (red).
		The data points labelled by the empty circles and the empty triangles  
		have been obtained with Surface Evolver through the two formulas in \eqref{FaSurfaceEvolver} respectively.
		The solid black curves for disks have been found numerically by employing Mathematica. 
		All the curves and the data points have been obtained by using (\ref{finite term}).
	}
	\label{Faplot_BH_ellipses}
\end{figure}

The finite term $F_A$ when the bulk metric is the black hole \eqref{BH_metric} depends also on $d_\theta$.
In Fig.\,\ref{Faplot_BH_ellipses} we show $F_A$ for ellipses having different eccentricity in terms of their perimeter $P_A$
for two different values of $d_\theta$ ($d_\theta=1.5$ in the left panel and $d_\theta=2.5$ in the right panel) and the same value of 
the Lifshitz parameter $\zeta=1.5$. 
Also in this case, the data points have been found by evaluating numerically \eqref{Fa_BH} on the 
approximated minimal surfaces obtained with Surface Evolver,
while the solid black curve has been obtained numerically by using Mathematica. 
The very good agreement between the various methods provides a highly non trivial check of the general formula \eqref{finite term}.  

A qualitative difference can be observed between the two panels in Fig.\,\ref{Faplot_BH_ellipses}.
Indeed, for very small regions the behaviour of $F_A$ depends on $d_\theta$. 
In particular, when $P_A \rightarrow 0$, we have that 
$F_A \rightarrow 0$ for $d_\theta<2$ while $F_A \rightarrow + \infty$ for $d_\theta>2$. 
This can be understood by observing that the finite term $F_A$ of small regions
(whose maximal penetration in the bulk is very far from the horizon) is not influenced by the occurrence of the horizon, 
hence it scales approximately as in \eqref{FascalingR}, which is valid in hvLif$_4$. 

%
%
%
%
%
%
%
%
%
%
%
%


\section{Conclusions}
\label{sec:conclusions}

In this manuscript we have explored the shape dependence of the holographic entanglement entropy in AdS$_4$/CFT$_3$
in the presence of Lifshitz scaling and hyperscaling violation.
Both static and time dependent backgrounds have been studied and, 
for the sake of simplicity, we restricted to smooth entangling curves 
and to the regime $1\leqslant d_\theta \leqslant 5$ for the hyperscaling parameter.
In the expansion of the holographic entanglement entropy as the UV cutoff $\varepsilon$ vanishes,
both the divergent terms and the finite term have been analysed.

Our main results are analytic expressions for the finite term that can be applied for any smooth entangling curve:
for static backgrounds, they are given by (\ref{finite term}) when $1< d_\theta <3$ and by (\ref{finiteminsurf2}) when $3< d_\theta <5$;
for time dependent backgrounds, we have obtained (\ref{Fa time-dep extremal surf}) when $1< d_\theta <3$.
In the regime $1< d_\theta <3$, the finite term for static and time dependent backgrounds
has been studied also for surfaces that intersect orthogonally the boundary along smooth curves,
finding the expressions (\ref{finitegeneralsurface}) and (\ref{Fa time-dep}) respectively. 
This class of surfaces include the extremal surfaces providing the holographic entanglement entropy.

When $d_\theta \in \{1,3,5\}$, a logarithmic divergence occurs in the expansion 
of the holographic entanglement entropy. 
The coefficient of this divergence is determined only by the geometry of the entangling curve
and its analytic expression for a generic smooth entangling curve 
has been reported in (\ref{leadinglog}), (\ref{2divlog}) and (\ref{log_div_5}) respectively.

The new results summarised above have been found by extending the analysis first performed in \cite{Babich}
and then further developed in \cite{Mazzeo,Fonda:2015nma,Seminara:2018pmr} 
for gravitational backgrounds having $d_\theta =2$.

We find it worth mentioning two other analytic results obtained in this manuscript. { For hvLif$_{d+1}$ spacetime
we showed that 
the finite term of the extremal surface can be expressed as an integral over the entangling surface, 
since  the background metric admits a conformal Killing vector generating dilatations. Moreover  we have  briefly discussed the
extension of this result to more general geometries.}
By applying this result to hvLif$_{d+1}$, the simple expression \eqref{Fa=udtheta} is found for the finite term, 
valid in any dimension and for any $d_\theta > 1$. 
Another result has been obtained for the asymptotically hvLif$_4$ black hole (\ref{BH_metric})
in the asymptotic regime given by $d_\theta =2$ and $\zeta \to \infty$,
where we have found the analytic expression of the minimal surface anchored to a disk 
and of the finite term in the expansion of its area.


For the static backgrounds given by the hvLif$_4$ spacetime (\ref{hyperscaling4bis})
and the asymptotically hvLif$_4$ black hole (\ref{BH_metric}), 
a numerical analysis has been performed by considering disks and ellipses.
Disks have been studied mainly through the standard Wolfram Mathematica,
while for the ellipses we have employed Surface Evolver \cite{brakke,brakke2},
a software that has been already used to explore the shape dependence 
of the holographic entanglement entropy for four dimensional gravitational backgrounds
\cite{Fonda:2014cca,Fonda:2015nma,Seminara:2017hhh,Seminara:2018pmr}.
A very good agreement between the analytic expressions in (\ref{finite term}) and (\ref{finiteminsurf2}) 
and the numerical data has been observed.


The results reported in this manuscript can be extended in various directions. 
We find it worth exploring $d_\theta > 5$ because other divergent terms occur and 
it is interesting to understand their dependence on the shape of the entangling curve.
Also the numerical approach employed in this manuscript deserves further studies. 
For instance, it is important to extend the application of Surface Evolver to time dependent backgrounds, 
both to check on non spherical finite regions 
the analytic expressions for the finite term in the expansions of the holographic entanglement entropy
found in \cite{Fonda:2015nma} and in Section\;\ref{sec:time-dependent} of this manuscript
and to improve the current understanding of the shape dependence 
of the holographic entanglement entropy.

\subsection*{Acknowledgments}

We thank Alexander Bobenko, Matthew Headrick, Veronika Hubeny, 
Hong Liu, Tatsuma Nishioka, Costantino Pacillo, Mukund Rangamani and Paola Ruggiero
for useful discussions. 
We are grateful to the Galileo Galilei Institute for Theoretical Physics in Florence, 
where part of this work has been done during the program {\it Entanglement in Quantum Systems}.
JS thanks the Center for Theoretical Physics at MIT for warm hospitality during part of this work
and the MIT-FVG project for financial support. 
JS and ET are grateful to the Yukawa Institute for Theoretical Physics at Kyoto University,
where this work was completed during the workshop YITP-T-19-03 {\it Quantum Information and String Theory 2019}.
%


%
\appendix
\section{Null Energy Condition}
\label{app:NEC}

In this appendix we discuss the constraints for the Lifshitz and the hyperscaling exponents 
imposed by the Null Energy Condition (NEC), that has been introduced in Section\;\ref{sec HEE}.

Let us consider spacetimes whose metric has the following form
\begin{equation}
\label{metric_class}
ds^{2}=e^{2 A(z)}\Bigl(-e^{2B(z)}f(z)dt^{2} +\dfrac{dz^{2}}{f(z)}+dx^{2}+dy^{2} \Bigr)
\end{equation}
for some $A(z)$, $B(z)$ and $f(z)$, being $z>0$ the holographic coordinate. 
In \cite{Dong:2012se}, it is shown that the NEC leads to the following constraints
\bea
\label{null_energy_ineq1}
& &
 (2 A'+3 B')f'+2 f(2 A' B'+B'^2+B'')+f'' \geqslant 0  
 \\
 \label{null_energy_ineq2}
 \rule{0pt}{.5cm}
 & &
  f(A'^2+A' B'-A'')\geqslant 0\,.
\eea

Since we are mainly interested in the black hole metric \eqref{BH_metric}, let us fix the functions 
$A(z)$, $B(z)$ and $f(z)$ as follows
\begin{equation}
\label{BH_choice}
A(z)= -\frac{d_\theta}{2} \log z 
\qquad 
B(z)=(1-\zeta)\log z 
\qquad 
f(z)= 1- \left( \frac{z}{z_h} \right)^{\chi_1}+ a \,z^{\chi_2}
\end{equation}
where $a$ is a constant. 
Plugging \eqref{BH_choice} into \eqref{null_energy_ineq1} and \eqref{null_energy_ineq2}, one obtains respectively
\bea
\label{null_energ_spec1}
& &\hspace{-.4cm}
d_\theta (d_\theta + 2 \zeta -4)f \geqslant 0
\\
\label{null_energ_spec2}
\rule{0pt}{.7cm}
& &\hspace{-.4cm}
2 (d_\theta+\zeta)(\zeta-1)+ \left(\frac{z}{z_{h}} \right)^{\chi_1}  (d_\theta+\zeta-\chi_1)(2-2\zeta +\chi_1) 
- a \,z^{\chi_2} \left( d_\theta + \zeta-\chi_2 \right)(2-2\zeta +\chi_2) \geqslant 0.
\nonumber
\\
&&
\eea

Restricting to the region of spacetime outside the horizon, where $f>0$, 
one observes that \eqref{null_energ_spec1} provides the same constraint holding in the hvLif$_4$, that is the first inequality in \eqref{NEC}. 
The constraint  \eqref{null_energ_spec2} is more involved because it depends on the coordinate $z$ in a non trivial way. 
Notice that the second inequality in \eqref{NEC} is recovered by taking $z \rightarrow 0$ in  \eqref{null_energ_spec2}.

Let us focus on the simple case given by $a=0$ and assume that $\chi_1 \geqslant 0$, 
in order to have an asymptotically hvLif$_4$ background (this class of metrics includes (\ref{BH_metric})).  
Taking the limit $z \rightarrow z_h$ in the inequality \eqref{null_energ_spec2} with $a=0$, one finds 
$ \chi_1 \leqslant d_\theta + 3\zeta- 2$.
Setting $\chi_1=d_\theta+\zeta \geqslant 0$ as in \eqref{BH_metric}, one obtains $\zeta-1 \geqslant 0$
corresponding to the first constraint in \eqref{NEC}.


\section{Expansion of the area near the boundary}
\label{appxUVbehav}
This appendix  is devoted to review the derivation of the expansion near the boundary 
of the area functional $\mathcal{A}[ \gamma_A]$ for two dimensional surfaces $\gamma_A$ 
that intersect orthogonally the boundary $\partial \mathcal{M}_3$.   
In the following we adapt the analysis reported in \cite{Mazzeo} to the gravitational backgrounds of our interest. 
Since  the structure of   this expansion  depends only on the local geometry of  $ \gamma_A  $ near  $\partial \mathcal{M}_3$, 
we may as well suppose that  $ \mathcal{M}_{3} $ is  conformally flat ({i.e.} $ \widetilde{\mathcal{M}}_{3}=\mathbb{R}^{3}$)  
and  that the form  (\ref{statichyperscaling}) of the metric is valid for any value of the coordinate $z$. 
 The analysis below can  be also adapted directly to spaces whose metric is  only asymptotically of the form \eqref{statichyperscaling}, 
 though the equations involve higher order correction terms and  the procedure becomes more complicated.
 
The boundary curve $ \partial\gamma_A \subset\partial\widetilde{\mathcal{M}}_{3}\equiv\mathbb{R}^{2} $ is taken to be smooth 
and its  parametric form  $ \boldsymbol{x}_A(s)$ is given by $(x(s),y(s)) $, 
being $s$ the affine parameter. 
At each non singular point of $ \partial\gamma_A $ the unit tangent  vector $ \widetilde{T}=\boldsymbol{x'}_A(s) $ 
and the  normal one $ \widetilde{N}$ provide  a basis for the boundary plane $\partial\widetilde{\mathcal{M}}_{3}$.  
Then, let us consider the vertical cylinder $ \Gamma\subset\widetilde{\mathcal{M}}_{3} $ 
constructed over the curve $\boldsymbol{x}_A(s) $, which is given by
$ \{(z,x,y)\in \mathcal{M}_{3} \, | \,(z,\boldsymbol{x}_A(s))\}$.
Near $ \partial \widetilde{\mathcal{M}}_{3}, $ i.e. close to the boundary  plane $ z=0, $ we can parametrize the surface $ \gamma_A $ 
as a horizontal graph over $\Gamma$. This means that we can introduce a scalar function $u(s,z)$ so that the embedding $E(s,z)$ 
of $\gamma_A$ takes the form
\begin{equation}
\label{embed}
E(s,z)=\bigl( \,z\,,\boldsymbol{x}_A(s)+u(s,z)\widetilde{N}\,\bigr)\,.
\end{equation}
The function $ u(s,z) $ in \eqref{embed} describes the displacement of $ \gamma_A $ from the vertical cylinder over $ \partial\gamma_A $. The boundary condition $ E(s,0)=\boldsymbol{x}_A(s)$ implies that $ u(s,0)=0,$  and thus  the partial derivative with respect to $ s $  at $ z=0$ vanishes as well, i.e. $ u_{s}(s,0)=0. $ 
From (\ref{embed}) one finds the two vectors tangent to the surface by taking the derivative with respect to $s$ and $z$
\begin{equation}
\label{tangentvectsurf}
t_{1}=E_{s}(s,z)=\bigl(0,w(s,z)\widetilde{T}+u_{s}\widetilde{N}\bigr) 
\;\;\qquad \;\;
t_{2}=E_{z}(s,z)=\bigl(1,u_{z}\widetilde{N}\bigr)
\end{equation}
where we have introduced $ w(s,z)=1-k(s)u(s,z)$, 
being $ k(s) $ the geodesic curvature of the entangling curve $ \boldsymbol{x}_A(s)$. 

The scalar product of the vectors in (\ref{tangentvectsurf}) provides the metric $\tilde{h}_{ab}$ (and the its inverse $\tilde{h}^{ab}$)  
induced  on the surface  by the embedding (\ref{embed}) 
\begin{equation}
\label{inducedmetric}
\tilde{h}_{ab}=
\begin{pmatrix}
w^{2}+u_{s}^{2} & u_{z}u_{s} \\
u_{z}u_{s} & 1+u_{z}^{2}
\end{pmatrix} 
\;\;\qquad \;\;
\tilde{h}^{ab}=\frac{1}{\tilde{h}}
\begin{pmatrix}
1+u_{z}^{2} & -u_{z}u_{s} \\
-u_{z}u_{s} & w^{2}+u_{s}^{2}
\end{pmatrix}
\end{equation}
where $\tilde{h} = \textrm{det}(\tilde{h}_{ab})=u_{s}^{2}+w^{2}(1+u_{z}^{2})$.
The inward unit normal vector $ \tilde{n}_{\mu} $ can be  evaluated by taking the  normalized wedge product of $ t_{1} $ and $t_2$,
finding that
\begin{equation}
\label{normal}
\tilde{n}^{\mu}
=\dfrac{\bigl( t_{1}\wedge t_{2}\bigr)^{\mu}}{|t_{1}\wedge t_{2}|}
=\frac{1}{\sqrt{\tilde{h}}} \, \Bigl(-\, u_{z}w \, ,- \, u_{s} \,\widetilde{T}+w\widetilde{N}\Bigr)\,.
\end{equation}

In order to study the behaviour of  the minimal surface  $\hat{\gamma}_A$ near the boundary $z=0$, 
we expand the function $u=u(s,z)$ in a power series of $z$ about $z=0$ as follows
\begin{equation}
\label{svil}
u(s,z)\!=\! 
\frac{U_{2}(s)}{2}\,z^{2}+\frac{U_3(s)}{3!}\,z^{3}+\frac{U_{4}(s)}{4!}\,z^{4}
+ \dots + z^\alpha \left[ \, \mathcal{U}_{\alpha}(s) +
\mathcal{U}_{\alpha+1}(s)\,z+  \mathcal{U}_{\alpha+2}(s)\,\frac{z^2}{2!}+\dots \,\right]
\end{equation}
where we have assumed that  this expansion may contain both an analytic and a non analytic part,  
in order to be consistent with the non analytic behaviour of the bulk metric near the boundary. 
The  non analytic component is controlled by a real exponent $\alpha$.  
The boundary condition $u(s,0)=0$ has been employed to set  $U_0(s)=0$ in \eqref{svil}.
Instead, the requirement that $\gamma_A$ intersects orthogonally the plane $z=0$  leads to $U_1(s)=0$ and $\alpha>1$.
In fact, if we use  the expression in \eqref{tangentvectsurf} for $t_2^\mu$, 
we immediately recognize that this condition translates into $u_z(s,0)=0$, which in turn entails the  above two constraints. 
In the following  we shall adopt the stronger requirement  $\alpha \geqslant d_\theta+1$.
This  ensures that the structure of the divergences is determined only by the analytical part of the expansion 
and, moreover, it is automatically satisfied by a minimal surface, as discussed below.  

From \eqref{inducedmetric}, we can easily write  the regularized area functional as follows
\begin{equation}
\label{area1}
\mathcal{A}[\gamma_{A,\varepsilon}]
=
\int_{\gamma_{A,\varepsilon}}\dfrac{1}{z^{d_{\theta}}} \, \sqrt{\tilde{h}}\,d\Sigma\,
=
\int_{\gamma_{A,\varepsilon}}\dfrac{1}{z^{d_{\theta}}} \, \sqrt{u_{s}^{2}+w^{2}(1+u_{z}^{2})}\;ds\, dz
\end{equation}
where $\gamma_{A,\varepsilon} \equiv \gamma_A \cap \{z \geqslant \varepsilon\}$.
Assuming that the embedding function $u(s,z)$ can be expanded as in 
\eqref{svil} (with  $\alpha \geqslant d_\theta+1$), 
for the leading contributions as $z\to 0$ we obtain
\bea
\label{expansionarea}
\mathcal{A}[\gamma_{A,\varepsilon}] 
&=&
\int_{\partial\gamma_{A,\varepsilon}} \!\!
ds \int_\varepsilon^{z_\text{\tiny{max}}} \!\!
\, \frac{1}{z^{d_\theta}} \,
\bigg[ \, 1+\frac{z^2}{4}  \big(\! -2 k(s) \,U_2(s)+U_2'(s)+2\, U_2(s)^2\big)
\\
&&\hspace{3.8cm}
+\,\frac{z^3}{12}  \big(\!  -2 k(s) \,U_3(s)+6 \,U_2(s) U_3(s)+U_3'(s)\big)+\mathcal{O}\big(z^4\big)\, \bigg] \,dz
\nonumber
\eea
which contains  divergent terms only if $d_\theta\geqslant 1$. 
The integration of the first term within the expansion between square bracket provides the leading divergence \eqref{divercont1},
where  the perimeter $P_A$ of the entangling curve comes from the integration over $s$.
The  subleading terms are obtained  by performing the integration over $z$ in the remaining terms  in the expansion  \eqref{expansionarea}. 
This leads to
	\bea
\label{A_exp_ortog}
\mathcal{A}[\gamma_A]
&= & 
\frac{P_A}{(d_\theta-1)\varepsilon^{d_\theta-1}}
+ \frac{1}{2 (d_\theta-3)\varepsilon^{d_\theta-3}}\int_{\partial A} 
\!\! \left[ U_2(s) - k(s) \right]U_2(s) \, ds
\\
& & + \; 
\frac{1}{6 (d_\theta-4)\varepsilon^{d_\theta-4}}\int_{\partial A} 
\!\! \left[3 U_2(s)- k(s)\right] U_3(s) \, ds
+ \mathcal{O}\big( \text{max}\big\{  1/\varepsilon^{d_\theta-5}, 1 \big\} \big)
\hspace{.6cm}
d_\theta \notin \mathbb{N}\,.
\nonumber
\eea
When $d_\theta=n \in \mathbb{N}$ is a positive integer, 
this expansion still holds except for a crucial modification of the $O( \varepsilon^{n-d_\theta })$ term,
where $1/[(d_\theta - n) \varepsilon^{d_\theta - n}]$ has to be replaced with $\log \varepsilon$.
For instance, when $d_\theta=3$ we obtain
\begin{equation}
\label{A_exp_ortog_dtheta3}
\mathcal{A}[\gamma_A]
= 
\frac{P_A}{2\,\varepsilon^{2}}- \frac{\log \varepsilon }{2} \int_{\partial A} ds \, \left[
U_2(s) - k(s) \right]U_2(s) + \mathcal{O}(1)\,.
\end{equation}  

In the above analysis, we  considered surfaces $\gamma_A$ whose smooth boundary is $\partial \gamma_A =\partial A$,
that intersect orthogonally the boundary plane $z=0$ and which are not necessarily minimal.
Moreover, we have assumed that the embedding function $u(s,z)$ defined in \eqref{embed}  admits an expansion of the form
\eqref{svil} close to $z=0$ with $\alpha\geqslant 0$. 
In the following we specialize to surfaces $\hat{\gamma}_A$ that are extrema of the area functional \eqref{areafunctional}, 
namely to surfaces whose mean curvature vanishes everywhere (see \eqref{extremalcondit}) or, equivalently, 
which obey \eqref{traceminimalcondition}.

In terms of the parameterisation introduced in (\ref{embed}),  the second fundamental form $ \widetilde{K}_{ab} $ reads
\begin{equation}
\label{extrcurv}
\widetilde{K}_{ab}=-\,\tilde{h}^{-1}
\begin{pmatrix}
w(u_{ss}\!+\!kw) \!- \! u_{s}(w_{s} \!- \!k u_{s}) \;\;& w u_{zs}\!+\!k u_{z}u_{s} \\
w u_{zs}+ku_{z}u_{s} & w u_{zz}
\end{pmatrix}\,.
\end{equation}
Taking  the trace of (\ref{extrcurv}), we can translate the  extremality condition (\ref{extremalcondit}) 
into the following second order partial differential equation for  $u(s,z)$
\bea
\label{eqmin}
& &
 (1 + u_{z}^{2})\big[w(u_{ss} + k\,w) - u_{s}(w_{s} - k\,u_{s})\big]
 - 2\, u_{z} \, u_{s} \big(w\,u_{zs} + k\,u_{z}u_{s}\big)
 + 
 w \, u_{zz}\big(w^{2} + u_{s}^{2}\big)
\nonumber
\\
\rule{0pt}{.6cm}
& &=\,
d_{\theta}\dfrac{u_{z}w}{z}\, \big[u_{s}^{2} + w^{2}(1 + u_{z}^{2})\big]
\eea
with the boundary conditions $ u(s,0)=0 $.

We can employ the expansion \eqref{svil} to solve the equation \eqref{eqmin} order by order in $z$. 
Even if $U_1(s)=0$ is not assumed in \eqref{svil},  
 the vanishing of the leading term  in the sector of the expansion of \eqref{eqmin} with integer powers implies $U_1(s)=0$.
In other words, an extremal surface is necessarily  orthogonal to the boundary. 
Instead, the vanishing of the leading term  in the non analytic sector of the expansion of \eqref{eqmin}, where the powers depends on $\alpha$,  determines the value of $\alpha$ to be $d_\theta+1$.
The associated coefficient $ \mathcal{U}_{\alpha}(s) $  in \eqref{svil} cannot be determined through this  local analysis  near the boundary because it encodes global properties of $ \hat{\gamma}_{A} $. 
On the other hand, \eqref{eqmin} allows us to determine recursively the analytical part of the expansion \eqref{svil}.
For the lowest coefficients of an extremal surface $ \hat{\gamma}_{A} $, we find
\begin{subequations}
\label{coefficients}
\begin{align}
\label{coefficientsu}
& U_{2}(s)=\frac{k(s)}{d_{\theta}-1} & d_{\theta}\neq 1 \\
& U_{3}(s)=0 & d_{\theta}\neq 2   \label{coefficientsu3}\\
& U_{4}(s)=\dfrac{3k''(s)}{(d_{\theta}-1)(d_{\theta}-3)}
+\dfrac{3(d_{\theta}^{2}-2d_{\theta}-1)}{(d_{\theta}-1)^{3}\,(d_{\theta}-3)} \,k^{3}(s) & d_{\theta}\neq 1,3 \\
& U_{5}(s)=0 & d_{\theta}\neq 4\,.
\end{align}
\end{subequations}

The integer values of $ d_{\theta} $ require a separate analysis. 
For even values of $ d_{\theta} $, the  non analytical sector in  (\ref{svil}) disappears and in general 
the odd coefficients $U_{d_\theta+2n +1}(s)$ (with $n\geqslant 0$) can be non vanishing. 
In particular, this local analysis leaves $U_{d_\theta+1}(s)$ undetermined, as above.
When $ d_{\theta} $ is an odd integer, it is necessary to introduce terms of the form 
$ z^{d_{\theta}+1+n}\log z $ in the expansion (\ref{svil}) in order to satisfy the extremality condition \eqref{eqmin}. 
However, these additional terms do not contribute to the divergent part of $\mathcal{A}[\gamma_A]$, 
hence they can be neglected in the present discussion.

Finally, by plugging the expressions in \eqref{coefficients} into the expansions \eqref{A_exp_ortog} and \eqref{A_exp_ortog_dtheta3}, 
one obtains the subleading divergent contributions in \eqref{2divlog} and \eqref{2div}.

\subsection{Asymptotic hvLif$_4$ black hole}
\label{App_BH_analysis}

	In the above analysis we have investigated the UV divergent terms in the expansion of the holographic entanglement entropy 
	when the bulk metric $ \tilde{g}_{\mu\nu} $ of $\widetilde{\mathcal{M}}_3$ is flat. 
	However, since the leading divergence in \eqref{divercont1} 
	is completely determined  by the value of $ \sqrt{\tilde{h}} $ on the boundary curve $\partial\hat{\gamma}_A$, i.e. $ \tilde{h}\vert _{z=0}=1$, 
	the expansion of the area of the minimal surface is given by \eqref{divercont1} 
	for any metric $ g_{\mu\nu} $ satisfying (\ref{statichyperscaling}).  
	Instead, the subleading divergent terms in the expansion  \eqref{divercont1} can be different from the ones occurring for the hvLif$ _{4} $ spacetime. 
	Thus, in the expansion $ g_{\mu\nu}(z,\boldsymbol{x})=g_{\mu\nu}^{\text{hvLif}}(\boldsymbol{x})+\delta g_{\mu\nu}^{(1)}(\boldsymbol{x})z+\delta g_{\mu\nu}^{(2)}(\boldsymbol{x})z^{2}+\dots$ of the metric near the plane $ z=0$,
	the occurrence of the terms $ \delta g_{\mu\nu}^{(n)}$ might lead to important modifications of the analysis presented above 
	(e.g. \eqref{coefficients} are expected to be modified).
	In this appendix we address this issue in a concrete example where the asymptotic behaviour of the metric near the boundary 
	is given by a black hole geometry with hyperscaling violation.

Considering the general metric \eqref{metric_class} with $A(z),B(z)$ and $f(z)$ given by \eqref{BH_choice},
the induced metric $ g_{\mu\nu} $ on $ \mathcal{M}_{3} $ reads
\begin{equation}
\label{BHtimeslice}
ds^{2}=\dfrac{1}{z^{d_{\theta}}}
\left(\, \dfrac{dz^{2}}{f(z)}+dx^{2}+dy^{2} \right)
\;\; \qquad \;\;
f(z)= 1- ( z/z_h)^{\chi_1} + a \,z^{\chi_2}\,.
\end{equation}
The parametrization \eqref{embed}  for $ \hat \gamma_A\subset \widetilde{\mathcal{M}}_{3} $
allows to write the unit normal vector as follows
\begin{equation}
\label{normalBH}
\tilde{n}^{\mu}= \frac{1}{\sqrt{u_{s}^{2}+w^{2} [1+u_{z}^{2} f(z)]}} 
\, \Bigl(-\,u_{z}\,w f(z) \,, - \,u_{s} \,\widetilde{T}+w\widetilde{N} \,\Bigr)\,.
\end{equation}
By expressing $\tilde{n}^\mu$  in terms of the unit normal vector $ \tilde{n}^{\mu}_{\textrm{\tiny hvLif}} $ corresponding to  $ f(z)\equiv 1$,
one finds
\begin{equation}
\label{normalBH2}
\tilde{n}^{\mu}
=
C \bigl(\, \tilde{n}_{\textrm{\tiny hvLif}}^{z} \,f(z) \,,\,\tilde{n}^{\boldsymbol{x}}_{\textrm{\tiny hvLif}} \, \bigr)
\;\; \qquad \;\; 
C\equiv \frac{\sqrt{\tilde{h}_{\textrm{\tiny hvLif}}}}{\sqrt{u_{s}^{2}+w^{2} [1+u_{z}^{2} f(z)]}}
\end{equation}
where $\tilde{h}_{\textrm{\tiny hvLif}}$ is the determinant of the induced metric for hvLif$_4$.
Thus, for the trace of the second fundamental form we have
\bea
\label{TraceBH}
\textrm{Tr} \widetilde{K}
=
\widetilde{\nabla}_{\alpha}\tilde{n}^{\alpha}
&=&
C^{-1}\tilde{n}^\alpha\partial_\alpha C+C\,\widetilde{\nabla}_{\alpha}\bigl(C^{-1}\tilde{n}^{\alpha}\bigr)
\\
&=&
C^{-1}\tilde{n}^\alpha\partial_\alpha C+C\bigl(\partial_{\boldsymbol{x}}\tilde{n}^{\boldsymbol{x}}_{\textrm{\tiny hvLif}}+\partial_{z}\tilde{n}^{z}_{\textrm{\tiny hvLif}}f(z)+\dfrac{1}{2}\tilde{n}^{z}_{\textrm{\tiny hvLif}}f'(z)\bigr)
\nonumber
\eea
where we used that, for the metric (\ref{BHtimeslice}), the following result holds
\begin{equation}
\Gamma^{\alpha}_{\alpha\mu}\,\tilde{n}^{\mu} = - \,\frac{C}{2} \,f'(z)\,\tilde{n}^{z}_{\textrm{\tiny hvLif}}\,. 
\end{equation}

The extremal surfaces $\hat{\gamma}_{A} $ fulfil (\ref{traceminimalcondition}), which can be written as 
\begin{equation}
\label{extremaleqBH}
C^{-2} \,\tilde{n}^\alpha\partial_\alpha C+\partial_{\boldsymbol{x}}\tilde{n}^{\boldsymbol{x}}_{\textrm{\tiny hvLif}}
+f(z) \,\partial_{z}\tilde{n}^{z}_{\textrm{\tiny hvLif}}
+\frac{1}{2} \,f'(z)\,\tilde{n}^{z}_{\textrm{\tiny hvLif}}
=
d_{\theta} \, \frac{f(z)}{z}\, \tilde{n}^{z}_{\textrm{\tiny hvLif}}\,.
\end{equation} 

Specialising \eqref{extremaleqBH} to the expression of $f(z)$ given in \eqref{BHtimeslice}, 
we find that the equation solved by extremal surfaces in hvLif$_4$ 
gets modified by $O(z^{\chi_1})$ and $O(z^{\chi_2})$ terms. 
Thus, for arbitrary exponents $\chi_1$ and $\chi_2$, 
the divergent terms in $\mathcal{A}[\hat \gamma_{A,\varepsilon}]$ are different from the ones discussed in Section\;\ref{sec:UVstructure}.
However, in the following we show that, for black hole geometries,
new divergencies do not occur because of the NEC.

The black hole geometry corresponds to $a=0$ and $\chi_1=d_\theta+\zeta$ in \eqref{BHtimeslice}. 
In this case the NEC inequalities in \eqref{null_energ_spec1} and \eqref{null_energ_spec2} reduce to ones in \eqref{NEC}.
Since $ d_{\theta}+\zeta\geqslant 0, $ we also have $ \zeta\geqslant 1 $;
hence for the cases of interest, where $d_{\theta}>1$, we can assume $ d_{\theta}+\zeta> 2$.
Now we are ready to analyze the behaviour of the solution of  (\ref{extremaleqBH}) for small  $z$. 
Since the leading behaviour of  $ \tilde{n}^{z}_{\textrm{\tiny hvLif}}$ for $z\to 0$ 
(see (\ref{svil}) and (\ref{normal})) is given by
$ \tilde{n}^{z}_{\textrm{\tiny hvLif}}\simeq -\,U_{2}\,z+O(z^{3}) $,
the extremality  equation (\ref{extremaleqBH}) in a black hole geometry differs from (\ref{eqmin}) by $O(z^{d_{\theta}+\zeta})$ terms. 
This implies that the putative  expansion for the function $ u(s,z)$, which solves (\ref{extremaleqBH}), must also  contain  terms of  the form $ z^{d_{\theta}+\zeta+n} $ with $n\in \mathbb{N}$.
An explicit calculation shows that the first new non vanishing term occurs for $n=2$ and its coefficient reads
\begin{equation}
\dfrac{d_{\theta}-\zeta-2}{2(d_{\theta}-1)(d_{\theta}+\zeta+2)(d_{\theta}+\zeta+1)} \, k(s)\,.
\end{equation}
These new terms, which scale at least like $ z^{d_{\theta}+\zeta+2}$,  
cannot contribute to the divergent part of the holographic entanglement entropy.
Thus, the analysis performed for hvLif$_4$ remains valid also for the black hole geometry.

\section{On the finite term}
\label{appxderivaionfa2}

In this appendix we describe the details of the derivation of the results presented in Section\;\ref{fine_term}.

Considering a constant time slice $ \mathcal{M}_{3} $ of an asymptotically hvLif$ _{4} $ spacetime endowed with the metric $ g_{\mu\nu} $, 
the asymptotically flat metric $\tilde{g}_{\mu\nu}$ of the conformally equivalent space $\widetilde{\mathcal{M}}_3$ is related to $ g_{\mu\nu} $ through the relation $g_{\mu\nu}=e^{2\varphi}\tilde{g}_{\mu\nu}$. 
In  \cite{Fonda:2015nma} it was shown that, 
for any  surface (not necessarily anchored to  a curve on the boundary) the following identity holds
\begin{equation}
\label{identity1}
\Bigl(\widetilde{\mathcal{D}}^{2}\varphi\!-\!\widetilde{\nabla}^{2}\varphi
+
\tilde{n}^{\mu}\tilde{n}^{\nu}\widetilde{\nabla}_{\mu}\widetilde{\nabla}_{\nu}\varphi
-
(\tilde{n}^{\lambda} \partial_{\lambda} \varphi)^{2} 
- 
\frac{1}{4}(\textrm{Tr} \widetilde{K})^{2}\Bigr) d \tilde{\mathcal{A}}
+
\frac{1}{4}(\textrm{Tr} K)^{2}d\mathcal{A}
\,=\,0
\end{equation}
where the tilded  quantities are evaluated considering $\widetilde{\mathcal{M}}_3$ as embedding space, 
while $\mathcal{M}_3$ is the embedding space for the untilded ones.
In particular, $\textrm{Tr} {K}$ and $\textrm{Tr} \widetilde{K}$ are the  mean  curvatures of $\gamma_A$ computed in the two embedding spaces,  while $d\mathcal{A}$ and $d\widetilde{\mathcal{A}}$ are the two area elements. 
Denoting by $\tilde{n}^{\nu}$ the versor perpendicular  to  the surface $\gamma_A$ viewed  as a submanifold of  $\widetilde{\mathcal{M}}_{3}$, 
the covariant derivative $\widetilde{\nabla}$ is the one
defined in $\widetilde{\mathcal{M}}_3$ while $\widetilde{\mathcal{D}}$ is the one induced on the surface $\gamma_A$ by the embedding space $\widetilde{\mathcal{M}}_{3}$.

Let us focus on surfaces $\gamma_A$ anchored orthogonally to $\partial A$, that are not necessarily extremal surfaces.  
The first term in the left hand side of \eqref{identity1} is a total derivative;
hence it yields a boundary term when integrated over $\gamma_A$. 
As we will discuss in detail later in this Appendix, 
the main  step to construct a finite area functional is to  multiply both sides of \eqref{identity1} by a suitable term that makes this total derivative the only source of the type of divergences  discussed in Section\;\ref{sec:UVstructure} when the integration over $\gamma_A$ is carried out.
Our analysis  follows slightly different paths, depending on the ranges of $d_\theta$. 
In particular, we consider  separately the ranges $ 1<d_{\theta}<3 $ and $ 3<d_{\theta}<5$.
The special cases $d_{\theta}=3$ and $d_{\theta}=5$, where a logarithmic divergence occurs,
can be studied as limiting cases. 


\subsection{Regime $ 1<d_{\theta}<3 $}
\label{appendix_1-3}

In order to find the finite term in the expansion (\ref{leadingdiver}) of the area of the
surfaces $\gamma_A$ anchored orthogonally to $\partial A$ (not necessarily extremal),
first we multiply the identity \eqref{identity1} by a factor $c_1 e^{2\phi}$, 
where $\phi$ is a function of the coordinates and $c_1$ is a numerical constant to be determined. 
Then, integrating the resulting expression over the surface $ {\gamma_{A,\varepsilon}} \equiv\gamma_A\cap \{z\geqslant\varepsilon\}$,
one finds
\bea
\label{identity2}
0
&= & 
c_1\int_{\gamma_{A,\varepsilon} }e^{2\phi}\Bigl(\widetilde{\mathcal{D}}^{2}\varphi-\widetilde{\nabla}^{2}\varphi+\tilde{n}^{\mu}\tilde{n}^{\nu}\widetilde{\nabla}_{\mu}\widetilde{\nabla}_{\nu}\varphi-(\tilde{n}^{\lambda} \partial_{\lambda} \varphi)^{2} - \frac{1}{4}(\textrm{Tr} \widetilde{K})^{2}\Bigr)d \tilde{\mathcal{A}}
\nonumber \\
& & +\;c_1\int_{\gamma_{A,\varepsilon}} e^{2\phi}\frac{1}{4}(\textrm{Tr} K)^{2}d\mathcal{A}\,.
\eea
By adding the area functional of $\gamma_A$ to both sides of this identity, we get
\bea
\label{area1bis}
\mathcal{A}[{\gamma_{A,\varepsilon}}]
&= & 
c_1\int_{\gamma_{A,\varepsilon}} \!\!\! e^{2\phi}\Bigl(\widetilde{\mathcal{D}}^{2}\varphi-\widetilde{\nabla}^{2}\varphi+\tilde{n}^{\mu}\tilde{n}^{\nu}\widetilde{\nabla}_{\mu}\widetilde{\nabla}_{\nu}\varphi-(\tilde{n}^{\lambda} \partial_{\lambda} \varphi)^{2} - \frac{1}{4}(\textrm{Tr} \widetilde{K})^{2}\Bigr)d \tilde{\mathcal{A}}
\nonumber \\
& & 
+\int_{\gamma_{A,\varepsilon}} \!\!\!  e^{2\varphi}d\mathcal{\widetilde{A}}\,
+\frac{c_1}{4} \int_{\gamma_{A,\varepsilon}} \!\!\! e^{2\phi} (\textrm{Tr} K)^{2}\, d\mathcal{A}\,.
\eea
The first term of the first integrand can be arranged as a divergence minus a term that does not contain second derivatives as follows
\begin{equation}
\label{intbypart}
e^{2\phi} \, \widetilde{\mathcal{D}}^{2}\varphi
=
\widetilde{\mathcal{D}}^{\mu}(e^{2\phi}\partial_{\mu}\varphi)
-2\,e^{2\phi}\tilde{h}^{\mu\nu}\partial_{\nu}\phi \, \partial_{\mu}\varphi\,.
\end{equation}
At this point, Stokes' theorem can be employed  to transform the integration over the divergence in \eqref{intbypart}  into a integral over the boundary of ${\gamma_{A,\varepsilon}}$. Thus, \eqref{area1bis} becomes
\bea
\label{area2}
\mathcal{A}[\gamma_{A,\varepsilon}]
&= & 
c_1\int_{\partial \gamma_{A,\varepsilon}} \!\! e^{2\phi} \, \tilde{b}^{\mu}\partial_{\mu}\varphi \,d\tilde{s}\,
+\int_{\gamma_{A,\varepsilon}}  \!\!  e^{2\varphi}d\mathcal{\widetilde{A}}\,
+\frac{c_1}{4}\int_{\gamma_{A,\varepsilon}}  \!\!  e^{2\phi} (\textrm{Tr} K)^{2}d\mathcal{A}
\\
& & -\; c_1
\int_{\gamma_{A,\varepsilon} }  \!\!\!
e^{2\phi}
\Bigl(2\tilde{h}^{\mu\nu}\partial_{\nu}\phi\partial_{\mu}\varphi
+\widetilde{\nabla}^{2}\varphi
-\tilde{n}^{\mu}\tilde{n}^{\nu}\widetilde{\nabla}_{\mu}\widetilde{\nabla}_{\nu}\varphi
+(\tilde{n}^{\lambda} \partial_{\lambda} \varphi)^{2} 
+ \frac{1}{4}(\textrm{Tr} \widetilde{K})^{2}\Bigr)d \tilde{\mathcal{A}}
\nonumber
\eea
where $\tilde{b}^{\mu}$ is the outward pointing unit vector normal to the boundary curve.
The function $\phi$ and the constant $c_1$  can be fixed by requiring that the divergence originating from the boundary term in \eqref{area2} 
as $\varepsilon \to 0$ matches the divergence in \eqref{leadingdiver}. 
The limit $\varepsilon \to 0$ of the remaining terms provides the finite contribution $\mathcal{F}_A$ in \eqref{leadingdiver}.

As for the vector $ \tilde{b}^{\mu} $ normal to the boundary of $\gamma_{A,\varepsilon}$,
it has the same direction of the vector $t^{\mu}_{2}$ in \eqref{tangentvectsurf}.
This gives
\begin{equation}
\label{vectorb}
\tilde{b}^{\mu}= \frac{-1}{\sqrt{1+u_{z}^{2}}} \,\bigl(1,u_{z}\tilde{N}\bigr)
\end{equation}
whose expansion as $\varepsilon \to 0$ reads
\begin{equation}
\label{vectorbalpha}
\tilde{b}^{\mu}
=\Bigl(- 1 + \frac{\varepsilon^{2}}{2}\,U_{2}^{2} + O(\varepsilon^{4}), - \,U_{2} \,\widetilde{N}\, \varepsilon + O \big(\varepsilon^{3}\big)\Bigr)\,.
\end{equation}
This expansion can be used to determine the behaviour of  the boundary term in \eqref{area2}, finding
\begin{equation}
\label{boundaryterm}
c_1\int_{\partial\gamma_A,\varepsilon} \!\!\!
e^{2\phi} \, \tilde{b}^{\mu}\partial_{\mu}\varphi \,d\tilde{s} 
 = 
 - \frac{c_1 d_\theta P_{A}}{2\varepsilon}\, e^{2\phi(\varepsilon)}  +O(\varepsilon^{a})
\end{equation}
where 
\be
\label{phitorto}
\varphi=- \frac{d_{\theta}}{2} \log z
\ee
and $a$ is determined by the specific choice of $\phi$. 
By imposing  consistency between the leading divergence in \eqref{leadingdiver} and \eqref{boundaryterm},
one obtains
\begin{equation}
\label{alpha}
\phi=\frac{2-d_\theta}{2}\, \log z   +O(z^2)     
\;\; \qquad  \;\;
 c_1=\frac{2}{d_\theta(d_\theta-1)}\,.
\end{equation} 
By considering the expressions of $\varphi$ in \eqref{phitorto} and of $\phi$ in \eqref{alpha}, together with the expansion in \eqref{vectorbalpha}, 
the integral \eqref{boundaryterm} leads to $a= 3-d_\theta$.
Notice that the leading singular behaviour of  $\phi$ vanishes identically when $d_\theta=2$. 
The sum of  the remaining terms in \eqref{area2} must be finite; 
hence we can safely remove the cutoff $\varepsilon$, obtaining the expression  \eqref{finitegeneralsurface} for the finite term.

We remark that \eqref{finitegeneralsurface} holds for surfaces $\gamma_A$ that intersect orthogonally $\partial \mathcal{M}_3$
and that this class includes the extremal surfaces. 
For extremal surfaces, \eqref{extremalcondit} and \eqref{traceminimalcondition} 
can be employed to simplify \eqref{finitegeneralsurface}, which reduces to \eqref{finite term}.
In the special case of $d_{\theta}=2$, the expression \eqref{finite term} simplifies further to the formula valid for the asymptotically AdS$_4$ backgrounds  
found in \cite{Fonda:2015nma}.

\subsection{Regime $ 3<d_{\theta}<5 $}
\label{app_deriv_finte_3-5}

%
In this range of  $d_\theta$ we limit our analysis to the case of extremal surfaces 
because the condition of orthogonal intersection with the boundary does not fix completely the structure of the divergences.
Instead,  for  extremal surfaces anchored to $\partial A$ we can have only two types of divergences as $\varepsilon \to 0$ 
and they are of the form occurring in \eqref{2div}. 
 To single out these singular terms, we multiply both sides  of the identity \eqref{identity1}  by the following factor 
\begin{equation}
c_1 e^{2\phi}+c_2 e^{2\psi}(\textrm{Tr} \widetilde{K})^{2}
\end{equation}
where $c_1 $ ans $c_2$ are numerical coefficients and $e^{2\phi}$ and $e^{2\psi}$ are functions of the coordinates to be determined. 
Integrating the resulting expression over $\hat  \gamma_{A,\varepsilon} $ and then adding 
the area $ \mathcal{A}[\hat\gamma_{A,\varepsilon}] $ to both sides, we obtain
\bea
\label{secondinvFa2}
\mathcal{A}[\hat \gamma_A]
&=& 
\int_{\hat \gamma_{A,\varepsilon}}
\! \Bigl(c_1 e^{2\phi}\!+\! c_2 e^{2\psi}(\textrm{Tr} \widetilde{K})^{2}\Bigr)\Bigl(\widetilde{\mathcal{D}}^{2}\varphi\!-\!\widetilde{\nabla}^{2}\varphi\!+\!\tilde{n}^{\mu}\tilde{n}^{\nu}\widetilde{\nabla}_{\mu}\widetilde{\nabla}_{\nu}\varphi\!-\!(\tilde{n}^{\lambda} \partial_{\lambda} \varphi)^{2} \!-\! \frac{1}{4}(\textrm{Tr} \widetilde{K})^{2}\Bigr)d \tilde{\mathcal{A}}
\nonumber\\
& &
+ \int_{\hat \gamma_{A,\varepsilon}} \!\! e^{2\varphi} \, d\tilde{\mathcal{A}}
\eea
where the equation of motion $\text{Tr} K=0$ has been used. 
As done in Section\;\ref{appendix_1-3},
let us rewrite the term proportional to  $ \widetilde{\mathcal{D}}^{2}\varphi $ as a total divergence minus residual contributions.
In particular, we have
\bea
\Big(c_1 e^{2\phi}+c_2 e^{2\psi}(\textrm{Tr} \widetilde{K})^{2} \Big)\widetilde{\mathcal{D}}^{2}\varphi
&=& 
\widetilde{\mathcal{D}}^{\mu} 
\big[ \, c_1\,e^{2\phi}\partial_{\mu}\varphi+c_2\,e^{2\psi} \big(\textrm{Tr} \widetilde{K}\big)^{2}\partial_{\mu}\varphi \, \big]
-2\, c_1\, e^{2\phi}\tilde{h}^{\mu\nu}\partial_{\mu}\phi\partial_{\nu}\varphi 
\hspace{1cm}\nonumber\\
& & \hspace{-0cm}
- \;2 \,c_2 \,e^{2\psi}(\textrm{Tr} \widetilde{K})^{2}\tilde{h}^{\mu\nu}\partial_{\mu}\psi\partial_{\nu}\varphi 
-2 \,c_2 \,e^{2\psi} \big(\textrm{Tr} \widetilde{K} \big)\,
\tilde{h}^{\mu\nu}\partial_{\mu}\bigl(\textrm{Tr} \widetilde{K}\bigr)\partial_{\nu}\varphi\,.
\nonumber
\eea
%
%
%
Plugging this expression back into (\ref{secondinvFa2}), we can write the area of $ \hat \gamma_{A,\varepsilon} $ in the following form
\begin{equation}
\label{pinodaniele}
\mathcal{A}[\hat \gamma_{A,\varepsilon}]
=
\int_{\hat\gamma_{A,\varepsilon} } 
\!\!\! \widetilde{\mathcal{D}}^{\mu}J_{\mu} \, d\tilde{\mathcal{A}}
-\mathscr{F}_{A,\varepsilon}
\end{equation}
where
\begin{equation}
J_{\mu}
=
c_1\, e^{2\phi}\partial_{\mu}\varphi+c_2 \,e^{2\psi} \big(\textrm{Tr} \widetilde{K}\big)^{2}\partial_{\mu}\varphi  
\end{equation}
and $ \mathscr{F}_{{A,\varepsilon}}$ contains all the remaining terms. 
By Stokes' theorem, the integral of the divergence turns into a line integral over the boundary curve 
\begin{equation}
\label{Stokes}
\int_{{\hat\gamma_{A,\varepsilon}}}
\!\!\! \widetilde{\mathcal{D}}^{\mu}J_{\mu} \, d\tilde{\mathcal{A}}
\,=
\int_{\partial \hat \gamma_{A,\varepsilon}}
\!\!\!  \tilde{b}^{\mu}J_{\mu}d\tilde{s}
\, = 
\int_{\partial {\hat\gamma_{A,\varepsilon}}}\!
\Bigl(c_1 \,e^{2\phi} \, \tilde{b}^{\mu}\partial_{\mu}\varphi
+ c_2 \,e^{2\psi}\big(\textrm{Tr} \widetilde{K}\big)^{2} \, \tilde{b}^{\mu}\partial_{\mu}\varphi\Bigr) d\tilde{s}\,.
\end{equation}
The first term occurs also in \eqref{boundaryterm} and it contains the leading divergence of $\mathcal{A}[\hat \gamma_{A,\varepsilon}]$. 
Thus, we must choose $e^{2\phi}$ and $c_1$ as in  \eqref{alpha}. Then we fix $c_2$ and $e^{2\psi}$ so that the boundary term \eqref{Stokes} reproduces also the subleading divergence in \eqref{2div}. Specifically, if we use the explicit expressions of $c_1$, of $e^{2\phi}$ and 
the extremal equation $\eqref{traceminimalcondition}$, we can rewrite the above boundary term as follows
\be
\label{boundarytermbis}
 \int_{\partial\hat \gamma_{A,\varepsilon}}\!\!\!
 \tilde{b}^{\mu}J_{\mu}d\tilde{s}
\,=\,
-\int_{\partial \hat \gamma_{A,\varepsilon}} \!\!\!\!
\tilde b^z  
\left(\frac{\varepsilon^{1-d_{\theta}}}{d_\theta-1}  +c_2\, e^{2\psi} d_{\theta}^{3} \,\frac{ (\tilde{n}^{z})^{2}}{2\varepsilon^3}\right)d\tilde{s}\,.
\ee
From the analysis reported in Appendix \ref{sec:UVstructure}, 
we obtain the following expansions as $z \to 0$
\begin{subequations}
\label{vettori}
\begin{align}
& \tilde{b}^{z}= -\,1+\frac{U_{2}(s)^2}{2}\,z^{2}+O(z^{4})
\\
\label{normal}
\rule{0pt}{.5cm}
& \tilde{n}^{z}= -\,U_{2}(s)\, z+O(z^{3})\\
\label{metrbordo}
\rule{0pt}{.7cm}
& d\tilde{s}= \biggl(1-\dfrac{k(s)\,U_{2}(s)}{2} \, z^{2}+O(z^{4})\biggr)ds
\end{align}
\end{subequations}
where $U_2(s)$ is given in \eqref{coefficientsu}.
Plugging  \eqref{vettori} into \eqref{boundarytermbis} and collecting the terms containing $k(s)^2$, 
we get
\bea
\label{boundaryterm1}
\int_{\partial\hat \gamma_{A,\varepsilon}}\tilde{b}^{\mu}J_{\mu} \, d\tilde{s}\,
&=&
\int_{\partial \hat \gamma_{A,\varepsilon}} \!\!
 \left( 1-\frac{U_2^2 }{2} \,\varepsilon^2 \right) 
 \left(\frac{ \varepsilon^{1-d_{\theta}}}{d_\theta-1} 
 +c_2 \, e^{2\psi} d_{\theta}^{3} \,\frac{ U_2^2}{2\varepsilon} \,\right) 
 \left( 1-\frac{U_2\, k}{2} \, \varepsilon^2 \right)d s
 \\
& =&
\frac{P_A}{(d_\theta-1)\,\varepsilon^{d_\theta-1}}
- \int_{\partial \hat \gamma_{A,\varepsilon}}   \!\!
\left( \,\frac{\varepsilon^{3-d_\theta} }{2(d_\theta-1)^3} 
-  \frac{c_2\,d^3_\theta \,e^{2\psi}}{2(d_\theta-1)^2 \varepsilon} 
+ \frac{ \varepsilon^{3-d_\theta}}{2(d_\theta-1)^2}\right) k^2 \,ds
\nonumber\\
& =&
\dfrac{P_A}{(d_{\theta}-1)\,\varepsilon^{d_{\theta}-1}}
+\frac{1}{2 (d_{\theta}-1)^2\, \varepsilon^{d_{\theta}-3}}
\left(c_2 d_{\theta}^{3} e^{2\psi}\varepsilon^{d_\theta-4}-\dfrac{d_{\theta}}{d_{\theta}-1}\right)
\int_{\partial \hat\gamma_{A,\varepsilon}} \!\! \!\! k^{2}\,ds\,.
\nonumber
\eea
	The simplest choice to obtain the right	subleading divergence in \eqref{2div} is given by
	\begin{equation}
	\label{beta}
	c_2=-\dfrac{2}{d_{\theta}^{3}(d_{\theta}-3)(d_{\theta}-1)}\qquad e^{2\psi}=z^{4-d_{\theta}}\big(1+O(z^2)\big)\,.
	\end{equation}
	%
Since the boundary integral \eqref{boundaryterm1} with the substitutions \eqref{beta} yields  all the correct divergences of the area as $\varepsilon\to 0$, the sum of the remaining terms is  finite in this limit and provides the finite contribution $\mathscr{F}_A$ to 
$\mathcal{A}[\hat \gamma_{A,\varepsilon}]$. 
After some simple algebraic  manipulations, $\mathscr{F}_A$ can be expressed as in \eqref{finiteminsurf2}.

The procedure to subtract  the divergences and consequently to write down a finite functional  $F_A$ is not unique.  Instead of adding a second
exponential  weighted by the $(\mathrm{Tr}K)^2$,  we could have achieved the same result by tuning the subleading  in
the  expansion of $\phi$. For instance if we choose
\be
	\label{phi func general}
	\phi=\frac{2-d_\theta}{2}\,\log z-\frac{k(s)^2}{(d_\theta-3)(d_\theta-1)^2}  \,z^2  +O(z^4)
	\ee
the functional \eqref{finite term} would produce the correct result in the entire interval $1<d_\theta<5$.  
It would be interesting to find a geometrical interpretation of (\ref{phi func general}).

\subsection{HvLif$_4$}
\label{hvLif}

In hvLif$_{4}$, we have that $ \tilde{g}_{\mu\nu}=\delta_{\mu\nu}$
and this leads to drastic simplifications in \eqref{finite term} and \eqref{finiteminsurf2}.

As for $F_A$ in \eqref{finite term},
we observe that the following combination of terms vanishes identically (for any $d_\theta$)
\begin{equation}
\widetilde{\nabla}^{2}\varphi
+2\,\tilde{g}^{\mu\nu}\,\partial_{\nu}\phi \, \partial_{\mu}\varphi
-\frac{d_{\theta}(d_{\theta}-1)}{2}\,e^{2(\varphi-\phi)}
=
\frac{1}{2z^{2}}\Bigl(d_{\theta}+d_{\theta}(d_{\theta}-2)-d_{\theta}(d_{\theta}-1)\Bigr)=0\,.
\end{equation}
The remaining terms can be written through $\tilde n^z$ as follows
\begin{equation}
\label{simplifications}
\tilde{n}^{\mu}\tilde{n}^{\nu}\widetilde{\nabla}_{\mu}\widetilde{\nabla}_{\nu}\varphi = d_{\theta}\,\dfrac{(\tilde{n}^{z})^{2}}{2z^{2}} 
\;\qquad \;
(\textrm{Tr} \widetilde{K})^{2}=d_{\theta}^{2}\,\frac{(\tilde{n}^{z})^{2}}{z^{2}}
\;\qquad \;
\tilde{n}^{\mu}\tilde{n}^{\nu}\partial_{\nu}\phi \, \partial_{\mu}\varphi
=
d_{\theta}(d_{\theta}-2)\,\frac{(\tilde{n}^{z})^{2}}{4z^{2}}\,.
\end{equation}
The above observations allow to write $ F_{A} $ in the form \eqref{Fa1} or \eqref{Fa2}.

Next, we show that $\mathscr{F}_A$ in \eqref{finiteminsurf2}  simplifies to  \eqref{Fa2vacuum} for the  hvLif$ _4 $ geometry. 
First, we find it useful to decompose $f$ in \eqref{f} as the following sum
\begin{equation}
f=f_0+f_n
\end{equation}
where $f_0$ includes the  terms that do not contain the vector $\tilde{n}^\mu$, namely 
\begin{equation}
f_0=-\,\widetilde{\nabla}^{2}\varphi-2\,\tilde{g}^{\mu\nu}\partial_{\mu}\psi \, \partial_{\nu}\varphi
\end{equation}
while the terms containing $\tilde{n}^\mu$ are collected into $f_n$.
Then, the combination
\be
F_{A}-c_2\int_{\hat \gamma_A}e^{2\psi}(\textrm{Tr} \widetilde{K})^{2} f_0 \,d\tilde{\mathcal{A}}
\ee
in  $\mathscr{F}_A$  can be shown to vanish identically  when $ \tilde{g}_{\mu\nu}=\delta_{\mu\nu}$ with the help of 
 \eqref{Fa1} and  \eqref{simplifications}. In fact, we find 
\be
F_{A}-c_2\int_{\hat \gamma_A}e^{2\psi}f_0(\textrm{Tr} \widetilde{K})^{2}d\tilde{\mathcal{A}} 
=
\dfrac{1}{d_{\theta}-1}\int_{\hat \gamma_A}\dfrac{(\tilde{n}^{z})^{2}}{z^{d_{\theta}}} \, d\tilde{\mathcal{A}}
+
c_2 \dfrac{d_{\theta}^{3} (d_{\theta}-3)}{2}\int_{\hat \gamma_A}\dfrac{(\tilde{n}^{z})^{2}}{z^{d_{\theta}}} \, d\tilde{\mathcal{A}}
\;= \,0
\ee
%
where in the last equality we used the value of $ c_2 $ in \eqref{beta}. Thus the functional \eqref{finiteminsurf2} for $ \mathscr{F}_A $ collapses to
\begin{equation}
\label{fuffa}
\mathscr{F}_A
=
-\,c_2 \int_{\hat \gamma_{A}} e^{2\psi}
\Bigl(\big(\textrm{Tr} \widetilde{K}\big)^{2} f_n -2 (\textrm{Tr} \widetilde{K})\tilde{h}^{\mu\nu}\partial_{\mu}(\textrm{Tr} \widetilde{K})\partial_{\nu}\varphi\Bigr)d\tilde{\mathcal{A}}
\end{equation}
with
\begin{equation}
f_n=\tilde{n}^{\mu}\tilde{n}^{\nu} \,\widetilde{\nabla}_{\mu}\widetilde{\nabla}_{\nu}\varphi
-2(\tilde{n}^{\lambda} \partial_{\lambda} \varphi)^{2}+2\tilde{n}^{\mu}\tilde{n}^{\nu}\partial_{\mu}\psi\partial_{\nu}\varphi
\end{equation}
and reduces to \eqref{Fa2vacuum} when $\tilde{g}_{\mu\nu}$ is the flat metric.
We can also explicitly verify that the result (\ref{Fa2vacuum}) is finite in the limit $ \varepsilon\rightarrow 0 $. If we use  the near boundary expansion  \eqref{normal} of the normal vector, we can easily check that the integrand in first   term of \eqref{Fa2vacuum} 
is of order $ z^{4-d_{\theta}} $ and it is convergent for $ d_{\theta}<5$. 
Then, assuming the parametrization \eqref{embed},  for the integrand in the the second term of \eqref{Fa2vacuum} one gets
\begin{equation}
\dfrac{\tilde{n}^{z}}{z^{d_{\theta}-2}} \, \tilde{h}^{z\mu}\,
\partial_{\mu} \biggl(\dfrac{\tilde{n}^{z}}{z}\biggr)
=
\dfrac{\tilde{n}^{z}}{z^{d_{\theta}-2}}
\,\tilde{h}^{zz} \, \partial_{z}\biggl(\dfrac{\tilde{n}^{z}}{z}\biggr)
+
\dfrac{\tilde{n}^{z}}{z^{d_{\theta}-2}} \, \tilde{h}^{zs}\,\partial_{s}\biggl(\dfrac{\tilde{n}^{z}}{z}\biggr)\,.
\end{equation}
From (\ref{inducedmetric}) we know that near $ z\!=\!0 $ the inverse metric components are $ \tilde{h}^{zz}\!=\! 1+O(z^{2}) $ and $ \tilde{h}^{zs}= O(z^{3})$, so that we have the following behaviours
\be
\dfrac{\tilde{n}^{z}}{z^{d_{\theta}-2}} \, \tilde{h}^{zz} \,
\partial_{z}\biggl(\dfrac{\tilde{n}^{z}}{z}\biggr)
\propto
\dfrac{1}{z^{d_{\theta}-3}} \, \partial_{z}
\biggl(\dfrac{U_{2}z+O(z^{3})}{z}\biggr)
\propto z^{4-d_{\theta}}
\;\;\qquad \;\;
\dfrac{\tilde{n}^{z}}{z^{d_{\theta}-2}} \, \tilde{h}^{zs} \, 
\partial_{s}\biggl(\dfrac{\tilde{n}^{z}}{z}\biggr)\propto z^{6-d_{\theta}}
\ee
and both scalings provide convergent integrals for $ d_{\theta}<5$.

	\subsubsection{Consistency check of $\mathscr{F}_A$ for the strip}
\label{app strip 3-5}

In this section we show that the functional $\mathscr{F}_A$ in 
\eqref{Fa2vacuum} gives the expected result when $\hat \gamma_A$ 
is the extremal surface anchored to the infinite strip discussed in \ref{sec_strip_hvLif},
when the gravitational background is \eqref{hyperscaling4bis} with $3<d_\theta < 5$.

By employing the parametrization of Section \ref{sec_strip_hvLif}, we find that \eqref{Fa2vacuum} becomes
\bea
\label{Fa3-5strip}
\mathscr{F}_A
&=&
\,\frac{4}{(d_{\theta}-1)(d_{\theta}-3)}\int^{L/2}_{0}\int_{0}^{\ell /2}
\left[\, \frac{2}{z^{d_{\theta}-2}} \,
\bigg(1-\frac{1}{1+(z')^2}\bigg)\frac{1}{z'}\,\partial_{x}\bigg(\dfrac{1}{z\sqrt{1+(z')^2}}\bigg)
\right. \\
& & \hspace{9.7cm}	
\left. -\,\frac{3}{z^{d_{\theta}}}\frac{1}{(1+(z')^2)^{\frac{3}{2}}}
\,\right]dxdy
\nonumber
\eea
where
$ \tilde{h}^{z\mu}\partial_\mu=\tilde{h}^{zz}\partial_z+\tilde{h}^{zy}\partial_y=(1-\tilde{n}^z\tilde{n}^z)(1/z')\partial_x \,$
has been used. 
The conserved quantity \eqref{conservedmomentum} allows to rewrite the \eqref{Fa3-5strip} as
\begin{equation}
\label{Fa3-5strip2}
\mathscr{F}_A
=
-\,\frac{4}{(d_{\theta}-1)(d_{\theta}-3)}\int^{L/2}_{0} \!\! \int_{0}^{\ell /2}
\left[\,
\frac{3}{z_*^{d_{\theta}}(1+(z')^2)}
-\frac{2(d_\theta-1)\, (z')^2}{z_*^{d_\theta}(1+(z')^2)}
\,\right]dxdy
\end{equation}
which can be further simplified by eliminating $z'$ with the help of \eqref{conservedmomentum}:
\begin{equation}
\label{Fa3-5strip3}
\mathscr{F}_A
=
-\,\frac{2L \, (2d_\theta+1)}{(d_{\theta}-1)(d_{\theta}-3)\, z_*^{3d_\theta}}
\int_{0}^{\ell /2}z^{2d_{\theta}}dx
+\,\frac{2L\ell}{(d_{\theta}-3)\, z_*^{d_\theta}}\,.
\end{equation}
Now we perform the integral in \eqref{Fa3-5strip3} 
\begin{equation}
\label{intstrip}
\int_{0}^{\ell /2}z^{2d_{\theta}}dx
=\int_{0}^{z_{\ast}}\dfrac{z^{2d_{\theta}}dz}{\sqrt{(z_{\ast}/z)^{2d_{\theta}}-1}}
=\frac{\sqrt{\pi}\,\Gamma\Bigl(\frac{3}{2}+\frac{1}{2d_{\theta}}\Bigr)}{
	2d_\theta\,\Gamma\Bigl(2+\frac{1}{2d_{\theta}}\Bigr)}z_{\ast}^{2d_{\theta}+1}
=\dfrac{\ell(d_\theta+1)}{2(2d_\theta+1)}\;z_*^{2d_{\theta}}
\end{equation}
where in the first step we changed integration variable first and then we used \eqref{conservedmomentum} again, 
while in the last step we employed the expression \eqref{stripwidth} for $\ell /2$. 
Finally, by plugging \eqref{intstrip} in \eqref{Fa3-5strip3} we obtain the r.h.s. of \eqref{Fastrip}.

We stress that the same result can be achieved by starting from the more general functional \eqref{finiteminsurf2}. 
Since the functional $F_A$ in \eqref{finiteminsurf2} is the same as the one in \eqref{finite term}, it is sufficient to show that the remaining integral in \eqref{finiteminsurf2} vanishes. 
This can be shown through a calculation similar to the one performed in this section.


\section{On the finite term as an integral along the entangling curve}
\label{app:integral entangling curve}

This appendix is devoted to an alternative and more field theoretical 
derivation of the expression \eqref{Fa=udtheta} for the finite term
written as an integral along the entangling curve. 
The method employed below is also discussed in \cite{Ogawa:1998pm}.

Let us denote with $\hat \gamma $ an extremal $m$ dimensional hypersurface embedded in $ \mathcal{M}_{d}$ 
with tangent vectors $ t^{\mu}_{a} $, where $ a=1\cdots m$. 
The area of $\hat \gamma $ is the integral 
%
\begin{equation}
\label{area:secHvLif}
\mathcal{I}
=
\int_{\hat \gamma}\mathcal{L}[x^{\mu}(\sigma),\partial_{b}x^{\mu}({\sigma})]d^m\sigma 
\qquad 
\mathcal{L}[x^{\mu}(\sigma),\partial_{b}x^{\mu}({\sigma})]\equiv \sqrt{h} 
\end{equation}
where $ \sigma $ is a set of local coordinates on $ \hat \gamma $ and $ h=\text{det}(t^{\mu}_{a}t^{\nu}_{b}g_{\mu\nu}) $.
Next we assume that  the   metric $ g_{\mu\nu} $ is  endowed with a conformal Killing vector $ V^{\mu}$, 
namely a vector field obeying the equation
\begin{equation}
\label{confkillingeq}
\nabla_{\mu}V_{\nu}+\nabla_{\nu}V_{\mu}
=\frac{2}{d} \,g_{\mu\nu} \nabla_{\rho}V^{\rho} \,.
\end{equation}
This vector generates  the infinitesimal coordinate transformation 
$ x^{\mu}\rightarrow x^{\mu}+\epsilon V^{\mu}$,
under which  the volume form on $\hat \gamma $ transforms as
\begin{equation}
\label{variationinducedmetric}
\delta \sqrt{h}
=
\dfrac{1}{2}\sqrt{h} \, h^{ab} \, \delta h_{ab}
=
\dfrac{1}{2}\sqrt{h} \, h^{ab} \, t^{\mu}_{a}t^{\nu}_{b} \, \delta g_{\mu\nu}\,.
\end{equation}
The variation of the metric $ g_{\mu\nu} $ is  given by $ \delta g_{\mu\nu}=\epsilon \, 
g_{\mu\nu} \nabla_\rho V^\rho  $, hence the variation  (\ref{variationinducedmetric}) can be rewritten as
\begin{equation}
\delta \sqrt{h}=\dfrac{\epsilon }{2}\sqrt{h} \,h^{ab}h_{ab}\nabla_
\rho V^\rho= \, \epsilon \; \dfrac{m\,(2-d_\theta)}{2} \,\sqrt{h}\,.
\end{equation}
Let us now suppose that the divergence of the vector $V^{\mu} $ is a  constant $c$.  
The transformation law of the area of $ \hat \gamma$ becomes
\begin{equation}
\label{transformationaction}
\delta \mathcal{I}=\, \epsilon \; \dfrac{m\,c}{2}\,\mathcal{I}\,.
\end{equation}
The left hand side of \eqref{transformationaction} can be cast into a total divergence as follows
\bea
\label{deltaS}
\delta \mathcal{I}
& =&
\int_{\hat \gamma}\bigg[ \dfrac{\delta\mathcal{L}}{\delta x^{\mu}}\delta x^{\mu}
+\dfrac{\delta\mathcal{L}}{\delta\partial_{a} x^{\mu}} \, \delta \partial_{a}x^{\mu}\biggr]d^m\sigma 
\nonumber \\
& =&
\int_{\hat \gamma}
\biggl[ \biggl(\dfrac{\delta\mathcal{L}}{\delta x^{\mu}}-\partial_{a}\dfrac{\delta\mathcal{L}}{\delta \partial_{a}x^{\mu}}\biggr)\delta x^{\mu}+\partial_{a}\biggl(\dfrac{\delta\mathcal{L}}{\delta \partial_{a}x^{\mu}}\,\delta x^{\mu}\biggr)\biggr]d^m\sigma 
\\
&=&\int_{\hat \gamma}\partial_{a}
\biggl(\dfrac{\delta\mathcal{L}}{\delta \partial_{a}x^{\mu}} \,\delta x^{\mu}\biggr)d^m\sigma
=\epsilon\int_{\hat \gamma}\partial_{a}\biggl(\dfrac{\delta\mathcal{L}}{\delta \partial_{a}x^{\mu}} \,V^{\mu}\biggr)d^m\sigma
\nonumber
\eea
where  the equations of motions and $ \delta x^{\mu}=\epsilon\, V^{\mu} $ have been used. 
By employing the Stokes' theorem, we can write \eqref{deltaS} as the following integral over  $ \partial\hat \gamma$
\begin{equation}
\label{deltaS2}
\delta \mathcal{I}=\epsilon\int_{\partial \hat\gamma}b_{a}\biggl(\dfrac{\delta\mathcal{L}}{\delta \partial_{a}x^{\mu}} \,V^{\mu}\biggr) d^{m-1}s
\end{equation}
where $ b^{a} $ is the unit normal vector to $ \partial\hat \gamma$.
Finally, by plugging \eqref{deltaS2} into \eqref{transformationaction}, we get
\begin{equation}
\label{S=boundaryintegral}
\mathcal{I}=\dfrac{2}{m\,c}\int_{\partial\hat \gamma}b_{a}\biggl(\dfrac{\delta\mathcal{L}}{\delta \partial_{a}x^{\mu}} \,V^{\mu}\biggr) d^{m-1}s\,.
\end{equation}
This result tells us that the area of an extremal hypersurface can be expressed as a boundary integral whenever the ambient metric exhibits a conformal Killing vector with constant divergence.

Let us now specialize \eqref{S=boundaryintegral} to our case of interest, namely to a two dimensional extremal surface $\hat \gamma_A$ anchored to $\partial A$ embedded into $\mathcal{M}_3$ with metric $ g_{\mu\nu} $ given by \eqref{hyperscaling4bis}
(thus, $m=2$ and $d=3$).
This metric has a conformal Killing vector $ V^{\mu}=x^{\mu} $ with constant divergence that generates scale transformations $ x^{\mu}\rightarrow \lambda x^{\mu} $. Under dilation the metric acquires an overall factor $ {g_{\mu\nu}\rightarrow \lambda^{2-d_{\theta}}g_{\mu\nu}} $, i.e. $c=2-d_\theta$. Thus, in the case of hvLif$_4$ geometry we can rewrite \eqref{S=boundaryintegral}  as
\begin{equation}
\label{S=boundaryintegralHvLIf}
\mathcal{I}=\dfrac{1}{2-d_{\theta}}\int_{\partial \hat\gamma_A}b_{a}\biggl(\dfrac{\delta\mathcal{L}}{\delta \partial_{a}x^{\mu}} x^{\mu}\biggr)ds\,.
\end{equation}

The expression \eqref{S=boundaryintegralHvLIf} can be further simplified by employing the parametrization \eqref{embed} 
for the minimal surface $ \hat \gamma_{A}$; hence $\sigma=\left\lbrace z,s \right \rbrace$.
The derivative of $ \mathcal{L}=\sqrt{h}=e^{2\varphi}\sqrt{\tilde{h}} $ yields
\begin{equation}
\label{lagrangianderivative:secHvLif}
\dfrac{\delta\mathcal{L}}{\delta \partial_{a}x^{\mu}}
=
\dfrac{e^{2\varphi}}{2} \, \sqrt{\tilde{h}} \, \tilde{h}^{bc}\dfrac{\delta\tilde{h}_{bc}}{\delta \partial_{a}x^{\mu}}
=
e^{2\varphi}\sqrt{\tilde{h}} \, \tilde{h}^{ab}\partial_{b}x^{\nu}\tilde{g}_{\mu\nu}\,. 
\end{equation}
In order to compute the vector $ b_{a} $ we remind that the integral \eqref{S=boundaryintegralHvLIf} is defined on\footnote{Notice that. the index $a$ in $ b_{a} $ is not associated with the metric on $\hat \gamma_{A}$ but with the metric of $\mathbb{R}^{2}$.} $ \mathbb{R}^{2}, $ so it is simply the normal vector to the boundary of the coordinate domain of the surface $\hat \gamma_{A}$. 
The integral is divergent and therefore we need to introduce a cutoff. 
In particular, this means the line integral \eqref{S=boundaryintegralHvLIf} has to be performed over the curve  $ \partial \hat \gamma_{A,\varepsilon}=\{z=\varepsilon\}\cap \hat \gamma_A $.
Finally, by plugging \eqref{lagrangianderivative:secHvLif} into \eqref{S=boundaryintegralHvLIf}, using the explicit expression of $\tilde h^{ab}$ in \eqref{inducedmetric} and $\tilde g_{\mu\nu}=\delta_{\mu\nu}$, 
for the area of extremal surfaces in hvLif$_{4}$ in terms of the function $ u(z,s) $ we obtain
\begin{equation}
\label{S=boundaryH2}
\mathcal{I}
=
\dfrac{1}{d_{\theta}-2}\int_{ \partial \hat \gamma_{A,\varepsilon}}
\!\!
\dfrac{
	(w^2+u_{s}^{2})(z+u_{z}\, \boldsymbol{x}_A\cdot\widetilde{N}+u_{z}u)
	-u_{z}u_{s}(w\,\tilde{T}\cdot\partial\gamma+u_{s}\, \boldsymbol{x}_A\cdot\widetilde{N}+u_{s}\,u)
}{
	z^{d_{\theta}}\,\sqrt{u_{s}^{2}+w^{2}(1+u_{z}^{2})}\,.
}\; ds
\end{equation}
Although this form is not very illuminating,  
it is interesting to observe that, once we expand the integrand near to $z=0,$ only the term 
$u_{z}\, \boldsymbol{x}_A\cdot\widetilde{N}$ gives a finite contribution to $ \mathcal{I}$. 
By writing the area of the regularized extremal surface $\gamma_{A,\varepsilon}$ in the following form
\begin{equation}
\mathcal{A}[\hat \gamma_{A,\varepsilon}]= P_{A}(\varepsilon)-F_{A}+\mathcal{O}(\varepsilon)
\end{equation}
where $P_{A}(\varepsilon)$ is a shorthand for all the divergent terms in \eqref{S=boundaryH2},
and employing the expansion of $u(z,s)$ given in (\ref{svil}), we find \eqref{Fa=udtheta}.

\section{Time dependent backgrounds}
\label{Apptime}

In this  appendix we derive the expressions \eqref{Fa time-dep} and \eqref{Fa time-dep extremal surf}, 
which generalize the results found in the Appendix\;\ref{appendix_1-3} to time dependent backgrounds. 

Let us consider a two dimensional spacelike surface $\gamma_A$ embedded in a four dimensional Lorentzian spacetime $\mathcal{M}_4,$ endowed with the metric $g_{MN}.$ Given the two unit vectors $n^{(i)}$ (with $i=1,2$) normal to $\gamma_A$ and orthogonal between them, the induced metric (the projector) on the surface is 
\begin{equation}
\label{time dependent projector}
h_{MN}=g_{MN}-\sum_{i=1}^{2}\epsilon_i \,n^{(i)}_M n^{(i)}_N
\end{equation}
where $\epsilon_i=g^{MN} n^{(i)}_M n^{(i)}_N$ is either $+1$ or $-1.$ The surface $\gamma_A$ is now a codimension two surface in the full spacetime $\mathcal{M}_4$ and we can compute its two extrinsic curvatures as
\begin{equation}
\label{extrinsic curvatures}
K^{(i)}_{MN}= \tensor{h}{_M^A}\tensor{h}{_N^B} \, \nabla_A n^{(i)}_B\,.
\end{equation}
We introduce an auxiliary conformally equivalent four dimensional space
$\widetilde{\mathcal{M}}_{4}$ given by $\mathcal{M}_4$ with the same boundary at $z=0$, but equipped 
with the metric $ \tilde{g}_{MN}$, 
which is asymptotically flat as $z\to 0$ and Weyl related to $g_{MN}$, i.e.
\begin{equation}
g_{MN}=e^{2\varphi}\,\tilde{g}_{MN}
\end{equation}
where $ \varphi $ is a function of the coordinates.  Within this framework, in \cite{Fonda:2015nma} 
the following identity was shown to hold for any surface (not necessarily anchored to  a curve on the boundary)
\bea
\label{identity time-dep}
0
&=&
\biggl[\,
\widetilde{\mathcal{D}}^{2}\varphi+\sum_{i=1}^{2}\epsilon_i \widetilde{N}^{(i)M} \tilde{n}^{(i)N}\Bigl(\widetilde{D}_M \widetilde{D}_N\varphi-\widetilde{D}_M\varphi \widetilde{D}_N\varphi\Bigr)-\widetilde{D}^2\varphi- \frac{1}{4}\sum_{i=1}^2\epsilon_i\bigl(\textrm{Tr} \widetilde{K}^{(i)}\bigr)^{2}\,
\biggr] \,d \tilde{\mathcal{A}}
\nonumber \\
& &
+\,\frac{1}{4}\sum_{i=1}^2\epsilon_i\bigl(\textrm{Tr} K^{(i)}\bigr)^{2}d\mathcal{A}
\eea
where the tilded  quantities are evaluated considering $\widetilde{\mathcal{M}}_4$ as embedding space, 
while for the untilded ones the embedding space is $\mathcal{M}_4$.
In particular $\textrm{Tr} {K}^{(i)}$ and $\textrm{Tr} \widetilde{K}^{(i)}$ are the  mean  curvatures of the surface computed in the two embedding spaces,  
while $d\mathcal{A}$ and $d\widetilde{\mathcal{A}}$ are the two area elements. 
The vectors $\tilde{n}^{(i)M}$ are versors perpendicular  to  the surface viewed  as a submanifold of  $\widetilde{\mathcal{M}}_{4}$. The covariant derivative $\widetilde{\nabla}$ is the one
defined in $\widetilde{\mathcal{M}}_4$ while $\widetilde{\mathcal{D}}$ is the one induced on the surface by the embedding space $\widetilde{\mathcal{M}}_{4}$.

At this point, let us consider the surfaces $\gamma_A$ anchored to some smooth entangling curve $\partial A$ and orthogonal to the boundary.
 Similarly to the static case considered in Section\,\ref{appendix_1-3}, we multiply \eqref{identity time-dep} by $c_1 e^{2\phi}$,  integrate over $\gamma_{A,\varepsilon}$ and add the regularized area function to both sides of \eqref{identity time-dep}. Thus, we obtain 
\bea
\label{area time-dep}
\mathcal{A}[\gamma_{A,\varepsilon}]
&= &
c_1\int_{\gamma_{A,\varepsilon} } \!\!\!
e^{2\phi}
\biggl[ \, \widetilde{\mathcal{D}}^{2}\varphi
+\sum_{i=1}^{2}\epsilon_i \tilde{n}^{(i)M} \tilde{n}^{(i)N}\Bigl(\widetilde{D}_M \widetilde{D}_N\varphi-\widetilde{D}_M\varphi \widetilde{D}_N\varphi\Bigr)-\widetilde{D}^2\varphi
\\ 
& &\hspace{1.8cm}
- \,\frac{1}{4}\sum_{i=1}^2\epsilon_i\bigl(\textrm{Tr} \widetilde{K}^{(i)}\bigr)^{2} \,\biggr] \,d \tilde{\mathcal{A}}
+\int_{\gamma_{A,\varepsilon}}e^{2\varphi}d\mathcal{\widetilde{A}}+\frac{c_1}{4}\sum_{i=1}^2\epsilon_i
\int_{\gamma_{A,\varepsilon}} \!\!\! e^{2\phi}\bigl(\textrm{Tr} K^{(i)}\bigr)^{2}d\mathcal{A}\,.
\nonumber
\eea
When we evaluate the first term in the r.h.s. of \eqref{area time-dep} over $\gamma_{A,\varepsilon}$ with the same procedure of the static case, 
it provides the divergent contribution to $\mathcal{A}[\gamma_{A,\varepsilon}] $.
Thus, the expansion \eqref{leadingdiver} is obtained, with $\mathcal{F}_A$ given by \eqref{Fa time-dep}. 

For non static geometries the holographic entanglement entropy of a region $A$  belonging to the asymptotic boundary of $\mathcal{M}_4$ can be computed by employing the prescription \cite{Hubeny:2007xt}. 
One has to  compute the area of the minimal surface $\hat{\gamma}_A$ anchored to the boundary of the region $A$. 
Since $\hat{\gamma}_A$ has codimension two,  
we have the following two extremality conditions 
\begin{equation}
\label{extremalcondition time-dep}
\textrm{Tr}K^{(i)}=0\qquad \Longleftrightarrow\qquad  \bigl(\textrm{Tr}\widetilde{K}^{(i)}\bigr)^2=4\bigl(\tilde{n}^{(i)M}\partial_M\varphi\bigr)^2\,.
\end{equation}
By specialising \eqref{Fa time-dep} to an extremal surface $\hat \gamma_A$, 
we find the expression \eqref{Fa time-dep extremal surf} for the finite term in the expansion of the area.

For scale invariant theories, where $d_\theta=2,$ 
the first term in \eqref{Fa time-dep extremal surf} vanishes because $\phi$ can be set to $0$;
hence the expression for $F_A$ reduces to \cite{Fonda:2015nma}
\begin{equation}
F_A= \int_{\hat\gamma_A } 
\biggl[\,\widetilde{D}^2\varphi-\sum_{i=1}^{2}\epsilon_i \tilde{n}^{(i)M} \tilde{n}^{(i)N}\widetilde{D}_M \widetilde{D}_N\varphi-e^{2\varphi} +\frac{1}{2}\sum_{i=1}^2\epsilon_i\bigl(\textrm{Tr} \widetilde{K}^{(i)}\bigr)^{2} \,\biggr] \,d \tilde{\mathcal{A}}\,.
\end{equation}

We shall now briefly discuss how to recover the result \eqref{finite term} for the static cases from \eqref{Fa time-dep extremal surf}.  
The most general static metric can be written as 
\begin{equation}
\label{staticmetric}
ds^2= - \,N^2 dt^2 + g_{\mu\nu} dx^\mu dx^\nu
\end{equation}
where $N$ and $g_{\mu\nu}$ are functions of the spatial coordinates $x^\mu=(z,\boldsymbol x)$ only. 
In this background metric, the two unit normal vectors can be written as $n^{(1)}_M=(N,0,\boldsymbol 0)$ and $n^{(2)}_M=(0,n_\mu)$.
With the choice of coordinates  \eqref{staticmetric}, the only non vanishing Christoffel symbols are 
\begin{equation}
\label{dec_cristoffel}
\Gamma^{t}_{\mu t}=\frac{1}{2 N^2}\partial_\mu N \qquad \Gamma^{\mu}_{tt}=\frac{1}{2} g^{\mu \nu}\partial_\nu N \qquad \Gamma^{\mu}_{\nu\rho}={}^{(3)}\Gamma^{\,\mu}_{\nu\rho}
\end{equation}
where ${}^{(3)}\Gamma^\mu_{\nu\rho}$ denotes the Christoffel computed with the three dimensional metric $g_{\mu\nu}$ of the constant time hypersurface. Combining \eqref{dec_cristoffel} with the observation that the time components  $h_{t M}$ of the projector \eqref{time dependent projector} vanish, 
we easily conclude  that the extrinsic curvature in the timelike direction $K^{(1)}_{MN}$ is zero.  
Thus, the first equation of motion in \eqref{extremalcondition time-dep} is identically satisfied.  
Instead the second equation of motion in \eqref{extremalcondition time-dep} reduces to  \eqref{extremalcondit} because only the spatial components of the  extrinsic curvature  $K^{(2)}_{MN}$   are non vanishing; hence $\text{Tr} K^{(2)}=\text{Tr} K$. 
Similar conclusions can be reached for the tilded quantities:  $\widetilde K_{MN}^{(1)}=0$, $\widetilde K^{(2)}_{\mu\nu}=\widetilde K_{\mu\nu}$ and  $\widetilde K^{(2)}_{tt}=0$, being $\varphi$ independent of $t$. 
Finally, due to \eqref{dec_cristoffel}, ${\tilde n^{(2)\,M} \tilde n^{(2)\,N}\widetilde D_{M}\widetilde D_{N}\varphi=\widetilde \nabla_M\widetilde \nabla_N \varphi}$, while the Laplacian $\widetilde D ^2 \varphi$ and the term $\tilde n^{(1)\,M}\tilde n^{(1)\,N} \widetilde D_M  \widetilde D_N  $ 
sum to $\widetilde \nabla^2 \varphi $.

	\section{On the analytic solution for a disk when $d_\theta =2$ and $\zeta \to \infty$}
\label{appendix_hard_wall_solution}

	In this appendix we analytically study minimal surfaces $\hat{\gamma}_A$ 
	anchored to circular regions $A$ in spacetimes equipped with the metric \eqref{BH_metric} 
	in the limit $\zeta \rightarrow +\infty$ and for $d_\theta=2$.
	The background metric becomes the AdS$_4$ metric for $z \leqslant z_h$ with an event horizon located at $z=z_h$. The only effect of the horizon is to forbid the minimal surface enters the region $z > z_h$. As discussed below, for regions large enough, the minimal surfaces reach and stick to the horizon sharing  a portion of surface with it.

	For small regions $A$, the minimal surfaces do not reach the horizon and their profile is the same as in AdS$_4$ case, i.e. it is given by the hemisphere: 
	$z(\rho)=\sqrt{R^{2}-\rho^{2}}$. This  occurs as long as the surface does not intersect the horizon, namely for $R<z_h$.  
	For  $R=z_h$ the hemisphere is tangent to the event horizon at the point $(z,\rho)=(z_h,0)$. 
	As the radius $R$ increases further, a certain portion of the dome   would cross the horizon;   
	hence  in this regime the hemispheres cannot be  the extremal surfaces.  
	The actual minimal surfaces consist of two parts: a flat disk that lies on the horizon 
	and a non trivial surface connecting the conformal boundary to the horizon. 
	The aim of the following discussion is to find analytically the latter one. 
	
	Let us consider
	the most general solution of the differential equation \eqref{eqmotosphere} for $d_\theta=2$. 
	Following \cite{Fonda:2014cca,Seminara:2018pmr} (see also \cite{Zarembo:1999bu, Olesen:2000ji, Drukker:2005cu, Dekel:2013kwa}), we replace $\rho$  with the variable $u$ and $ z(\rho)$ with the function $\hat z(u)$, defined as follows
	%
	\begin{equation}
	\label{variables}
	\rho=e^u\qquad \hat z(u)=\frac{z(\rho)}{\rho}={e^{-u}}{z(e^u)}\,. 
	\end{equation}
	%
	The minimality condition in AdS$_4$ gives \eqref{eqmotosphere} and for $d_\theta=2$ it becomes
	\begin{equation}
	\hat{z} \left(\hat{z}_u+\hat{z}_{uu}\right)
	+ \left[ 1+ (\hat z+\hat z_u)^2 \right] \left[ 2 + \hat z (\hat z+\hat z_u) \right]=0
	\end{equation}
	which can be integrated over $\hat z$ to yield
	\begin{equation}
	\label{first_int}
	\hat z_{u,\pm}= -\frac{1+\hat z^2}{\hat z}\left[ 1 \pm \frac{\hat z}{\sqrt{k(1+\hat z^2)-\hat z^4}}  \right]^{-1} \qquad k>0
	\end{equation}
	where $k$ is an integration constant.   
	The differential equation \eqref{first_int} can be integrated again, finding
	\begin{equation}
	\log \rho \,=\,
	 -\int \, \frac{\hat z}{1+\hat z^2} \left(  1 \pm \frac{\hat z}{\sqrt{k(1+\hat z^2)-\hat z^4}}   \right) d\hat z 
	 \,\equiv \,
	 -\,q(\hat z)_{\pm,k}+C
	\end{equation}
	where $C$ is a second  integration constant and 
	\begin{equation}
	\label{q_+-_explicit}
	\begin{split}
	q_{\pm,k} (\hat z) & \equiv \int_0^{\hat z} \, \frac{\lambda}{1+\lambda^2} 
	\left(  1 \pm \frac{\lambda}{\sqrt{k(1+\lambda^2)-\lambda^4}}   \right) d\lambda  \qquad  0 \leqslant \hat z < \hat z_m\,.
	\end{split}
	\end{equation}
	The parameter  $\hat z_m^2= (k+\sqrt{k(k+4)}\,)/2$  solves the polynomial under the square root in \eqref{q_+-_explicit}. 
	The integral \eqref{q_+-_explicit} can be performed explicitly obtaining \eqref{qpm}.
	
	The two integration constants $k$ and $C$ are determined through the boundary conditions. 
	In particular, $C$ can be fixed by imposing $\rho=R$  at $z=0$. 
	Since ${q_{\pm,k}(\hat z=0)=0}$, we  get $C=\log R$ and the profile reads
	\begin{equation}
	\label{profile}
	\rho =R \, e^{-q_{\pm,k}(\hat z)}
	\end{equation}
	where the plus/minus ambiguity will be fixed below.
	
	Let us denote by $P_*=(\rho_*,z_h)$ the intersection point between (\ref{profile}) and the horizon.
	For $\rho< \rho_*$,  the minimal  surface is a disk lying  exactly on the horizon.
	 The position of $P_*$ and the constant $k$ are then determined by requiring that the solution is continuous and differentiable at $P_*$.
	Since the tangent vector to the surface for $\rho\geqslant \rho_*$ is 
	$t_\rho^\mu =(t_\rho^\rho,t_\rho^z)=(\rho',\rho+\hat z  \rho')$,
	%
	%
	the condition of being tangent to the horizon reads $\rho+\hat z  \rho'=0$. 
	Being $\rho'=-\rho \,q'_{\pm,k}$, we obtain 
	$\hat z_* \,q'(\hat z_*)_{\pm,k}=1 $, that implies 
	$\pm \hat z_*^3= \sqrt{k(1+\hat z_*^2)-\hat z_*^4}$;
	and this is meaningful only if the plus sign is chosen in \eqref{profile}.
	This choice, in turn, gives $\hat z_*=k^{1/4}$. 
	Finally, the value of $k$ is evaluated by imposing that $z=z_h$ when $\hat z=\hat z_*$. 
	This leads to \eqref{Rvsk}  which implicitly determines $k$ in terms of $R/z_h$. 
	The possibility of inverting  \eqref{Rvsk} is controlled by its derivative  with respect to $k$. 
	We find
	\begin{equation}
	\frac{d}{dk}\left( \frac{R}{z_h} \right)= - \frac{R}{z_h} \int_0^{k^{1/4}} 
	\!\!\! \frac{\lambda^2}{2 \left[ k(1+\lambda^2)-\lambda^4 \right]^{3/2}}
	\, \leqslant \,0\,.
	\end{equation} 
	Since $R/z_h$ is a monotonic function of $k$,  the condition 
	\eqref{Rvsk} has at most one solution for any value of $R/z_h$. 
	On the other hand, 
	in Section\,\ref{limiting_regimes_har_wall} we show that 
	$R/z_h \rightarrow +\infty$ for $k \rightarrow 0$,
	while $R/z_h \rightarrow 1 $ for $k \rightarrow +\infty$. 
	Thus \eqref{Rvsk} admits exactly one solution in the range $R/z_h \in (1,+\infty)$
	which leads to the profile \eqref{profile_har_wall}.
	Instead, let us remind that in the range $R/z_h \in(0,1]$ the solution is the hemisphere $z(\rho)=\sqrt{R^{2}-\rho^{2}}$.

	\subsection{Area}
	As for the  area of the minimal surface $\hat{\gamma}_A$, 
	when $R<z_h$ it is the area of the hemisphere $z(\rho)=\sqrt{R^{2}-\rho^{2}}$ regularised by the condition $z\geqslant \varepsilon$,
	namely
	\begin{equation}
	\mathcal{A} =\frac{2\pi R}{\varepsilon}-2\pi \qquad R<z_h\,.
	\end{equation}
	For $R>z_h$, the area is $\mathcal{A}=\mathcal{A}_1 + \mathcal{A}_2$,
	where $\mathcal{A}_1 $ corresponds to a flat disk located at $z_h$ and with radius $\rho_*=z_h/\hat z_*=k^{1/4}/z_h$;
	hence it reads
	\begin{equation}
	\label{area_1part}
	\mathcal{A}_1 = \frac{\pi \rho_*^2}{z^2_h}=\frac{\pi}{\sqrt{k}}\,.
	\end{equation}
	The contribution $\mathcal{A}_2$ is the area of the profile \eqref{profile} between $\hat z=0$ and $\hat z_*= k^{1/4}$. 
	In terms of the variables introduced in \eqref{variables}, the area functional \eqref{area_func_BH_disk} in the limit $\zeta \rightarrow +\infty$ and for $d_\theta=2$ reduces to
	\begin{equation}
	\label{func_2part}
	\mathcal{A}_2
	=
	2\pi  \,\displaystyle\int_{\varepsilon/R}^{\hat z_*} \, 
	\frac{d\lambda}{\lambda^2 \sqrt{1+\lambda^2-\lambda^4/k}}
	\end{equation}
	where we introduced the UV cutoff $\varepsilon$. 
	The primitive $\mathcal{F}_k(\lambda)$ of the integrand in \eqref{func_2part} can be written explicitly  in terms of 
	elliptic integrals and it has been reported in \eqref{F_A int_evaluated}. 
	%
	%
	%
	In order to single out the UV divergence, one employs its expansion as $\lambda\rightarrow 0^+$
	\begin{equation}
	\mathcal{F}_k(\lambda)= \frac{1}{\lambda}+\frac{\lambda}{2}+ \mathcal{O}(\lambda^3)
	\end{equation}
	which gives
	\begin{equation}
	\label{area_2part}
	\mathcal{A}_2=\frac{2\pi R}{\varepsilon}- 2\pi \mathcal{F}_k(k^{1/4})+ \mathcal{O}(\varepsilon/R)
	\end{equation}
	where also $\hat z_*=k^{1/4}$ has been used. 
	By adding \eqref{area_2part} to \eqref{area_1part}, we find that the area of $\hat{\gamma}_A$ for $R>z_h$ reads
	\begin{equation}
	\label{areavsk}
	\mathcal{A}= \frac{2\pi R}{\varepsilon}-2\pi \left( \mathcal{F}_k(k^{1/4})-\frac{1}{2\sqrt{k}}  \right) \qquad R>z_h
	\end{equation}
	which provides \eqref{hardwall_FA}.
	\subsection{Limiting regimes}
	\label{limiting_regimes_har_wall}
	Let us consider the limit of \eqref{Rvsk} and \eqref{areavsk} for $R/z_h \rightarrow +\infty$, which corresponds to $k \rightarrow 0$. 
	The expansion of  \eqref{areavsk} is straightforward, and we find
	\begin{equation}
	\label{A_k_0}
	\mathcal{A}=
	\frac{2\pi R}{\varepsilon}
	-2\pi\left[
	-\frac{1 }{2\sqrt{k}}+\frac{\sqrt{2} \, \pi^{3/2}}{\Gamma(1 /4)^2  \sqrt[4]{k}}+\frac{1}{2} 
	\right]
	+\mathcal{O}\big(k^{1/4}\big)\,.
	\end{equation}
	In order to expand  \eqref{Rvsk} for small  $k$, we find  more convenient to use the integral representation \eqref{q_+-_explicit}. 
	First one performs the change of variable  $\lambda \rightarrow k^{1/4} \lambda$, 
	obtaining a definite integral between the two extrema in $\lambda=0$ and $\lambda=1$. 
	Then, we expand the integrand as $k \to 0$ and we integrate  term by term, finding 
	\begin{equation}
	\label{q_limit_hem}
	q_{+,k}(k^{1/4}) = \frac{\sqrt{2} \, \pi^{3/2}}{\Gamma(1 /4)^2} \,k^{1/4}+ \frac{\sqrt{k}}{2}+\dots
	\end{equation}
	that leads to
	\begin{eqnarray}
	\label{R_k_0}
	\frac{R}{z_h} &= & 
	\frac{1}{k^{1/4}}+\frac{\sqrt{2}\, \pi ^{3/2}}{\Gamma(1 /4)^2}
	+ 
	\left( \frac{ \pi ^{3}}{\Gamma(1/4)^4}+  \frac{1}{2} \right)  k^{1/4} +\dots\;. 
	\end{eqnarray}
	Now, by plugging  \eqref{R_k_0} into \eqref{A_k_0} we get
	\begin{equation}
	\label{pizzo}
	\mathcal{A}
	= \frac{2\pi R}{\varepsilon}+\left( \frac{\pi R^2}{z_h^2}+  \frac{4\pi \sqrt{2} \, \pi^{3/2} R}{\Gamma(1 /4)^2 \,z_h} \right)
	+ \mathcal{O}(1)
	\end{equation} 
	where the leading term in $R$ agrees with  \eqref{large_BH2}.  
	
	In the regime given by $k \rightarrow +\infty$,
	from the definition of $\hat z_m$ we have $\hat z_m \rightarrow +\infty$, 
	and therefore the surface reaches $\rho=0$. Moreover from  \eqref{q_+-_explicit} we obtain
	\begin{equation}
	\label{q_limit_large_K}
	q_{\pm,k}(\hat z)= \int^{\hat z}_0 \!\! \frac{\lambda}{1+\lambda^2} \, d\lambda = \frac{1}{2} \log(1+\hat z^2)  
	\end{equation}
	that gives the profile of the hemisphere $z(\rho)=\sqrt{R^{2}-\rho^{2}}$. 
	By means of \eqref{q_limit_large_K} we find that 
	$q_{+,k}(k^{1/4})=\log k^{1/4}+\dots$ as $k\rightarrow \infty$,
	which leads to $R /z_h \to 1$ in the same limit. 
	Notice that $R=z_h$ is the value of the radius corresponding to the transition between the two  minimal surfaces. 
	Since we showed that the solution reduces to the hemisphere with radius $R=z_h$ in this limit,
	we conclude that \eqref{areavsk} reduces to
	$\mathcal{A}\to 2\pi R / \varepsilon-2\pi $ as $k\rightarrow \infty$.
	In particular, this means that the function $F_A(R)$ given in \eqref{hardwall_FA} is continuous in $R$.


\newpage

\bibliographystyle{jhep}
\bibliography{bibliografia}

\providecommand{\href}[2]{#2}\begingroup\raggedright\begin{thebibliography}{100}

\bibitem{Amico:2007ag}
L.~Amico, R.~Fazio, A.~Osterloh and V.~Vedral, \emph{{Entanglement in many-body
  systems}}, \href{https://doi.org/10.1103/RevModPhys.80.517}{\emph{Rev. Mod.
  Phys.} {\bfseries 80} (2008) 517}
  [\href{https://arxiv.org/abs/quant-ph/0703044}{{\ttfamily
  quant-ph/0703044}}].

\bibitem{Eisert:2008ur}
J.~Eisert, M.~Cramer and M.~B. Plenio, \emph{{Area laws for the entanglement
  entropy - a review}},
  \href{https://doi.org/10.1103/RevModPhys.82.277}{\emph{Rev. Mod. Phys.}
  {\bfseries 82} (2010) 277} [\href{https://arxiv.org/abs/0808.3773}{{\ttfamily
  0808.3773}}].

\bibitem{specialissue}
P.~Calabrese, J.~Cardy and B.~Doyon, \emph{Entanglement entropy in extended
  quantum systems},
  \href{https://doi.org/10.1088/1751-8121/42/50/500301}{\emph{J.\ Phys.\ A {\bf
  42} (2009) special issue} }.

\bibitem{Solodukhin:2011gn}
S.~N. Solodukhin, \emph{{Entanglement entropy of black holes}},
  \href{https://doi.org/10.12942/lrr-2011-8}{\emph{Living Rev. Rel.} {\bfseries
  14} (2011) 8} [\href{https://arxiv.org/abs/1104.3712}{{\ttfamily
  1104.3712}}].

\bibitem{Rangamani:2016dms}
M.~Rangamani and T.~Takayanagi, \emph{{Holographic Entanglement Entropy}},
  \href{https://doi.org/10.1007/978-3-319-52573-0}{\emph{Lect. Notes Phys.}
  {\bfseries 931} (2017) pp.1}
  [\href{https://arxiv.org/abs/1609.01287}{{\ttfamily 1609.01287}}].

\bibitem{Islam1}
R.~Islam, R.~Ma, P.~Preiss, M.~Eric~Tai, A.~Lukin, M.~Rispoli et~al.,
  \emph{Measuring entanglement entropy in a quantum many-body system},
  \href{https://doi.org/10.1038/nature15750}{\emph{Nature} {\bfseries 528}
  (2015) 77}.

\bibitem{Kaufman794}
A.~M. Kaufman, M.~E. Tai, A.~Lukin, M.~Rispoli, R.~Schittko, P.~M. Preiss
  et~al., \emph{Quantum thermalization through entanglement in an isolated
  many-body system},
  \href{https://doi.org/10.1126/science.aaf6725}{\emph{Science} {\bfseries 353}
  (2016) 794}.

\bibitem{Lukin256}
A.~Lukin, M.~Rispoli, R.~Schittko, M.~E. Tai, A.~M. Kaufman, S.~Choi et~al.,
  \emph{Probing entanglement in a many-body{\textendash}localized system},
  \href{https://doi.org/10.1126/science.aau0818}{\emph{Science} {\bfseries 364}
  (2019) 256}.

\bibitem{Bombelli:1986rw}
L.~Bombelli, R.~K. Koul, J.~Lee and R.~D. Sorkin, \emph{{A Quantum Source of
  Entropy for Black Holes}},
  \href{https://doi.org/10.1103/PhysRevD.34.373}{\emph{Phys. Rev.} {\bfseries
  D34} (1986) 373}.

\bibitem{Srednicki:1993im}
M.~Srednicki, \emph{{Entropy and area}},
  \href{https://doi.org/10.1103/PhysRevLett.71.666}{\emph{Phys. Rev. Lett.}
  {\bfseries 71} (1993) 666}
  [\href{https://arxiv.org/abs/hep-th/9303048}{{\ttfamily hep-th/9303048}}].

\bibitem{Callan:1994py}
C.~G. Callan, Jr. and F.~Wilczek, \emph{{On geometric entropy}},
  \href{https://doi.org/10.1016/0370-2693(94)91007-3}{\emph{Phys. Lett.}
  {\bfseries B333} (1994) 55}
  [\href{https://arxiv.org/abs/hep-th/9401072}{{\ttfamily hep-th/9401072}}].

\bibitem{Holzhey:1994we}
C.~Holzhey, F.~Larsen and F.~Wilczek, \emph{{Geometric and renormalized entropy
  in conformal field theory}},
  \href{https://doi.org/10.1016/0550-3213(94)90402-2}{\emph{Nucl. Phys.}
  {\bfseries B424} (1994) 443}
  [\href{https://arxiv.org/abs/hep-th/9403108}{{\ttfamily hep-th/9403108}}].

\bibitem{Vidal:2002rm}
G.~Vidal, J.~I. Latorre, E.~Rico and A.~Kitaev, \emph{{Entanglement in quantum
  critical phenomena}},
  \href{https://doi.org/10.1103/PhysRevLett.90.227902}{\emph{Phys. Rev. Lett.}
  {\bfseries 90} (2003) 227902}
  [\href{https://arxiv.org/abs/quant-ph/0211074}{{\ttfamily
  quant-ph/0211074}}].

\bibitem{Calabrese:2004eu}
P.~Calabrese and J.~L. Cardy, \emph{{Entanglement entropy and quantum field
  theory}}, \href{https://doi.org/10.1088/1742-5468/2004/06/P06002}{\emph{J.
  Stat. Mech.} {\bfseries 0406} (2004) P06002}
  [\href{https://arxiv.org/abs/hep-th/0405152}{{\ttfamily hep-th/0405152}}].

\bibitem{Wolf:2006zzb}
M.~M. Wolf, \emph{{Violation of the entropic area law for Fermions}},
  \href{https://doi.org/10.1103/PhysRevLett.96.010404}{\emph{Phys. Rev. Lett.}
  {\bfseries 96} (2006) 010404}
  [\href{https://arxiv.org/abs/quant-ph/0503219}{{\ttfamily
  quant-ph/0503219}}].

\bibitem{Gioev:2006zz}
D.~Gioev and I.~Klich, \emph{{Entanglement Entropy of Fermions in Any Dimension
  and the Widom Conjecture}},
  \href{https://doi.org/10.1103/PhysRevLett.96.100503}{\emph{Phys. Rev. Lett.}
  {\bfseries 96} (2006) 100503}
  [\href{https://arxiv.org/abs/quant-ph/0504151}{{\ttfamily
  quant-ph/0504151}}].

\bibitem{PhysRevLett.35.1678}
R.~M. Hornreich, M.~Luban and S.~Shtrikman, \emph{Critical behavior at the
  onset of $\stackrel{\ensuremath{\rightarrow}}{\mathrm{k}}$-space instability
  on the $\ensuremath{\lambda}$ line},
  \href{https://doi.org/10.1103/PhysRevLett.35.1678}{\emph{Phys. Rev. Lett.}
  {\bfseries 35} (1975) 1678}.

\bibitem{PhysRevB.23.4615}
G.~Grinstein, \emph{Anisotropic sine-gordon model and infinite-order phase
  transitions in three dimensions},
  \href{https://doi.org/10.1103/PhysRevB.23.4615}{\emph{Phys. Rev. B}
  {\bfseries 23} (1981) 4615}.

\bibitem{PhysRevB.69.224415}
E.~Fradkin, D.~A. Huse, R.~Moessner, V.~Oganesyan and S.~L. Sondhi,
  \emph{Bipartite rokhsar--kivelson points and cantor deconfinement},
  \href{https://doi.org/10.1103/PhysRevB.69.224415}{\emph{Phys. Rev. B}
  {\bfseries 69} (2004) 224415}
  [\href{https://arxiv.org/abs/cond-mat/0311353}{{\ttfamily
  cond-mat/0311353}}].

\bibitem{PhysRevB.69.224416}
A.~Vishwanath, L.~Balents and T.~Senthil, \emph{Quantum criticality and
  deconfinement in phase transitions between valence bond solids},
  \href{https://doi.org/10.1103/PhysRevB.69.224416}{\emph{Phys. Rev. B}
  {\bfseries 69} (2004) 224416}
  [\href{https://arxiv.org/abs/cond-mat/0311085}{{\ttfamily
  cond-mat/0311085}}].

\bibitem{Ardonne:2003wa}
E.~Ardonne, P.~Fendley and E.~Fradkin, \emph{{Topological order and conformal
  quantum critical points}},
  \href{https://doi.org/10.1016/j.aop.2004.01.004}{\emph{Annals Phys.}
  {\bfseries 310} (2004) 493}
  [\href{https://arxiv.org/abs/cond-mat/0311466}{{\ttfamily
  cond-mat/0311466}}].

\bibitem{Fisher:1986zz}
D.~S. Fisher, \emph{{Scaling and critical slowing down in random-field Ising
  systems}}, \href{https://doi.org/10.1103/PhysRevLett.56.416}{\emph{Phys. Rev.
  Lett.} {\bfseries 56} (1986) 416}.

\bibitem{Maldacena:1997re}
J.~M. Maldacena, \emph{{The Large N limit of superconformal field theories and
  supergravity}}, \href{https://doi.org/10.1023/A:1026654312961,
  10.4310/ATMP.1998.v2.n2.a1}{\emph{Int. J. Theor. Phys.} {\bfseries 38} (1999)
  1113} [\href{https://arxiv.org/abs/hep-th/9711200}{{\ttfamily
  hep-th/9711200}}].

\bibitem{Witten:1998qj}
E.~Witten, \emph{{Anti-de Sitter space and holography}},
  \href{https://doi.org/10.4310/ATMP.1998.v2.n2.a2}{\emph{Adv. Theor. Math.
  Phys.} {\bfseries 2} (1998) 253}
  [\href{https://arxiv.org/abs/hep-th/9802150}{{\ttfamily hep-th/9802150}}].

\bibitem{Gubser:1998bc}
S.~S. Gubser, I.~R. Klebanov and A.~M. Polyakov, \emph{{Gauge theory
  correlators from noncritical string theory}},
  \href{https://doi.org/10.1016/S0370-2693(98)00377-3}{\emph{Phys. Lett.}
  {\bfseries B428} (1998) 105}
  [\href{https://arxiv.org/abs/hep-th/9802109}{{\ttfamily hep-th/9802109}}].

\bibitem{Aharony:1999ti}
O.~Aharony, S.~S. Gubser, J.~M. Maldacena, H.~Ooguri and Y.~Oz, \emph{{Large N
  field theories, string theory and gravity}},
  \href{https://doi.org/10.1016/S0370-1573(99)00083-6}{\emph{Phys. Rept.}
  {\bfseries 323} (2000) 183}
  [\href{https://arxiv.org/abs/hep-th/9905111}{{\ttfamily hep-th/9905111}}].

\bibitem{Kachru:2008yh}
S.~Kachru, X.~Liu and M.~Mulligan, \emph{{Gravity duals of Lifshitz-like fixed
  points}}, \href{https://doi.org/10.1103/PhysRevD.78.106005}{\emph{Phys. Rev.}
  {\bfseries D78} (2008) 106005}
  [\href{https://arxiv.org/abs/0808.1725}{{\ttfamily 0808.1725}}].

\bibitem{Balasubramanian:2008dm}
K.~Balasubramanian and J.~McGreevy, \emph{{Gravity duals for non-relativistic
  CFTs}}, \href{https://doi.org/10.1103/PhysRevLett.101.061601}{\emph{Phys.
  Rev. Lett.} {\bfseries 101} (2008) 061601}
  [\href{https://arxiv.org/abs/0804.4053}{{\ttfamily 0804.4053}}].

\bibitem{Taylor:2008tg}
M.~Taylor, \emph{{Non-relativistic holography}},
  \href{https://arxiv.org/abs/0812.0530}{{\ttfamily 0812.0530}}.

\bibitem{Charmousis:2010zz}
C.~Charmousis, B.~Gouteraux, B.~S. Kim, E.~Kiritsis and R.~Meyer,
  \emph{{Effective Holographic Theories for low-temperature condensed matter
  systems}}, \href{https://doi.org/10.1007/JHEP11(2010)151}{\emph{JHEP}
  {\bfseries 11} (2010) 151} [\href{https://arxiv.org/abs/1005.4690}{{\ttfamily
  1005.4690}}].

\bibitem{Gouteraux:2011ce}
B.~Gouteraux and E.~Kiritsis, \emph{{Generalized Holographic Quantum
  Criticality at Finite Density}},
  \href{https://doi.org/10.1007/JHEP12(2011)036}{\emph{JHEP} {\bfseries 12}
  (2011) 036} [\href{https://arxiv.org/abs/1107.2116}{{\ttfamily 1107.2116}}].

\bibitem{Huijse:2011ef}
L.~Huijse, S.~Sachdev and B.~Swingle, \emph{{Hidden Fermi surfaces in
  compressible states of gauge-gravity duality}},
  \href{https://doi.org/10.1103/PhysRevB.85.035121}{\emph{Phys. Rev.}
  {\bfseries B85} (2012) 035121}
  [\href{https://arxiv.org/abs/1112.0573}{{\ttfamily 1112.0573}}].

\bibitem{Ogawa:2011bz}
N.~Ogawa, T.~Takayanagi and T.~Ugajin, \emph{{Holographic Fermi Surfaces and
  Entanglement Entropy}},
  \href{https://doi.org/10.1007/JHEP01(2012)125}{\emph{JHEP} {\bfseries 01}
  (2012) 125} [\href{https://arxiv.org/abs/1111.1023}{{\ttfamily 1111.1023}}].

\bibitem{Dong:2012se}
X.~Dong, S.~Harrison, S.~Kachru, G.~Torroba and H.~Wang, \emph{{Aspects of
  holography for theories with hyperscaling violation}},
  \href{https://doi.org/10.1007/JHEP06(2012)041}{\emph{JHEP} {\bfseries 06}
  (2012) 041} [\href{https://arxiv.org/abs/1201.1905}{{\ttfamily 1201.1905}}].

\bibitem{Ryu:2006bv}
S.~Ryu and T.~Takayanagi, \emph{{Holographic derivation of entanglement entropy
  from AdS/CFT}},
  \href{https://doi.org/10.1103/PhysRevLett.96.181602}{\emph{Phys. Rev. Lett.}
  {\bfseries 96} (2006) 181602}
  [\href{https://arxiv.org/abs/hep-th/0603001}{{\ttfamily hep-th/0603001}}].

\bibitem{Ryu:2006ef}
S.~Ryu and T.~Takayanagi, \emph{{Aspects of Holographic Entanglement Entropy}},
  \href{https://doi.org/10.1088/1126-6708/2006/08/045}{\emph{JHEP} {\bfseries
  08} (2006) 045} [\href{https://arxiv.org/abs/hep-th/0605073}{{\ttfamily
  hep-th/0605073}}].

\bibitem{Hubeny:2007xt}
V.~E. Hubeny, M.~Rangamani and T.~Takayanagi, \emph{{A Covariant holographic
  entanglement entropy proposal}},
  \href{https://doi.org/10.1088/1126-6708/2007/07/062}{\emph{JHEP} {\bfseries
  07} (2007) 062} [\href{https://arxiv.org/abs/0705.0016}{{\ttfamily
  0705.0016}}].

\bibitem{Headrick:2007km}
M.~Headrick and T.~Takayanagi, \emph{{A Holographic proof of the strong
  subadditivity of entanglement entropy}},
  \href{https://doi.org/10.1103/PhysRevD.76.106013}{\emph{Phys. Rev.}
  {\bfseries D76} (2007) 106013}
  [\href{https://arxiv.org/abs/0704.3719}{{\ttfamily 0704.3719}}].

\bibitem{Wall:2012uf}
A.~C. Wall, \emph{{Maximin Surfaces, and the Strong Subadditivity of the
  Covariant Holographic Entanglement Entropy}},
  \href{https://doi.org/10.1088/0264-9381/31/22/225007}{\emph{Class. Quant.
  Grav.} {\bfseries 31} (2014) 225007}
  [\href{https://arxiv.org/abs/1211.3494}{{\ttfamily 1211.3494}}].

\bibitem{AbajoArrastia:2010yt}
J.~Abajo-Arrastia, J.~Aparicio and E.~Lopez, \emph{{Holographic Evolution of
  Entanglement Entropy}},
  \href{https://doi.org/10.1007/JHEP11(2010)149}{\emph{JHEP} {\bfseries 11}
  (2010) 149} [\href{https://arxiv.org/abs/1006.4090}{{\ttfamily 1006.4090}}].

\bibitem{Balasubramanian:2011ur}
V.~Balasubramanian, A.~Bernamonti, J.~de~Boer, N.~Copland, B.~Craps,
  E.~Keski-Vakkuri et~al., \emph{{Holographic Thermalization}},
  \href{https://doi.org/10.1103/PhysRevD.84.026010}{\emph{Phys. Rev.}
  {\bfseries D84} (2011) 026010}
  [\href{https://arxiv.org/abs/1103.2683}{{\ttfamily 1103.2683}}].

\bibitem{Balasubramanian:2011at}
V.~Balasubramanian, A.~Bernamonti, N.~Copland, B.~Craps and F.~Galli,
  \emph{{Thermalization of mutual and tripartite information in strongly
  coupled two dimensional conformal field theories}},
  \href{https://doi.org/10.1103/PhysRevD.84.105017}{\emph{Phys. Rev.}
  {\bfseries D84} (2011) 105017}
  [\href{https://arxiv.org/abs/1110.0488}{{\ttfamily 1110.0488}}].

\bibitem{Allais:2011ys}
A.~Allais and E.~Tonni, \emph{{Holographic evolution of the mutual
  information}}, \href{https://doi.org/10.1007/JHEP01(2012)102}{\emph{JHEP}
  {\bfseries 01} (2012) 102} [\href{https://arxiv.org/abs/1110.1607}{{\ttfamily
  1110.1607}}].

\bibitem{Callan:2012ip}
R.~Callan, J.-Y. He and M.~Headrick, \emph{{Strong subadditivity and the
  covariant holographic entanglement entropy formula}},
  \href{https://doi.org/10.1007/JHEP06(2012)081}{\emph{JHEP} {\bfseries 06}
  (2012) 081} [\href{https://arxiv.org/abs/1204.2309}{{\ttfamily 1204.2309}}].

\bibitem{Liu:2013iza}
H.~Liu and S.~J. Suh, \emph{{Entanglement Tsunami: Universal Scaling in
  Holographic Thermalization}},
  \href{https://doi.org/10.1103/PhysRevLett.112.011601}{\emph{Phys. Rev. Lett.}
  {\bfseries 112} (2014) 011601}
  [\href{https://arxiv.org/abs/1305.7244}{{\ttfamily 1305.7244}}].

\bibitem{Liu:2013qca}
H.~Liu and S.~J. Suh, \emph{{Entanglement growth during thermalization in
  holographic systems}},
  \href{https://doi.org/10.1103/PhysRevD.89.066012}{\emph{Phys. Rev.}
  {\bfseries D89} (2014) 066012}
  [\href{https://arxiv.org/abs/1311.1200}{{\ttfamily 1311.1200}}].

\bibitem{Hubeny:2013hz}
V.~E. Hubeny, M.~Rangamani and E.~Tonni, \emph{{Thermalization of Causal
  Holographic Information}},
  \href{https://doi.org/10.1007/JHEP05(2013)136}{\emph{JHEP} {\bfseries 05}
  (2013) 136} [\href{https://arxiv.org/abs/1302.0853}{{\ttfamily 1302.0853}}].

\bibitem{Hayden:2011ag}
P.~Hayden, M.~Headrick and A.~Maloney, \emph{{Holographic Mutual Information is
  Monogamous}}, \href{https://doi.org/10.1103/PhysRevD.87.046003}{\emph{Phys.
  Rev.} {\bfseries D87} (2013) 046003}
  [\href{https://arxiv.org/abs/1107.2940}{{\ttfamily 1107.2940}}].

\bibitem{Hubeny:2013gta}
V.~E. Hubeny, H.~Maxfield, M.~Rangamani and E.~Tonni, \emph{{Holographic
  entanglement plateaux}},
  \href{https://doi.org/10.1007/JHEP08(2013)092}{\emph{JHEP} {\bfseries 08}
  (2013) 092} [\href{https://arxiv.org/abs/1306.4004}{{\ttfamily 1306.4004}}].

\bibitem{Hubeny:2013gba}
V.~E. Hubeny, M.~Rangamani and E.~Tonni, \emph{{Global properties of causal
  wedges in asymptotically AdS spacetimes}},
  \href{https://doi.org/10.1007/JHEP10(2013)059}{\emph{JHEP} {\bfseries 10}
  (2013) 059} [\href{https://arxiv.org/abs/1306.4324}{{\ttfamily 1306.4324}}].

\bibitem{Headrick:2014cta}
M.~Headrick, V.~E. Hubeny, A.~Lawrence and M.~Rangamani, \emph{{Causality \&
  holographic entanglement entropy}},
  \href{https://doi.org/10.1007/JHEP12(2014)162}{\emph{JHEP} {\bfseries 12}
  (2014) 162} [\href{https://arxiv.org/abs/1408.6300}{{\ttfamily 1408.6300}}].

\bibitem{Hubeny:2007re}
V.~E. Hubeny and M.~Rangamani, \emph{{Holographic entanglement entropy for
  disconnected regions}},
  \href{https://doi.org/10.1088/1126-6708/2008/03/006}{\emph{JHEP} {\bfseries
  03} (2008) 006} [\href{https://arxiv.org/abs/0711.4118}{{\ttfamily
  0711.4118}}].

\bibitem{Tonni:2010pv}
E.~Tonni, \emph{{Holographic entanglement entropy: near horizon geometry and
  disconnected regions}},
  \href{https://doi.org/10.1007/JHEP05(2011)004}{\emph{JHEP} {\bfseries 05}
  (2011) 004} [\href{https://arxiv.org/abs/1011.0166}{{\ttfamily 1011.0166}}].

\bibitem{Headrick:2010zt}
M.~Headrick, \emph{{Entanglement Renyi entropies in holographic theories}},
  \href{https://doi.org/10.1103/PhysRevD.82.126010}{\emph{Phys. Rev.}
  {\bfseries D82} (2010) 126010}
  [\href{https://arxiv.org/abs/1006.0047}{{\ttfamily 1006.0047}}].

\bibitem{Faulkner:2013ana}
T.~Faulkner, A.~Lewkowycz and J.~Maldacena, \emph{{Quantum corrections to
  holographic entanglement entropy}},
  \href{https://doi.org/10.1007/JHEP11(2013)074}{\emph{JHEP} {\bfseries 11}
  (2013) 074} [\href{https://arxiv.org/abs/1307.2892}{{\ttfamily 1307.2892}}].

\bibitem{Calabrese:2009ez}
P.~Calabrese, J.~Cardy and E.~Tonni, \emph{{Entanglement entropy of two
  disjoint intervals in conformal field theory}},
  \href{https://doi.org/10.1088/1742-5468/2009/11/P11001}{\emph{J. Stat. Mech.}
  {\bfseries 0911} (2009) P11001}
  [\href{https://arxiv.org/abs/0905.2069}{{\ttfamily 0905.2069}}].

\bibitem{Calabrese:2010he}
P.~Calabrese, J.~Cardy and E.~Tonni, \emph{{Entanglement entropy of two
  disjoint intervals in conformal field theory II}},
  \href{https://doi.org/10.1088/1742-5468/2011/01/P01021}{\emph{J. Stat. Mech.}
  {\bfseries 1101} (2011) P01021}
  [\href{https://arxiv.org/abs/1011.5482}{{\ttfamily 1011.5482}}].

\bibitem{Coser:2013qda}
A.~Coser, L.~Tagliacozzo and E.~Tonni, \emph{{On Rényi entropies of disjoint
  intervals in conformal field theory}},
  \href{https://doi.org/10.1088/1742-5468/2014/01/P01008}{\emph{J. Stat. Mech.}
  {\bfseries 1401} (2014) P01008}
  [\href{https://arxiv.org/abs/1309.2189}{{\ttfamily 1309.2189}}].

\bibitem{DeNobili:2015dla}
C.~De~Nobili, A.~Coser and E.~Tonni, \emph{{Entanglement entropy and negativity
  of disjoint intervals in CFT: Some numerical extrapolations}},
  \href{https://doi.org/10.1088/1742-5468/2015/06/P06021}{\emph{J. Stat. Mech.}
  {\bfseries 1506} (2015) P06021}
  [\href{https://arxiv.org/abs/1501.04311}{{\ttfamily 1501.04311}}].

\bibitem{Coser:2015dvp}
A.~Coser, E.~Tonni and P.~Calabrese, \emph{{Spin structures and entanglement of
  two disjoint intervals in conformal field theories}},
  \href{https://doi.org/10.1088/1742-5468/2016/05/053109}{\emph{J. Stat. Mech.}
  {\bfseries 1605} (2016) 053109}
  [\href{https://arxiv.org/abs/1511.08328}{{\ttfamily 1511.08328}}].

\bibitem{Freedman:2016zud}
M.~Freedman and M.~Headrick, \emph{{Bit threads and holographic entanglement}},
  \href{https://doi.org/10.1007/s00220-016-2796-3}{\emph{Commun. Math. Phys.}
  {\bfseries 352} (2017) 407}
  [\href{https://arxiv.org/abs/1604.00354}{{\ttfamily 1604.00354}}].

\bibitem{Headrick:2017ucz}
M.~Headrick and V.~E. Hubeny, \emph{{Riemannian and Lorentzian flow-cut
  theorems}}, \href{https://doi.org/10.1088/1361-6382/aab83c}{\emph{Class.
  Quant. Grav.} {\bfseries 35} (2018) 10}
  [\href{https://arxiv.org/abs/1710.09516}{{\ttfamily 1710.09516}}].

\bibitem{Cui:2018dyq}
S.~X. Cui, P.~Hayden, T.~He, M.~Headrick, B.~Stoica and M.~Walter, \emph{{Bit
  Threads and Holographic Monogamy}},
  \href{https://arxiv.org/abs/1808.05234}{{\ttfamily 1808.05234}}.

\bibitem{Graham:1999pm}
C.~R. Graham and E.~Witten, \emph{{Conformal anomaly of submanifold observables
  in AdS/CFT correspondence}},
  \href{https://doi.org/10.1016/S0550-3213(99)00055-3}{\emph{Nucl. Phys.}
  {\bfseries B546} (1999) 52}
  [\href{https://arxiv.org/abs/hep-th/9901021}{{\ttfamily hep-th/9901021}}].

\bibitem{Solodukhin:2008dh}
S.~N. Solodukhin, \emph{{Entanglement entropy, conformal invariance and
  extrinsic geometry}},
  \href{https://doi.org/10.1016/j.physletb.2008.05.071}{\emph{Phys. Lett.}
  {\bfseries B665} (2008) 305}
  [\href{https://arxiv.org/abs/0802.3117}{{\ttfamily 0802.3117}}].

\bibitem{Hubeny:2012ry}
V.~E. Hubeny, \emph{{Extremal surfaces as bulk probes in AdS/CFT}},
  \href{https://doi.org/10.1007/JHEP07(2012)093}{\emph{JHEP} {\bfseries 07}
  (2012) 093} [\href{https://arxiv.org/abs/1203.1044}{{\ttfamily 1203.1044}}].

\bibitem{Astaneh:2014uba}
A.~F. Astaneh, G.~Gibbons and S.~N. Solodukhin, \emph{{What surface maximizes
  entanglement entropy?}},
  \href{https://doi.org/10.1103/PhysRevD.90.085021}{\emph{Phys. Rev.}
  {\bfseries D90} (2014) 085021}
  [\href{https://arxiv.org/abs/1407.4719}{{\ttfamily 1407.4719}}].

\bibitem{Babich}
M.~Babich and A.~Bobenko, \emph{Willmore tori with umbilic lines and minimal
  surfaces in hyperbolic space},
  \href{https://doi.org/10.1215/S0012-7094-93-07207-9}{\emph{Duke Mathematical
  Journal} {\bfseries 72} (1993) }.

\bibitem{Mazzeo}
S.~Alexakis and R.~Mazzeo, \emph{{Renormalized area and properly embedded
  minimal surfaces in hyperbolic 3-manifolds}},
  \href{https://doi.org/10.1007/s00220-010-1054-3}{\emph{Commun. Math. Phys.}
  {\bfseries 297} (2010) 621}.

\bibitem{Fonda:2015nma}
P.~Fonda, D.~Seminara and E.~Tonni, \emph{{On shape dependence of holographic
  entanglement entropy in AdS$_{4}$/CFT$_{3}$}},
  \href{https://doi.org/10.1007/JHEP12(2015)037}{\emph{JHEP} {\bfseries 12}
  (2015) 037} [\href{https://arxiv.org/abs/1510.03664}{{\ttfamily
  1510.03664}}].

\bibitem{Fonda:2014cca}
P.~Fonda, L.~Giomi, A.~Salvio and E.~Tonni, \emph{{On shape dependence of
  holographic mutual information in AdS$_{4}$}},
  \href{https://doi.org/10.1007/JHEP02(2015)005}{\emph{JHEP} {\bfseries 02}
  (2015) 005} [\href{https://arxiv.org/abs/1411.3608}{{\ttfamily 1411.3608}}].

\bibitem{brakke}
K.~A. Brakke, \emph{The surface evolver},
  \href{https://doi.org/10.1080/10586458.1992.10504253}{\emph{Experimental
  Mathematics} {\bfseries 1} (1992) 141}.

\bibitem{brakke2}
``\textit{Surface Evolver}.''
  \url{http://www.susqu.edu/brakke/evolver/evolver.html}.

\bibitem{Cardy:1986gw}
J.~L. Cardy, \emph{{Effect of Boundary Conditions on the Operator Content of
  Two-Dimensional Conformally Invariant Theories}},
  \href{https://doi.org/10.1016/0550-3213(86)90596-1}{\emph{Nucl. Phys.}
  {\bfseries B275} (1986) 200}.

\bibitem{Cardy:1989ir}
J.~L. Cardy, \emph{{Boundary Conditions, Fusion Rules and the Verlinde
  Formula}}, \href{https://doi.org/10.1016/0550-3213(89)90521-X}{\emph{Nucl.
  Phys.} {\bfseries B324} (1989) 581}.

\bibitem{Cardy:2004hm}
J.~L. Cardy, \emph{{Boundary conformal field theory}},
  \href{https://arxiv.org/abs/hep-th/0411189}{{\ttfamily hep-th/0411189}}.

\bibitem{Takayanagi:2011zk}
T.~Takayanagi, \emph{{Holographic Dual of BCFT}},
  \href{https://doi.org/10.1103/PhysRevLett.107.101602}{\emph{Phys. Rev. Lett.}
  {\bfseries 107} (2011) 101602}
  [\href{https://arxiv.org/abs/1105.5165}{{\ttfamily 1105.5165}}].

\bibitem{Fujita:2011fp}
M.~Fujita, T.~Takayanagi and E.~Tonni, \emph{{Aspects of AdS/BCFT}},
  \href{https://doi.org/10.1007/JHEP11(2011)043}{\emph{JHEP} {\bfseries 11}
  (2011) 043} [\href{https://arxiv.org/abs/1108.5152}{{\ttfamily 1108.5152}}].

\bibitem{Nozaki:2012qd}
M.~Nozaki, T.~Takayanagi and T.~Ugajin, \emph{{Central Charges for BCFTs and
  Holography}}, \href{https://doi.org/10.1007/JHEP06(2012)066}{\emph{JHEP}
  {\bfseries 06} (2012) 066} [\href{https://arxiv.org/abs/1205.1573}{{\ttfamily
  1205.1573}}].

\bibitem{Seminara:2017hhh}
D.~Seminara, J.~Sisti and E.~Tonni, \emph{{Corner contributions to holographic
  entanglement entropy in AdS$_{4}$/BCFT$_{3}$}},
  \href{https://doi.org/10.1007/JHEP11(2017)076}{\emph{JHEP} {\bfseries 11}
  (2017) 076} [\href{https://arxiv.org/abs/1708.05080}{{\ttfamily
  1708.05080}}].

\bibitem{Seminara:2018pmr}
D.~Seminara, J.~Sisti and E.~Tonni, \emph{{Holographic entanglement entropy in
  AdS$_{4}$/BCFT$_{3}$ and the Willmore functional}},
  \href{https://doi.org/10.1007/JHEP08(2018)164}{\emph{JHEP} {\bfseries 08}
  (2018) 164} [\href{https://arxiv.org/abs/1805.11551}{{\ttfamily
  1805.11551}}].

\bibitem{Goldstein:2009cv}
K.~Goldstein, S.~Kachru, S.~Prakash and S.~P. Trivedi, \emph{{Holography of
  Charged Dilaton Black Holes}},
  \href{https://doi.org/10.1007/JHEP08(2010)078}{\emph{JHEP} {\bfseries 08}
  (2010) 078} [\href{https://arxiv.org/abs/0911.3586}{{\ttfamily 0911.3586}}].

\bibitem{Gubser:2009qt}
S.~S. Gubser and F.~D. Rocha, \emph{{Peculiar properties of a charged dilatonic
  black hole in $AdS_5$}},
  \href{https://doi.org/10.1103/PhysRevD.81.046001}{\emph{Phys. Rev.}
  {\bfseries D81} (2010) 046001}
  [\href{https://arxiv.org/abs/0911.2898}{{\ttfamily 0911.2898}}].

\bibitem{Iizuka:2011hg}
N.~Iizuka, N.~Kundu, P.~Narayan and S.~P. Trivedi, \emph{{Holographic Fermi and
  Non-Fermi Liquids with Transitions in Dilaton Gravity}},
  \href{https://doi.org/10.1007/JHEP01(2012)094}{\emph{JHEP} {\bfseries 01}
  (2012) 094} [\href{https://arxiv.org/abs/1105.1162}{{\ttfamily 1105.1162}}].

\bibitem{Narayan:2012hk}
K.~Narayan, \emph{{On Lifshitz scaling and hyperscaling violation in string
  theory}}, \href{https://doi.org/10.1103/PhysRevD.85.106006}{\emph{Phys. Rev.}
  {\bfseries D85} (2012) 106006}
  [\href{https://arxiv.org/abs/1202.5935}{{\ttfamily 1202.5935}}].

\bibitem{Ammon:2012je}
M.~Ammon, M.~Kaminski and A.~Karch, \emph{{Hyperscaling-Violation on Probe
  D-Branes}}, \href{https://doi.org/10.1007/JHEP11(2012)028}{\emph{JHEP}
  {\bfseries 11} (2012) 028} [\href{https://arxiv.org/abs/1207.1726}{{\ttfamily
  1207.1726}}].

\bibitem{Bhattacharya:2012zu}
J.~Bhattacharya, S.~Cremonini and A.~Sinkovics, \emph{{On the IR completion of
  geometries with hyperscaling violation}},
  \href{https://doi.org/10.1007/JHEP02(2013)147}{\emph{JHEP} {\bfseries 02}
  (2013) 147} [\href{https://arxiv.org/abs/1208.1752}{{\ttfamily 1208.1752}}].

\bibitem{Alishahiha:2012qu}
M.~Alishahiha, E.~O~Colgain and H.~Yavartanoo, \emph{{Charged Black Branes with
  Hyperscaling Violating Factor}},
  \href{https://doi.org/10.1007/JHEP11(2012)137}{\emph{JHEP} {\bfseries 11}
  (2012) 137} [\href{https://arxiv.org/abs/1209.3946}{{\ttfamily 1209.3946}}].

\bibitem{Bueno:2012sd}
P.~Bueno, W.~Chemissany, P.~Meessen, T.~Ortin and C.~S. Shahbazi,
  \emph{{Lifshitz-like Solutions with Hyperscaling Violation in Ungauged
  Supergravity}}, \href{https://doi.org/10.1007/JHEP01(2013)189}{\emph{JHEP}
  {\bfseries 01} (2013) 189} [\href{https://arxiv.org/abs/1209.4047}{{\ttfamily
  1209.4047}}].

\bibitem{Gath:2012pg}
J.~Gath, J.~Hartong, R.~Monteiro and N.~A. Obers, \emph{{Holographic Models for
  Theories with Hyperscaling Violation}},
  \href{https://doi.org/10.1007/JHEP04(2013)159}{\emph{JHEP} {\bfseries 04}
  (2013) 159} [\href{https://arxiv.org/abs/1212.3263}{{\ttfamily 1212.3263}}].

\bibitem{Gouteraux:2012yr}
B.~Gouteraux and E.~Kiritsis, \emph{{Quantum critical lines in holographic
  phases with (un)broken symmetry}},
  \href{https://doi.org/10.1007/JHEP04(2013)053}{\emph{JHEP} {\bfseries 04}
  (2013) 053} [\href{https://arxiv.org/abs/1212.2625}{{\ttfamily 1212.2625}}].

\bibitem{Christensen:2013rfa}
M.~H. Christensen, J.~Hartong, N.~A. Obers and B.~Rollier, \emph{{Boundary
  Stress-Energy Tensor and Newton-Cartan Geometry in Lifshitz Holography}},
  \href{https://doi.org/10.1007/JHEP01(2014)057}{\emph{JHEP} {\bfseries 01}
  (2014) 057} [\href{https://arxiv.org/abs/1311.6471}{{\ttfamily 1311.6471}}].

\bibitem{Christensen:2013lma}
M.~H. Christensen, J.~Hartong, N.~A. Obers and B.~Rollier, \emph{{Torsional
  Newton-Cartan Geometry and Lifshitz Holography}},
  \href{https://doi.org/10.1103/PhysRevD.89.061901}{\emph{Phys. Rev.}
  {\bfseries D89} (2014) 061901}
  [\href{https://arxiv.org/abs/1311.4794}{{\ttfamily 1311.4794}}].

\bibitem{Hartong:2016nyx}
J.~Hartong, N.~A. Obers and M.~Sanchioni, \emph{{Lifshitz Hydrodynamics from
  Lifshitz Black Branes with Linear Momentum}},
  \href{https://doi.org/10.1007/JHEP10(2016)120}{\emph{JHEP} {\bfseries 10}
  (2016) 120} [\href{https://arxiv.org/abs/1606.09543}{{\ttfamily
  1606.09543}}].

\bibitem{Shaghoulian:2011aa}
E.~Shaghoulian, \emph{{Holographic Entanglement Entropy and Fermi Surfaces}},
  \href{https://doi.org/10.1007/JHEP05(2012)065}{\emph{JHEP} {\bfseries 05}
  (2012) 065} [\href{https://arxiv.org/abs/1112.2702}{{\ttfamily 1112.2702}}].

\bibitem{Alishahiha:2015goa}
M.~Alishahiha, A.~F. Astaneh, P.~Fonda and F.~Omidi, \emph{{Entanglement
  Entropy for Singular Surfaces in Hyperscaling violating Theories}},
  \href{https://doi.org/10.1007/JHEP09(2015)172}{\emph{JHEP} {\bfseries 09}
  (2015) 172} [\href{https://arxiv.org/abs/1507.05897}{{\ttfamily
  1507.05897}}].

\bibitem{Keranen:2011xs}
V.~Keranen, E.~Keski-Vakkuri and L.~Thorlacius, \emph{{Thermalization and
  entanglement following a non-relativistic holographic quench}},
  \href{https://doi.org/10.1103/PhysRevD.85.026005}{\emph{Phys. Rev.}
  {\bfseries D85} (2012) 026005}
  [\href{https://arxiv.org/abs/1110.5035}{{\ttfamily 1110.5035}}].

\bibitem{Alishahiha:2014cwa}
M.~Alishahiha, A.~Faraji~Astaneh and M.~R. Mohammadi~Mozaffar,
  \emph{{Thermalization in backgrounds with hyperscaling violating factor}},
  \href{https://doi.org/10.1103/PhysRevD.90.046004}{\emph{Phys. Rev.}
  {\bfseries D90} (2014) 046004}
  [\href{https://arxiv.org/abs/1401.2807}{{\ttfamily 1401.2807}}].

\bibitem{Fonda:2014ula}
P.~Fonda, L.~Franti, V.~Keränen, E.~Keski-Vakkuri, L.~Thorlacius and E.~Tonni,
  \emph{{Holographic thermalization with Lifshitz scaling and hyperscaling
  violation}}, \href{https://doi.org/10.1007/JHEP08(2014)051}{\emph{JHEP}
  {\bfseries 08} (2014) 051} [\href{https://arxiv.org/abs/1401.6088}{{\ttfamily
  1401.6088}}].

\bibitem{Ogawa:1998pm}
N.~Ogawa, \emph{{A Note on the scale symmetry and Noether current}},
  \href{https://arxiv.org/abs/hep-th/9807086}{{\ttfamily hep-th/9807086}}.

\bibitem{Zarembo:1999bu}
K.~Zarembo, \emph{{Wilson loop correlator in the AdS/CFT correspondence}},
  \href{https://doi.org/10.1016/S0370-2693(99)00717-0}{\emph{Phys. Lett.}
  {\bfseries B459} (1999) 527}
  [\href{https://arxiv.org/abs/hep-th/9904149}{{\ttfamily hep-th/9904149}}].

\bibitem{Olesen:2000ji}
P.~Olesen and K.~Zarembo, \emph{{Phase transition in Wilson loop correlator
  from AdS/CFT correspondence}},
  \href{https://arxiv.org/abs/hep-th/0009210}{{\ttfamily hep-th/0009210}}.

\bibitem{Drukker:2005cu}
N.~Drukker and B.~Fiol, \emph{{On the integrability of Wilson loops in $AdS_5
  \times S^5$: Some periodic ansatze}},
  \href{https://doi.org/10.1088/1126-6708/2006/01/056}{\emph{JHEP} {\bfseries
  01} (2006) 056} [\href{https://arxiv.org/abs/hep-th/0506058}{{\ttfamily
  hep-th/0506058}}].

\bibitem{Dekel:2013kwa}
A.~Dekel and T.~Klose, \emph{{Correlation Function of Circular Wilson Loops at
  Strong Coupling}}, \href{https://doi.org/10.1007/JHEP11(2013)117}{\emph{JHEP}
  {\bfseries 11} (2013) 117} [\href{https://arxiv.org/abs/1309.3203}{{\ttfamily
  1309.3203}}].

\end{thebibliography}\endgroup

\end{document}